\documentclass[superscriptaddress,nofootinbib,aps,prd,11pt]{revtex4-1}

\usepackage{graphicx,amsmath,array,verbatim,setspace,mathtools}
\usepackage{hyperref}
\usepackage{url}
\usepackage{amsfonts}
\usepackage[italicdiff]{physics}
\usepackage{color}
\usepackage{amssymb,amsthm}
\usepackage{here}
\usepackage{tikz-cd}
\usepackage{silence}
\allowdisplaybreaks

\newcommand{\bul}{\overset{\underset{\bullet}{}}}
\renewcommand{\arraystretch}{1}
\makeatletter
\@addtoreset{equation}{section}

\graphicspath{{./figures/}}

\begin{document}

\title{Exact-WKB Analysis of Two-level Floquet Systems}

\author{Toshiaki~Fujimori}
\email{toshiaki.fujimori018@gmail.com}
\affiliation{Department of Fundamental Education, Dokkyo Medical University, 880 Kitakobayashi, Mibu, Shimotsugagun, Tochigi 321-0293, Japan}
\affiliation{Department of Physics, Keio University, Kanagawa 223-8521, Japan}

\author{Syo~Kamata}
\email{skamata11phys@gmail.com}
\affiliation{Department of Physics, The University of Tokyo, Tokyo 113-0033, Japan}
\affiliation{Department of Physics, Keio University, Kanagawa 223-8521, Japan}

\author{Tatsuhiro~Misumi}
\email{misumi@phys.kindai.ac.jp}
\affiliation{Department of Physics, Kindai University, Osaka 577-8502, Japan}
\affiliation{Department of Physics, Keio University, Kanagawa 223-8521, Japan}

\author{Naohisa~Sueishi}
\email{sueishi@qunasys.com}
\affiliation{QunaSys Inc., 1-13-17 Hakusan, Bunkyo-ku, Tokyo 113-0001 Japan}
\affiliation{Department of Physics, Keio University, Kanagawa 223-8521, Japan}

\author{Hidetoshi~Taya}
\email{h$_$taya@keio.jp}
\affiliation{Department of Physics, Keio University, Kanagawa 223-8521, Japan}
\affiliation{iTHEMS, TRIP Headquarters, RIKEN, Wako 351-0198, Japan}

\begin{abstract}
\vspace{10mm}
We explore the application of the exact Wentzel-Kramers-Brillouin (WKB) analysis to two-level Floquet systems and establish a systematic procedure to calculate the quasi-energy and Floquet effective Hamiltonian.  We show that, in the exact-WKB analysis, the quasi-energy and Floquet effective Hamiltonian can be expressed in terms of cycle integrals (Voros symbol), which characterize monodromy matrices for Schr\"{o}dinger-type differential equations governing two-level Floquet systems.  We analytically evaluate the cycle integrals using the low-frequency expansion and derive both perturbative and non-perturbative corrections to the quasi-energy and Floquet effective Hamiltonian. 
To verify the accuracy of our results, we compare them with numerical calculations and analyze resonant oscillations, which reveal non-perturbative features that cannot be captured by the perturbative expansion.
\end{abstract}

\maketitle

\tableofcontents

\section{Introduction}
Floquet theory provides a powerful framework for understanding {\it Floquet systems}~\cite{Shirley:1965rgd,Sambe:1973cnm,Bukov:2015,Goldman:2014xja}, i.e., quantum systems subjected to periodic time-dependent driving, which arise in various physics contexts such as solid-state materials under laser fields~\cite{Oka1}, ultracold atoms in optical lattices~\cite{weitenberg2021tailoring}, and reheating of the Universe after inflation~\cite{Kofman:1994rk,Braden:2010wd}.  
It also provides insights into various physical phenomena, including the emergence of new energy levels (Floquet bands)~\cite{Lindner:2011guf}, quantum heating~\cite{DAlessio1}, localization~\cite{Seetharam1}, and coherent control in driven systems~\cite{Cayssol1}. 
  
One of the central tasks of the Floquet theory is to determine the so-called {\it Floquet effective Hamiltonian}~\cite{Bukov:2015,Eckardt:2016lof}, which encapsulates essential information of Floquet systems such as the quasi-energy~\cite{Goldman:2014xja}.  However, in general, the Floquet Hamiltonian cannot be determined exactly through analytical methods, necessitating the use of approximation techniques.  As such, high-frequency approaches (e.g., Floquet-Magnus expansion, van-Vleck expansion, and Brillouin-Wigner expansion) have been extensively developed and successfully applied to various Floquet systems over the past decade~\cite{Bukov:2015}.  These approaches are valid only when the driving frequency is sufficiently larger than the other energy scales in the system.  Consequently, alternative approaches are required in the low-frequency regime. Such low-frequency Floquet systems remain relatively unexplored, due to the lack of well-established approximation techniques (see Ref.~\cite{rodriguez2021low} for attempts). In particular, non-adiabatic processes such as the Landau-Zener transition~\cite{WEINBERG20171} can occur in this regime. These processes require non-perturbative treatments, making them inaccessible through naive perturbative methods.  

Motivated by the need to develop a systematic method for determining the Floquet Hamiltonian in low-frequency-driving regimes, we examine the exact Wentzel-Kramers-Brillouin (WKB) analysis, which is a powerful tool for analyzing the Stokes phenomenon of the Schr\"{o}dinger-type differential equation, 
\begin{align}
\big[-\hbar^2 \partial_t^2 + Q(t)\big] \psi(t) = 0, 
\label{eq:intro_schrodingereq}
\end{align}
where $Q(t)$ is a potential term, $\psi$ is a wave function, and $\hbar$ is a small parameter corresponding to the Planck constant. 
Originally developed in the field of mathematics~\cite{BPV, Voros1983, Silverstone, AKT1, AKT2, Takei1, DDP2, DP1, Takei2, Kawai1, Takei3, AKT3, AKT4, Iwaki1}, the exact-WKB analysis has recently  attracted significant attention in physics due to its applicability to a variety of problems.
For instance, the exact-WKB analysis provides insight into quantum mechanical systems, specifically the resurgent structure~\cite{Alvarez1,Alvarez2,Alvarez3,Jentschura:2004jg, Zinn-Justin:2004vcw,Zinn-Justin:2004qzw,
Jentschura:2010zza,Jentschura:2011zza,Basar:2013eka,Dunne:2013ada,Dunne:2014bca,Basar:2015xna,Misumi:2015dua,Behtash:2015loa,Gahramanov:2015yxk,Kozcaz:2016wvy,Dunne:2016jsr,Dunne:2016qix,Fujimori:2016ljw,Basar:2017hpr,Fujimori:2017oab,Sueishi:2019xcj,Cavusoglu:2023bai,Cavusoglu:2024usn,Ture:2024nbi}, which elucidates the nontrivial relationship between perturbative and non-perturbative contributions, as well as the associated Stokes phenomena~\cite{Sueishi:2020rug,Sueishi:2021xti,Kamata:2021jrs}.  
It also allows us to explore non-perturbative effects involving instanton and bion contributions appearing in various quantum systems~\cite{Zinn-Justin:1981qzi,Zinn-Justin:1983yiy,
Unsal:2007vu,Unsal:2007jx,Shifman:2008ja,Poppitz:2009uq,Anber:2011de,Poppitz:2012sw, Misumi:2014raa,Behtash:2018voa,Fujimori:2019skd,Misumi:2019upg,Fujimori:2020zka,Unsal:2020yeh,Pazarbasi:2021ifb,Pazarbasi:2021fey}.
Besides, the exact-WKB analysis has deep connections with 4d ${\cal N}=2$ gauge theories~\cite{Nekrasov:2009rc,Mironov:2009uv,
Kashani-Poor:2015pca,Kashani-Poor:2016edc, Ashok:2016yxz,Yan:2020kkb}, 
wall-crossing phenomena~\cite{Gaiotto:2012rg,Allegretti:2020dyt},
ODE/IM correspondence~\cite{Dorey:2001uw, Dorey:2007zx,Ito:2018eon,Ito:2019jio,Imaizumi:2020fxf}, TBA equations~\cite{Emery:2020qqu,Ito:2024nlt},
topological string theory~\cite{Grassi:2014cla, Grassi:2014zfa,Codesido:2017dns,Codesido:2017jwp,Hollands:2019wbr,Ashok:2019gee,Coman:2020qgf,Iwaki:2023cek}, as well as other subjects~\cite{Duan:2018dvj,Kuwagaki:2020pry,Taya:2020dco,Enomoto:2020xlf,Enomoto:2021hfv,Taya:2021dcz,Enomoto:2022mti,Enomoto:2022nuj,Imaizumi:2022dgj,vanSpaendonck:2022kit,Kamata:2023opn,Suzuki:2023slp,Bucciotti:2023trp,Kamata:2024tyb,vanSpaendonck:2024rin,Misumi:2024gtf}. 
Among these, its application to the vacuum pair production under external fields is particularly relevant to our present study~\cite{Taya:2020dco,Enomoto:2020xlf}.  It has been shown that the vacuum pair production can be regarded as a Stokes phenomenon of the Klein-Gordon equation, which is a special case of the Schr\"{o}dinger-type differential equation. In this framework, the number of produced pairs is shown to be given by a product of connection matrices in the exact-WKB analysis. 

The aim of this paper is to apply the exact-WKB analysis to two-level Floquet systems. 
The dynamics of these systems is governed by a time-dependent Schr\"{o}dinger equation, which is a first-order differential equation for the state vector. While the exact-WKB analysis for such a first-order differential equation with general multi-component vector functions is still under development (see, e.g., Ref.~\cite{WATT:2008} for a recent attempt in mathematics and Ref.~\cite{Suzuki:2023slp} for non-adiabatic transitions in multi-band Hamiltonians), in two-level systems, the Schr\"{o}dinger equation can be transformed into a second-order differential equation for a scalar function of the form of Eq.~\eqref{eq:intro_schrodingereq} and hence the well-established framework of the exact-WKB analysis can be applied directly.   

We will demonstrate that the exact-WKB analysis provides a powerful method for investigating non-perturbative phenomena in the low-frequency regime of Floquet systems. 
To achieve this, we will derive the Floquet effective Hamiltonian and quasi-energy from the time-evolution unitary matrix $U(T)$, which represents the linear transformation for the state vector over a single period $T$. As we will explain in detail, the unitary matrix $U(T)$ can be expressed in terms of the so-called {\it monodromy matrix} $M$, which encodes the mixing of the basis-state vectors over a single period. Determining the monodromy matrix reduces to analyzing the Stokes phenomena of the differential equation~\eqref{eq:intro_schrodingereq} and deriving a connection formula for {\it degenerate} Stokes curves in the Stokes graph. To the best of our knowledge, such a connection formula has not been established in the literature. We will show that it is given in terms of cycle integrals, known as {\it Voros symbols} in the exact-WKB analysis, and that these cycle integrals can be systematically evaluated using the low-frequency expansion (i.e., the large-period expansion). 
We will explicitly perform this evaluation and calculate the resulting quasi-energy and effective Hamiltonian. 
By comparing these analytical results with numerical solutions of the Schr\"{o}dinger equation, we will find a good agreement in the low-frequency regime. 
In particular, we will show that the exact-WKB analysis correctly captures resonant oscillations between the two levels that occur at specific frequencies. These resonant behaviors are the characteristic non-perturbative (non-adiabatic) effects that cannot be captured by a naive power-series expansion. Furthermore, we will find that the oscillation frequencies between the two levels, which are intrinsically non-perturbative, can be extracted from the monodromy matrix in the exact-WKB analysis. 

This paper is organized as follows.  
In Sec.~\ref{sec:2}, we introduce our setup for two-level Floquet systems and establish the formalism for applying the exact-WKB analysis. 
In Sec.~\ref{sec:EWKB}, we review the exact-WKB analysis and present a systematic procedure for computing the monodromy of the solutions to the Schr\"{o}dinger equation.
In Sec.~\ref{sec:4}~A, we apply the exact-WKB analysis to a simple Floquet system as a concrete demonstration of our approach, explicitly deriving the quasi-energies and Floquet effective Hamiltonian.  
In Sec.~\ref{sec:5}, we extend our analysis to general Floquet systems, providing a broader framework for applying the exact-WKB analysis.
Sec.~\ref{sec:6} is dedicated to a summary and discussion.
The appendices provide supporting mathematical details for the Borel resummation and median resummation techniques. We also present the details of derivations of key quantities, including the monodromy matrices and wave functions, as well as an extension to a more complex Floquet system is discussed.

\section{Setup} \label{sec:2}

In this paper, we study a general two-level Floquet system described by the time-dependent Schr\"{o}dinger equation:
\begin{align}
i\hbar \pdv{t} \Psi(t) = H(t) \Psi(t)
\ \ {\rm with}\ \ 
\Psi(t) = 
\left( \begin{array}{c} 
\psi_1(t) \\ \psi_2(t)    
\end{array} \right) .
\label{eq:Sch_eq}
\end{align}
The Hamiltonian $H(t)$ is assumed to be periodic in time as 
\begin{align}
H(t+T) = H(t). \label{eq:H_period}
\end{align}
Without loss of generality, we assume that the Hamiltonian is traceless, as the trace part of $H(t)$ can be eliminated through an appropriate redefinition of the overall phase of the state vector $\Psi(t)$. Under this assumption, the Hamiltonian can be expressed as an element of the $\mathfrak{su}(2)$ algebra:
\begin{align}
H(t) = \sum_{i=1}^3 f_i(t)\sigma_i, \label{eq:H_fsig}
\end{align}
where $f_i(t)$ ($i = 1, 2, 3$) are periodic functions satisfying
\begin{align}
f_i(t+T) = f_i(t),
\end{align}
and $\sigma_i$ ($i = 1, 2, 3 $) are the Pauli matrices defined as
\begin{align}
\sigma_1 = \left( \begin{array}{cc} 0 & 1 \\ 1 & 0 \end{array} \right), \hspace{5mm}
\sigma_2 = \left( \begin{array}{cc} 0 & -i \\ i & 0 \end{array} \right), \hspace{5mm}
\sigma_3 = \left( \begin{array}{cc} 1 & 0 \\ 0 & -1 \end{array} \right).
\end{align}

The central task in a Floquet system is to determine the so-called \textit{Floquet effective Hamiltonian}, which is given by the logarithm of the time-evolution unitary matrix over one period:
\begin{align}
H_{\rm eff} = \frac{i \hbar}{T} \log U(T). 
 \label{eq:Heff_U}
\end{align}
The time-evolution unitary $U(T)$ is determined by the following first-order differential equation and initial condition:
\begin{align}
i \hbar \partial_t U(t) = H(t) U(t), \hspace{10mm} U(0) = \left( \begin{array}{cc} 1 & 0 \\ 0 & 1 \end{array} \right).
\label{eq:1st_order_diffeq}
\end{align}
The eigenvalues of the Floquet effective Hamiltonian $H_{\rm eff}$ are referred to as \textit{quasi-energies}. 

Although determining the quasi-energies for a general Hamiltonian is difficult, their asymptotic form in the large period limit $T \rightarrow \infty$ can be obtained systematically.
This limit corresponds to the low-frequency regime $\omega = 2\pi/T \to 0$, where the Hamiltonian $H(t)$ varies slowly. Consequently, the adiabatic approximation becomes applicable in determining the time-evolution unitary matrix $U(t)$. According to the adiabatic theorem \cite{BornFock}, in the low-frequency limit, the state vectors evolving under the Schr\"odinger equation remain in the instantaneous eigenstates of $H(t)$. 
Under this approximation, the time-evolution unitary matrix takes the form:
\begin{align}
U(t) \ \approx \
\left(
\begin{array}{cc}
\xi_1^{+}(t) & \xi_1^{-}(t) \\
\xi_2^{+}(t) & \xi_2^{-}(t)
\end{array} \right) 
\left(
\begin{array}{cc}
e^{i \theta_0(t) + i \gamma(t)} & 0 \\
0 &e^{-i \theta_0(t) - i \gamma(t)}
\end{array} \right) 
\left(
\begin{array}{cc}
\xi_1^{+}(0) & \xi_1^{-}(0) \\
\xi_2^{+}(0) & \xi_2^{-}(0)
\end{array} \right) ^{-1}, \label{eq:U_adiabatic}
\end{align}
where the column vectors, $\xi^\pm(t) = (\xi_1^\pm(t), \xi_2^\pm(t))^{\rm T}$, are the instantaneous eigenvectors of $H(t)$, satisfying:
\begin{align}
H(t) \, \xi^\pm(t) = \pm \epsilon_0(t) \, \xi^\pm(t), \qquad \epsilon_0(t) = \sqrt{f_1(t)^2 + f_2(t)^2 + f_3(t)^2}.
\end{align}
From Eq.~\eqref{eq:U_adiabatic},
it follows that the quasi-energies in the low-frequency limit are given by the phase $\theta_0(t)+\gamma(t)$, 
where $\theta_0(t)$ represents the dynamical phase corresponding to the leading-order contribution that scales as $\mathcal{O}((\hbar\omega)^{-1})$:
\begin{align}
\theta_0(t) = -\frac{1}{\hbar} \int_0^t \epsilon_0(t') \, dt',
\label{eq:dynamical_phase}
\end{align}
while $\gamma(t)$ denotes the Berry (geometric) phase, corresponding to the subleading-order term that scales as  $\mathcal{O}((\hbar\omega)^{0})$: 
\begin{align}
\gamma(t) = \pm i \int_0^t \xi_\pm^\dagger(t') \partial_{t'} \xi_\pm(t') \, dt'.
\label{eq:geometric_phase}
\end{align}
Thus, at the leading order, the quasi-energies are given by the time-averaged instantaneous eigenvalues of the Hamiltonian over one period, while the first-order correction is determined by the Berry phase.
Higher-order corrections to this approximation can be systematically obtained by solving Eq.~\eqref{eq:1st_order_diffeq} order by order in $\hbar \omega = 2\pi \hbar/T$. Details of this expansion are summarized in Appendices \ref{subsec:WKB_1st_order} and \ref{subsec:WKB_2nd_order}. 

Note that in the Hamiltonian $H(t)$, the frequency $\omega$ always appears in combination with the time variable $t$. This allows us to introduce a rescaled time variable $s = \omega t$, in terms of which 
Eq.~\eqref{eq:1st_order_diffeq} takes the form:
\begin{align}
i \hbar \omega \partial_s \, {\cal U}(s) = {\cal H}(s) \, {\cal U}(s),
\end{align}
where we have defined ${\cal H}(s) = H(t)$ and ${\cal U}(s) = U(t)$. 
This reformulation shows that an expansion in $\omega$ is equivalent to an expansion in $\hbar$, which corresponds to the WKB expansion.
However, it is well known that such an expansion typically leads to a divergent asymptotic series, indicating the presence of non-perturbative (non-adiabatic) effects that cannot be fully captured by a power-series expansion in $\hbar \omega$. To systematically investigate these effects, we employ the exact-WKB analysis, which allows us to extract non-perturbative corrections beyond the conventional adiabatic approximation.

To apply the exact-WKB analysis to the two-level Floquet system, 
it is convenient to rewrite the time-dependent Schr\"odinger equation \eqref{eq:Sch_eq} as a second-order differential equation. 
This can be achieved by introducing an arbitrary linear combination of the state vector $\Psi(t) = (\psi_1(t),\psi_2(t))^T$:
\begin{align}
\psi(t, C) = \langle C(t), \Psi(t) \rangle = - c_2(t) \psi_1(t) + c_1(t) \psi_2(t),
\label{eq:linear_combination}
\end{align}
where $C(t) = (c_1(t), c_2(t))^T$ is an arbitrary time-dependent vector with components $c_1(t)$ and $c_2(t)$ and $\langle \cdot, \cdot \rangle$ denotes an anti-symmetric bilinear form for two-component vectors,  
\begin{align}
\langle A, B \rangle := -a_2 b_1 + a_1 b_2, \hspace{10mm}
A = 
\left( \begin{array}{c} a_1 \\ a_2 \end{array} \right), \quad
B = 
\left( \begin{array}{c} b_1 \\ b_2 \end{array} \right).
\end{align}

If $\Psi = (\psi_1(t), \psi_2(t))^T$ satisfies Eq.~\eqref{eq:Sch_eq}, 
it can be shown that the linear combination $\psi(t, C) = \langle C(t), \Psi(t) \rangle$ satisfies a second-order differential equation of the form:
\begin{align}
\qty[-\hbar^2\qty(\pdv{t}-\pdv{t}\lambda(t,\hbar))^2+Q(t,\hbar)] \psi = 0,
\label{eq:2nd_order_diffeq}
\end{align}
where $\lambda(t, \hbar)$ and $Q(t, \hbar)$ are defined as
\begin{align}
\lambda(t, \hbar) = \frac{1}{2} \log \langle C, \mathcal{D} C \rangle, \hspace{10mm}
Q(t, \hbar) = -\hbar^2 \left[ \frac{\langle \mathcal{D}C, \mathcal{D}^2C \rangle}{\langle C, \mathcal{D} C \rangle} + \partial_t^2 \lambda - (\partial_t \lambda)^2 \right],
\end{align}
and $\mathcal{D}$ is a derivative operator given by
\begin{align}
\mathcal{D} C = \left[ \partial_t - \frac{1}{i \hbar} H(t) \right] C.
\end{align}
In the following sections, we apply the exact-WKB analysis to the second-order differential equation \eqref{eq:2nd_order_diffeq}.

Conversely, once a solution $\psi(t, C)$ to the second-order differential equation \eqref{eq:2nd_order_diffeq} is obtained, 
it can be mapped to a solution of the original first-order differential equation \eqref{eq:1st_order_diffeq} as
\begin{align}
\Psi(t) = 
\begin{pmatrix}
\psi_1(t) \\ \psi_2(t)
\end{pmatrix}
= e^{-2\lambda} \Big[ \psi \, \mathcal{D} C - \partial_t \psi \, C \Big].
\label{eq:2nd_to_1st}
\end{align}
It can be directly verified that this expression satisfies Eq.~\eqref{eq:1st_order_diffeq}, or equivalently $\mathcal{D} \Psi(t) = 0$, as follows:
\begin{align}
\mathcal{D} \Psi(t) = \frac{1}{\hbar^2 \langle \mathcal{D} C, C \rangle} 
\qty[-\hbar^2 \qty(\pdv{t} - \pdv{t} \lambda(t, \hbar))^2 + Q(t, \hbar)] \psi \, C = 0.
\end{align}
Using the relation \eqref{eq:2nd_to_1st}, we can construct a basis $G(t) = (\Psi^+(t), \Psi^-(t))$ for the solutions to the original first-order differential equation \eqref{eq:1st_order_diffeq} as
\begin{align}
G(t) = (\Psi^+(t), \Psi^-(t)) = e^{-2\lambda} \Big[ \mathcal{D} C \, \boldsymbol{\psi} - C \, \partial_t \boldsymbol{\psi} \Big],
\label{eq:1st_to_2nd}
\end{align}
where $\boldsymbol{\psi}(t,C) = (\psi^+(t,C), \psi^-(t,C))$ is a basis for the solutions to the second-order differential equation \eqref{eq:2nd_order_diffeq}. 
From this basis $G(t)$, the time-evolution unitary matrix $U(t)$, which is the solution to Eq.~\eqref{eq:1st_order_diffeq}, can be constructed as
\begin{align}
U(t) = G(t) G(0)^{-1}.
\label{eq:unitary_g}
\end{align}
Thus, the time-evolution unitary matrix can be obtained by solving the second-order differential equation~\eqref{eq:2nd_order_diffeq}. 

To determine the quasi-energies and Floquet effective Hamiltonian, we need to obtain $U(T)$, 
which can be derived by analyzing the evolution of $G(t)$ over one period. 
The Floquet theorem states that for a basis of the solutions to the second-order differential equation \eqref{eq:2nd_order_diffeq}, there exists a monodromy matrix $M$ such that
\begin{align}
\left( \psi^+(t+T), \psi^-(t+T) \right) = \left( \psi^+(t), \psi^-(t) \right) M,
\label{eq:monodromy}
\end{align}
or equivalently
\begin{align}
G(t+T) = G(t) M. 
\end{align}
Using the monodromy matrix $M$, the time-evolution unitary matrix $U(T)$ can be expressed as
\begin{align}
U(T) = G(T) G(0)^{-1} = G(0) M G(0)^{-1}.
\label{eq:unitary_monodromy}
\end{align}
Thus, the quasi-energies and Floquet effective Hamiltonian can be determined by computing the monodromy matrix $M$
and the initial basis $G(0)$. 
Notably, it can be shown that $U(T)$ constructed through this procedure is independent of the choices of the coefficients $C = (c_1, c_2)$ in the linear combination. 

The effective Hamiltonian can be obtained from $M$ and $G(0)$ using Eqs.\,\eqref{eq:Heff_U} and \eqref{eq:unitary_monodromy} as
\begin{align}
H_{\rm eff} = \frac{i \hbar}{T} \, G(0) \log M \, G(0)^{-1}. 
\end{align}
The quasi-energies $\epsilon$ are determined as the solutions to the following characteristic equation:
\begin{align}
\det(M - e^{-i \epsilon T / \hbar}) = 0.
\label{eq:detM}
\end{align}
In the following, we employ the exact-WKB analysis to compute the monodromy matrix $M$, which allows us to derive both effective Hamiltonian $H_{\rm eff}$ and quasi-energies $\epsilon$. 

\section{A review of exact-WKB analysis} 
\label{sec:EWKB}

In this section, we provide an overview of the exact-WKB analysis, which offers a systematic framework for calculating the monodromy matrix through the Borel resummation. To illustrate the exact-WKB analysis, we first review the construction of wave functions and the connection formula for the Airy-type differential equation in Sec.\,\ref{sec:3-1}. Following this, in Sec.\,\ref{sec:3-2}, we outline the construction of the monodromy matrix involving degenerate Stokes curves.

\subsection{Construction of wave function via Borel resummation and connection formula} 
\label{sec:3-1}

We begin with the second-order differential equation of the form,
\begin{align}     
\qty[ -\hbar^2 \frac{\partial^2}{\partial t^2} + Q(t,\hbar) ] \psi(t)=0 .
\label{eq:Sch_eq_WKB}
\end{align}
Since this problem is equivalent to a time-independent Schr\"odinger equation, 
we refer to $\psi(t)$ as the wave function and 
$Q(t,\hbar)$ as the potential.
We assume that the potential term $Q$ can be expanded as 
\begin{align}
Q = Q_0 + Q_1 \hbar + Q_2 \hbar^2 + \cdots.
\end{align}  
Note that the second-order differential equation Eq.~\eqref{eq:2nd_order_diffeq} can be rewritten in the form of Eq.(\ref{eq:Sch_eq_WKB}) by rescaling the function $\psi(t) \rightarrow \exp(\lambda(t)) \psi(t)$.

In the WKB analysis, we employ the WKB ansatz:
\begin{align}
\psi(t) & = \frac{1}{\sqrt{\partial_t W(t,h)}} \exp \left(  \frac{W(t,h)}{\hbar} \right) .
\label{WKBpsi}
\end{align} 
Substituting this ansatz into Eq.~\eqref{eq:Sch_eq_WKB}, 
we obtain the differential equation for $W(t,\hbar)$: 
\begin{align}
&(\partial_t W(t,\hbar))^2-\frac{\hbar^2}{2} \qty{ W(t,\hbar), t} = Q(t,\hbar), \label{eq:Riccati} 
\end{align}
where $\{W,t\}$ denotes the Schwarzian derivative, defined as
\begin{align}
\{W,t\} = \frac{\partial_t^3 W}{\partial_t W} - \frac{3}{2} \qty(\frac{\partial_t^2 W}{\partial_t W} )^2.
\end{align}
This equation can be solved formally by expanding $W(t,\hbar)$ and $Q(t,\hbar)$ in powers of $\hbar$ and determining the expansion coefficients order by order. The leading-order equation takes the form of the Hamilton-Jacobi equation. Since it is a quadratic equation for $\partial_t W$, there exist two solutions:
\begin{align}
\partial_t W(t,\hbar) = \sqrt{Q_0(t)} + \mathcal O(\hbar) \qquad \mbox{or} \qquad 
\partial_t W(t,\hbar) = - \sqrt{Q_0(t)} + \mathcal O(\hbar), 
\label{eq:S_odd_leading}
\end{align}
where $Q_0(t)$ is the $O(\hbar^0)$ term in $Q(t,\hbar)$. 
In general, if $W_{\rm sol}(t,\hbar)$ is a solution, then $-W_{\rm sol}(t,\hbar)$ also satisfies the differential equation, as Eq.~\eqref{eq:Riccati} is invariant under the sign flip $W(t,\hbar) \rightarrow - W(t,\hbar)$.
Once a specific branch is chosen from the two solutions given in Eq.~\eqref{eq:S_odd_leading}, 
the higher-order corrections to $W_{\rm sol}$ can be uniquely determined by recursively solving the equations obtained by expanding Eq.~\eqref{eq:Riccati} (see Appendix~\ref{appendix:recursion} for the details of the WKB expansion). 
For example, the leading and subleading terms of $W_{\rm sol}(t,\hbar)$ in the expansion $W_{\rm sol}(t,\hbar) = W_0 + W_1 \, \hbar + W_2 \, \hbar^2 + \cdots$ are given by\footnote{The leading order term $W_0(t)$ is Hamilton's characteristic function satisfying the (reduced) Hamilton-Jacobi equation.}
\begin{align}
W_0(t,t_0) = \int_{t_0}^t \sqrt{Q_0(t')} \, dt', \qquad 
W_1(t,t_0) = \int_{t_0}^t \frac{Q_1(t)}{2\sqrt{Q_0(t')}} \, dt',  \qquad \cdots
\label{eq:W0_W1}
\end{align} 
where the base point $t_0$ of the integration can be an arbitrary point on the complex $t$-plane. 
It can be shown that $\partial_t W_{\rm sol}$ is a ``local quantity" in the sense that it depends only on $Q_i(t)$ and their derivatives at $t$ and does not depend on the values of $Q_i(t')$ with $t \not = t'$. 
For notational simplicity, we denote this local quantity as $S_{\rm odd}$: 
\begin{align}
S_{\rm odd}(t,\hbar) = \frac{\partial_t W_{\rm sol}(t,\hbar)}{\hbar}. 
\label{eq:Sodd_def}
\end{align}
Then, the solution to Eq.~\eqref{eq:Riccati} can be obtained by integrating $S_{\rm odd}$ as
\begin{align}
W_{\rm sol}(t,\hbar) = \hbar \int_{t_0}^t S_{\rm odd} \, dt.
\label{eq:W_Sodd}
\end{align}
We refer to the base point $t_0$ as \textit{the normalization point}, since its choice determines the overall constant factor of the wave function $\psi(t)$. 

Once $W_{\rm sol}(t,\hbar)$ is obtained, we can construct the following two solutions $\psi_{\rm sol}^{\pm}$ to the original differential equation \eqref{eq:2nd_order_diffeq}:
\begin{align}
\psi_{\rm sol}^{\pm}(t) & = \frac{1}{\sqrt{\partial_t W_{\rm sol}(t,h)}} \exp \left( \pm \frac{W_{\rm sol}(t,h)}{\hbar} \right).
\label{WKBpsisol}
\end{align}
These solutions are referred to as \textit{WKB solutions}. 
Since their Wronskian is a non-zero constant,
\begin{align}
\psi_{\rm sol}^{-} \partial_t \psi_{\rm sol}^{+} - \psi_{\rm sol}^{+} \partial_t \psi_{\rm sol}^{-} = \frac{2}{\hbar},
\label{eq:WKB_Wronskian}
\end{align}
the solutions $\psi_{\rm sol}^{\pm}(t)$ are linearly independent, and thus the pair $(\psi_{\rm sol}^{+}(t),\psi_{\rm sol}^{-}(t))$ forms a basis of the solutions. 
Although obtaining the explicit closed-form expressions for $\psi_{\rm sol}^{\pm}(t)$ is generally difficult, 
a formal power-series expression can be derived by substituting the power-series solution 
$W_{\rm sol}(t,\hbar)= W_0 + W_1 \, \hbar + W_2 \, \hbar^2 + \cdots$ given in Eq.~\eqref{eq:W0_W1} into Eq.~\eqref{WKBpsisol}.
Schematically, the formal power-series expressions of $\psi_{\rm sol}^{\pm}(t)$ take the form:
\begin{align}
\psi_{\rm P}^\pm(t,t_0) &= \frac{1}{Q_0(t)^{\frac{1}{4}}} \exp \qty( \pm \frac{1}{\hbar} W_0(t,t_0) ) \sum_{n=0}^{\infty} \psi_{n}^\pm(t,t_0) \, \hbar^n ,
\label{eq:powerseriessolution}
\end{align}
where the subscript ${\rm P}$ indicates that these are power-series solutions. These are the standard WKB power-series solutions found in the literature. As is well known, $\psi_{\rm P}^\pm$ is a {\it formal} asymptotic series, since its coefficients exhibit factorial divergence $\psi_n^{\pm} \sim n!$ [see Eq.~\eqref{eq:formal_PS_airy} for an explicit example]. 
Consequently, an additional treatment is required to obtain a well-defined solution. 

The central idea of the exact-WKB analysis is to utilize the Borel resummation to make the divergent series well-defined. The Borel resummation $\mathcal{S}$ of the series $\psi_{\rm P}^\pm$ is expressed as:
\begin{align}
\mathcal{S}[\psi_{\rm P}^\pm(t,t_0)] = \frac{1}{\hbar} \frac{1}{Q_0(t)^{\frac{1}{4}}} \exp \qty( \pm \frac{1}{\hbar} W_0(t,t_0) ) \int_0^\infty ds \, e^{-\frac{s}{\hbar}} B^\pm(s,t,t_0),
\label{eq:Borel_psi}
\end{align}
where $B^\pm(s,t,t_0)$ are the Borel transforms of the series $\sum_{n=0}^\infty \psi_n^\pm(t,t_0) \, \hbar^n$, defined as:
\begin{align}
B^\pm(s,t,t_0) = \sum_{n=0}^\infty \psi_n^\pm(t,t_0) \frac{s^n}{n!}.
\label{eq:Borel_transf}
\end{align}
The Borel transform $B^\pm(s,t,t_0)$ is a function of the auxiliary variable $s$, defined on the complex $s$-plane, also referred to as the \textit{Borel plane}. If $B^\pm(s,t,t_0)$ have no singularities on the positive real axis of the Borel plane, the Laplace transform in Eq.~\eqref{eq:Borel_psi} can be performed unambiguously, and the Borel resummation $\mathcal{S}$ yields a finite, well-defined function. In such cases, the series $\psi_{\rm P}^\pm$ is said to be \textit{Borel summable}.
In particular, if the power series $\psi_{\rm P}^\pm$ is convergent, the Borel resummation reproduces the correct result. This can be shown by exchanging the order of integration and summation in Eqs.~\eqref{eq:Borel_psi} and \eqref{eq:Borel_transf}.

We here make a comment on the Borel resummation:
The Borel resummation \( \mathcal{S} \) maps a trans-series, whose definition is given in (\ref{eq:trans-series_f0pm}) in App.~\ref{sec:Quant_DDP_Stokes_story}, to a function \cite{Costin2008}.
Thus, $\mathcal{S}[\psi_{\rm P}(t, t_0)]$ is no longer a trans-series but a well-defined function. The exact-WKB analysis is formulated using these Borel-resummed wave functions, which are genuine functions. The term ``Borel-resummed form" in our discussions refers to a function reconstructed via the Borel resummation, rather than a trans-series.

Even when $\psi_{\rm P}^\pm$ is a divergent series, the Borel resummation still yields a finite and meaningful solution, provided that $B^\pm(s,t,t_0)$ has no singularities along the integration path in Eq.~\eqref{eq:Borel_psi}, i.e., the positive real axis of the Borel plane.
In general, the Borel transform $B^\pm(s,t,t_0)$, defined in Eq.~\eqref{eq:Borel_transf}, has a finite radius of convergence and exhibits singularities at certain points on the Borel plane that move with the parameter $t$. These \textit{movable singularities} cross the integration path (the positive real axis) on the Borel plane when $t$ varies across certain curves in the complex $t$-plane, known as \textit{Stokes curves}. 
Stokes curves are paths on the complex $t$-plane that originate from turning points. Their locations are determined by a condition: 
\begin{align}
\mathrm{Im} \left[ \frac{1}{\hbar} W_0(t,t_0) \right] = \mathrm{Im} \left[ \frac{1}{\hbar} \int_{t_0}^t \sqrt{Q_0(t')} \, dt' \right] = 0, 
\label{eq:def_curve}
\end{align}
where $W_0(t,t_0)$ is the leading-order part of $W_{\rm sol}(t,t_0)$. 

For example, if $Q_0(t)$ has a simple turning point at $t = t_0$, 
\begin{align}
Q_0(t) = c(t - t_0) + \mathcal{O}((t - t_0)^2)\quad \mbox{with} \quad c \neq 0,
\end{align}
three Stokes curves emanate from the turning point. These curves can be determined by solving the following condition for $t$:
\begin{align}
0 = \mathrm{Im} \left[ \frac{1}{\hbar} \int_{t_0}^t \sqrt{Q_0(t')} \, dt' \right] = \mathrm{Im} \left[ \frac{c}{\hbar} (t - t_0)^{\frac{3}{2}} + \cdots \right].
\label{eq:def_curve1}
\end{align}

When $t$ lies on one of these Stokes curves, at least one of the movable singularities of the Borel transforms $B^\pm(s,t)$ appears on the real axis of the Borel plane. As a result, the Borel resummations $\mathcal{S}[\psi_{\rm P}^\pm(t,t_0)]$ exhibit discontinuous jumps as $t$ crosses a Stokes curve. Consequently, the Borel resummation provides distinct solutions in each \textit{Stokes region}, which are regions on the complex $t$-plane separated by the Stokes curves.

Although the Borel resummation produces distinct pairs of solutions depending on the Stokes regions, these solutions are not linearly independent; rather, they are related through \textit{connection formulas}. Let $(\psi_{\rm I}^+(t, t_0), \psi_{\rm I}^-(t, t_0))$ and  $(\psi_{\rm II}^+(t, t_0), \psi_{\rm II}^-(t, t_0))$ denote bases of solutions obtained by applying the Borel resummation to the WKB power-series solutions in one region (e.g., Region I) and an adjacent region (e.g., Region II), respectively.\footnote{The functions  $\psi_{\rm I}^\pm$ and $\psi_{\rm II}^\pm$ are analytic continuations of $\mathcal{S}[\psi_{\rm P}^\pm(t, t_0)]$ from Regions I and II, respectively.} If the basis $(\psi_{\rm I}^+(t, t_0), \psi_{\rm I}^-(t, t_0))$ is analytically continued into Region II, the two bases are related by a connection formula of the form:
\begin{align}
\mqty( \psi_{\rm I}^+(t, t_0) & \psi_{\rm I}^-(t, t_0) ) 
= \mqty( \psi_{\rm II}^+(t, t_0) & \psi_{\rm II}^-(t, t_0) ) \, T_{\rm II, I}(t_0).
\label{eq:connection_12}
\end{align}
Here, the $2 \times 2$ matrix $T_{\rm II, I}(t_0)$, referred to as \textit{connection matrix}, relates the bases in Regions I and II.

The connection matrix depends on the choice of the normalization point $t_0$. If the normalization point is changed, the WKB solutions transform as:
\begin{align}
\mqty(\psi_{\rm I}^+(t, t_1) & \psi_{\rm I}^-(t, t_1)) = \mqty(\psi_{\rm I}^+(t, t_0) & \psi_{\rm I}^-(t, t_0)) \, \mathcal{N}_{t_0, t_1},
\label{eq:normalization_matrix}
\end{align}
where the normalization matrix $\mathcal{N}_{t_0, t_1}$ is a $2 \times 2$ matrix that relates the WKB solutions normalized at the points $t_0$ and $t_1$. Explicitly, $\mathcal{N}_{t_0, t_1}$ can be expressed in terms of $S_{\rm odd}$, defined in Eq.~\eqref{eq:Sodd_def}, as:
\begin{align}
\mathcal{N}_{t_0, t_1} = 
\begin{pmatrix}
 \exp \left( + \int_{t_1}^{t_0} S_{\rm odd}(t') dt' \right) & 0 \\ 
 0 &  \exp \left( - \int_{t_1}^{t_0} S_{\rm odd}(t') dt' \right)
\end{pmatrix}.
\label{eq:normt0t1}
\end{align}
When the normalization point is changed from  $t_0$ to $t_1$, the connection matrix $T_{\rm II, I}$ transforms as:
\begin{align}
T_{\rm II, I}(t_1) = \mathcal{N}_{t_0, t_1}^{-1} \, T_{\rm II, I}(t_0) \, \mathcal{N}_{t_0, t_1}.
\end{align}
As we will see, the monodromy matrix $M$ and the time-evolution unitary matrix $U$, given in Eqs.~\eqref{eq:monodromy} and \eqref{eq:unitary_monodromy}, can be expressed in terms of the connection matrices and the normalization matrices.

\subsubsection{Connection formula for Airy-type differential equation} \label{sec:connection_Airy_diff}

\begin{figure}[t]
\begin{minipage}{0.45\linewidth}
\centering
\fbox{
\includegraphics[width=60mm, page=1]{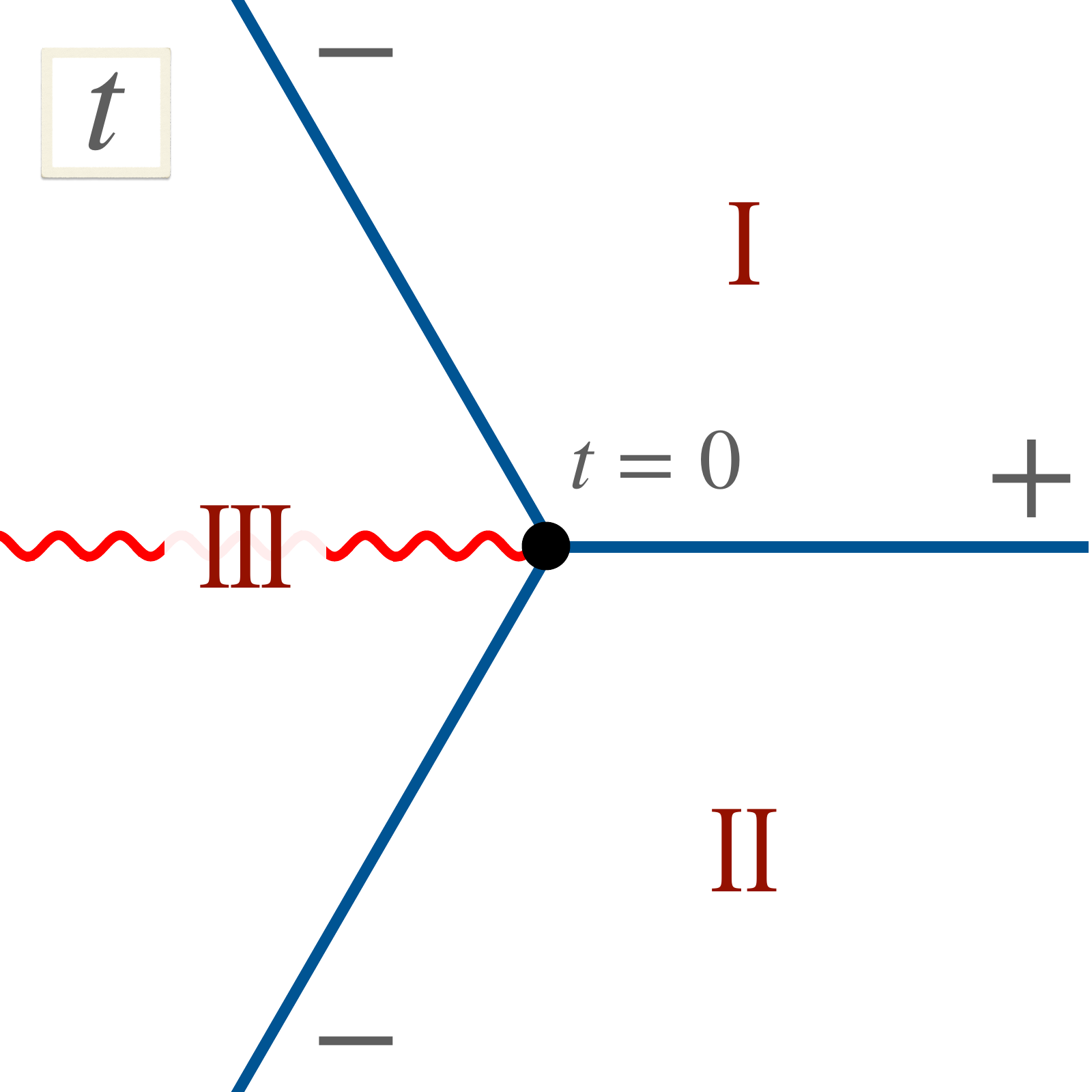}}
\caption{Stokes graph for $Q(t) = at$. The wavy line represents the branch cut of $W_0(t)$. The signs ($\pm$) attached to each Stokes curve indicate whether $W_0(t)/\hbar$ is increasing ($+$) or decreasing ($-$) along that line towards infinity $|t| \rightarrow \infty$.}
\label{fig:airy_stokes}
\end{minipage}
~~~~
\begin{minipage}{0.45\linewidth}
\centering
\fbox{
\includegraphics[width=60mm, page=24]{floquet_fig4.pdf}}
\caption{Integration contours for Airy functions. The integral converges as long as the contours extend to the infinity in the unshaded region. \vspace{8mm}}
\label{fig:contour_airy}
\end{minipage}
\end{figure}

As a basic example, we illustrate the connection formula for non-degenerate Stokes curves emanating from a simple turning point using the Airy equation, which corresponds to Eq.~\eqref{eq:Sch_eq_WKB} with $Q(t) = at$:
\begin{align}
\Big[\! - \! \hbar^2 \partial_t^2 + a t \, \Big] \, \psi(t) = 0. 
\label{eq:Airy_eq}
\end{align}
The Stokes graph for $Q(t) = at$ is shown in Fig.~\ref{fig:airy_stokes}. The turning point is located at $t=0$, from which three Stokes curves extend in the directions $\arg t = 0$ and $\pm 2\pi/3$. Correspondingly, there are three Stokes regions:
\begin{alignat}{2}
&\text{Region I} & & : \{ t \in \mathbb{C} \mid 0 < \arg t < 2\pi / 3 \}, \\
&\text{Region II} & & : \{ t \in \mathbb{C} \mid -2\pi / 3 < \arg t < 0 \}, \\
&\text{Region III} & & : \{ t \in \mathbb{C} \mid - 4\pi / 3 < \arg t < -2\pi / 3 \}.
\end{alignat}

To understand how the Borel-resummed WKB solutions are related across these regions, we begin with the formal power-series WKB solutions to the Airy equation~\eqref{eq:Airy_eq}, given by
\begin{align}
\psi_{\rm P}^\pm(t, t_0=0) = \frac{1}{(at)^{\frac{1}{4}}} \exp \left( \pm \frac{2}{3} \frac{a^{\frac{1}{2}}}{\hbar} t^{\frac{3}{2}} \right) 
\sum_{n=0}^\infty \frac{\Gamma\qty(n + \frac{1}{6}) \Gamma\qty(n + \frac{5}{6})}{2\pi \Gamma(n + 1)} 
\qty(\pm \frac{3}{4} \frac{\hbar}{a^{\frac{1}{2}} t^{\frac{3}{2}}})^n,
\label{eq:formal_PS_airy}
\end{align}
where the normalization point is chosen at the turning point $t_0 = 0$. 
The Borel transforms of the power series are given by
\begin{align}
B^\pm(s, t) = \sum_{n=0}^\infty \frac{\Gamma\qty(n + \frac{1}{6}) \Gamma\qty(n + \frac{5}{6})}{2\pi \Gamma(n + 1)^2} 
\qty(\pm \frac{3}{4} \frac{s}{a^{\frac{1}{2}} t^{\frac{3}{2}}})^n 
= \frac{\Gamma\qty(\frac{1}{6}) \Gamma\qty(\frac{5}{6})}{2\pi} \, {}_2F_1 \qty(\frac{1}{6}, \frac{5}{6}, 1, \frac{3}{4} \frac{s}{a^{\frac{1}{2}} t^{\frac{3}{2}}}),
\end{align}
where ${}_2F_1(a, b, c, z)$ denotes the hypergeometric function. Performing the Laplace transformation in Eq.~\eqref{eq:Borel_psi} in Regions I, II and III (see Fig.~\ref{fig:airy_stokes}), 
we obtain the Borel-resummed solutions:
\begin{align}
{\renewcommand{\arraystretch}{1.3}
{\setlength{\arraycolsep}{3.5mm}
\begin{array}{c|cc}
\text{Region} & \psi_{\rm I}^+(t,0) & \psi_{\rm I}^-(t,0) \\ \hline
\text{I} & \dfrac{C}{2}\Big[ \mathrm{Bi}(\alpha t) - i\,\mathrm{Ai}(\alpha t) \Big] & C\,\mathrm{Ai}(\alpha t) \\[1mm]
\text{II} & \dfrac{C}{2}\Big[ \mathrm{Bi}(\alpha t) + i\,\mathrm{Ai}(\alpha t) \Big] & C\,\mathrm{Ai}(\alpha t) \\[1mm]
\text{III} & \dfrac{C}{2}\Big[ \mathrm{Bi}(\alpha t) + i\,\mathrm{Ai}(\alpha t) \Big] & \dfrac{C}{2}\Big[ \mathrm{Ai}(\alpha t) + i\,\mathrm{Bi}(\alpha t) \Big]
\end{array}}}
\label{eq:airy_resummed}
\end{align}
where $C$ and $\alpha$ are constants given by
\begin{align}
C = \sqrt{\pi} / (a \hbar)^{\frac{1}{6}}, \qquad 
\alpha = a^{\frac{1}{3}} / \hbar^{\frac{2}{3}}. 
\end{align}
The functions $\mathrm{Ai}(y)$ and $\mathrm{Bi}(y)$, known as the Airy functions, are defined as integrals along the contours $\gamma_\pm$ shown in Fig.~\ref{fig:contour_airy}:
\begin{align}
f(y) = \frac{1}{2\pi} \int_{\gamma} dz \, \exp \qty[i \left( \frac{z^3}{3} + yz \right)], \qquad 
\gamma = \left\{ 
\begin{array}{cc}
\gamma_+ - \gamma_- & \text{for $\mathrm{Ai}(y)$}, \\
\gamma_+ + \gamma_- & \text{for $\mathrm{Bi}(y)$}.
\end{array} \right..
\end{align}

The Borel-resummed solutions in Eq.~\eqref{eq:airy_resummed} indicate that on the Stokes curve between Regions~I and II ($t \in \mathbb{R}_{\geq 0}$), the growing WKB solution $\psi^+$ exhibits a discontinuity. The connection matrix between Regions~I and II, which relates $\psi_{\rm I}^{\pm}$ and $\psi_{\rm II}^{\pm}$ as in Eq.~\eqref{eq:connection_12}, is given by
\begin{align}
T_{\rm II, I}(0) = 
\begin{pmatrix}
1 & 0 \\
-i & 1
\end{pmatrix}.
\label{eq:connection_matrix1}
\end{align}
Similarly, the connection matrices for the other Stokes curves and the branch cut on the negative real axis are determined as
\begin{align}
T_{\rm III, II}(0) = 
\begin{pmatrix}
1 & -i \\
0 & 1
\end{pmatrix}, \quad
T_{\rm I, III'}(0) = 
\begin{pmatrix}
1 & -i \\
0 & 1
\end{pmatrix}, \quad
T_{\rm III', III}(0) = 
\begin{pmatrix}
0 & i \\
i & 0
\end{pmatrix}.
\label{eq:connection_matrix2}
\end{align}
The Region~III$'$ corresponds to the sector $2\pi/3 <\arg t < 4\pi/3$, which is the counterpart of Region~III on the second Riemann sheet\footnote{
The branch cut of $W_0(t) \propto t^{\frac{1}{2}}$ is placed on the negative real axis. When $t$ crosses this branch cut from Region~III to Region~III$'$, the solutions $\psi^+$ and $\psi^-$ are swapped and multiplied by a factor of $i$.
}.

In general, the Airy-type connection matrices given in Eqs.~\eqref{eq:connection_matrix1} and \eqref{eq:connection_matrix2} can be used to derive connection formulas for any \textit{non-degenerate} Stokes curves, i.e., curves that do not overlap with other curves and emanate from simple turning points. 

Consider a simple turning point $t = t_0$, around which the potential function can be approximated as
\begin{align}
Q(t) = c (t - t_0) + \cdots.
\end{align}
Let $\mathcal{C}$ denote one of the Stokes curves emanating from the turning point $t_0$ and separating  Regions~I and II. 
When the variable $t$ crosses the Stokes curve $\mathcal{C}$ from Region I to II in a clockwise direction around the turning point, the Borel-resummed solutions normalized at $t_0$ satisfy the connection formula:
\begin{align}
\Big( \psi_{\rm I}^+(t, t_0), \psi_{\rm I}^-(t, t_0) \Big) = 
\Big( \psi_{\rm II}^+(t, t_0), \psi_{\rm II}^-(t, t_0) \Big) \, T_{\rm II, I}(t_0),
\label{eq:connection_nondegenerate}
\end{align}
where the connection matrix $T_{\rm II, I}(t_0)$ depends on the asymptotic behavior of the leading-order part $W_{\rm sol}(t,\hbar)$:
\begin{align}
T_{\rm II, I}(t_0)= 
T_+ &= 
\begin{pmatrix}
1 & 0 \\ 
-i & 1 
\end{pmatrix} \quad
\mbox{if $\displaystyle \, \frac{W_0}{\hbar} = \frac{1}{\hbar} \int_{t_0}^t \sqrt{Q_0(t')} \, dt'$ is increasing along ${\cal C}$}, \label{eq:Tplus} \\
T_{\rm II, I}(t_0) 
= T_- &= 
\begin{pmatrix}
1 & -i \\ 
0 & 1 
\end{pmatrix} \quad
\mbox{if $\displaystyle \, \frac{W_0}{\hbar} = \frac{1}{\hbar} \int_{t_0}^t \sqrt{Q_0(t')} \, dt'$ is decreasing along ${\cal C}$}. \label{eq:Tminus}
\end{align}
If $t$ crosses the Stokes curve $\mathcal{C}$ in a counterclockwise direction around the turning point, the connection matrix is given by the inverse of $T_{\pm}$.

\subsection{Connection formula for degenerate Stokes curves}\label{sec:3-2}
The Airy-type connection formula Eq.~\eqref{eq:connection_nondegenerate} can be used to relate the Borel-resummed WKB solutions in adjacent Stokes regions separated by a non-degenerate Stokes curve.
However, to apply the exact-WKB analysis to a two-level system of the form~\eqref{eq:Sch_eq}, it is necessary to establish the connection formula for \textit{degenerate Stokes curves}. 
To illustrate this, 
let us consider the Landau-Zener problem, i.e., a two-level system governed by the Schr\"odinger equation:
\begin{align}
i \hbar \partial_t \begin{pmatrix} \psi_1 \\ \psi_2 \end{pmatrix} = H \begin{pmatrix} \psi_1 \\ \psi_2 \end{pmatrix} =  
\begin{pmatrix}
\Delta & v t \\
v t & - \Delta  
\end{pmatrix} \begin{pmatrix} \psi_1 \\ \psi_2 \end{pmatrix},
\label{eq:LZ_Hamiltonian}
\end{align}
where $v$ and $\Delta$ are real constants.
This equation corresponds to Eq.~\eqref{eq:Sch_eq} with $f_1(t) =v t,~f_2(t)=0,~f_3(t)=\Delta$. 
Although this Hamiltonian is not periodic, it shares several key characteristics with Floquet systems in the context of the exact-WKB analysis (see also Ref.~\cite{Enomoto:2021hfv} for further discussion on the connection between the Landau-Zener problem and the exact-WKB analysis). In the following, we analyze the transition rate between the two instantaneous eigenstates using the exact-WKB analysis. 

\begin{figure}[t]
\centering
\fbox{
\includegraphics[width=60mm, page=2]{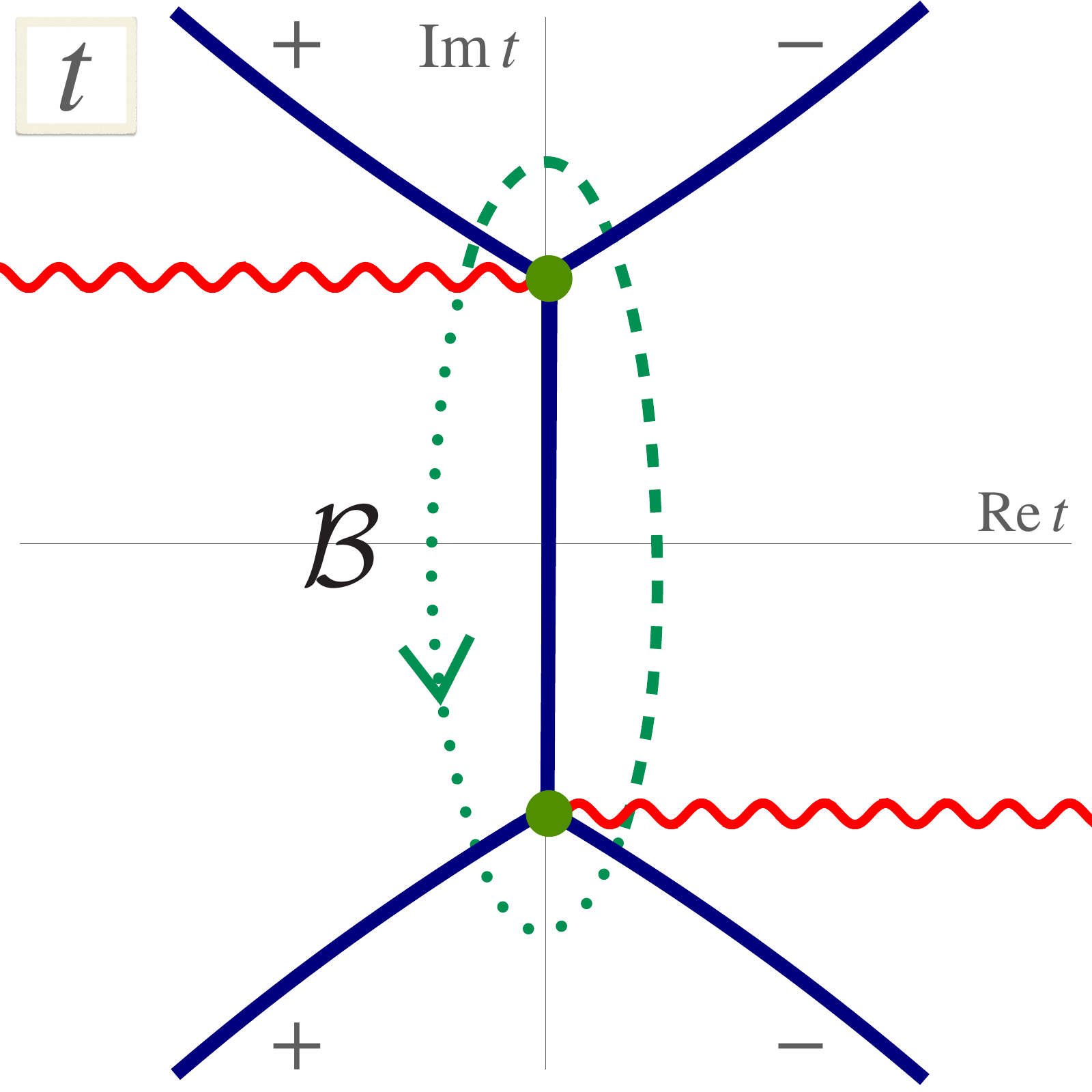}}
\caption{Stokes graph for the Landau-Zener problem.}
\label{fig:LZ_graph}
\end{figure}

\begin{figure}[t]
\begin{minipage}{0.45\linewidth}
\centering
\fbox{
\includegraphics[width=60mm, page=3]{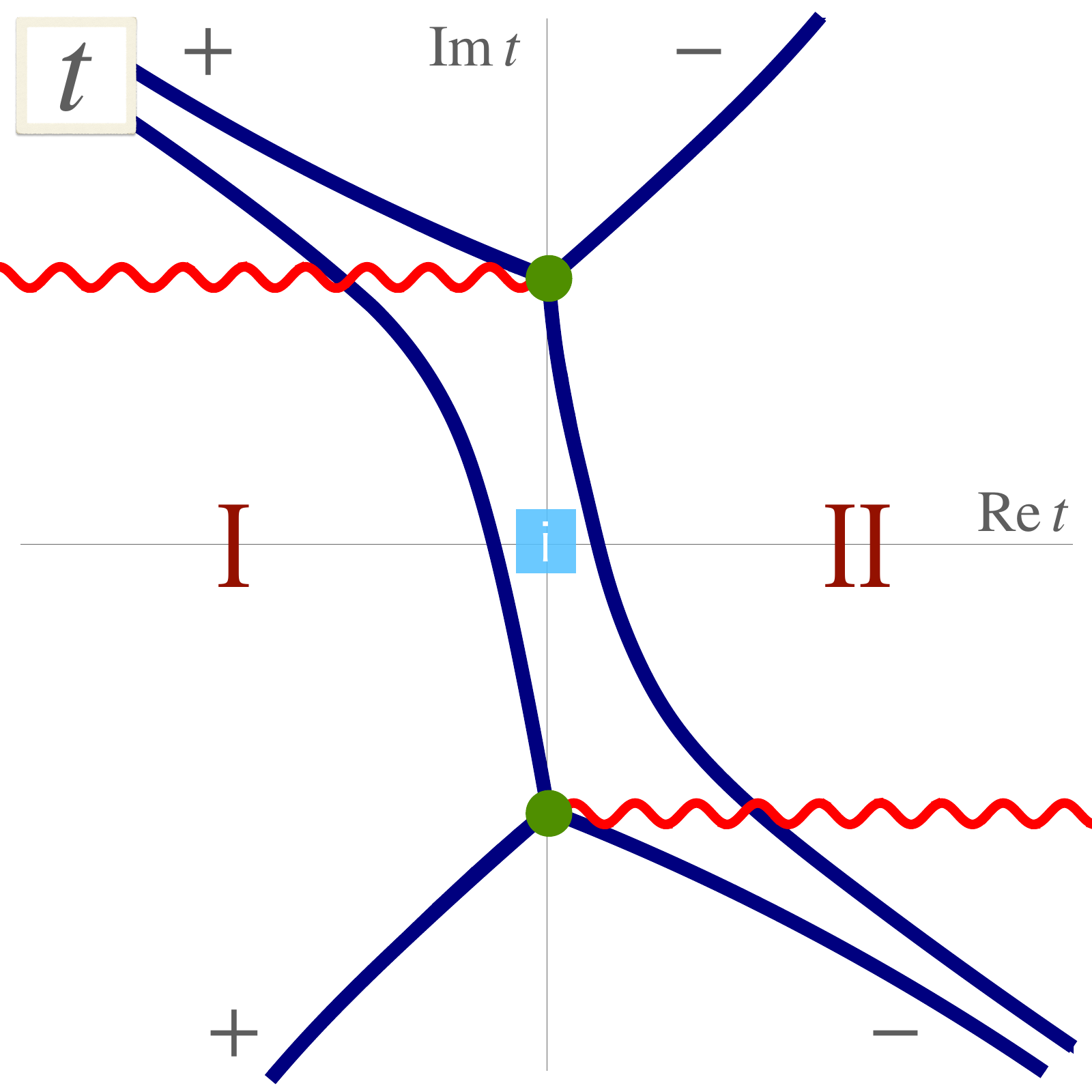}}
\caption{Stokes graph for $\arg \hbar > 0$}
\label{fig:LZ_graph_complex_p}
\end{minipage}
~~~~
\begin{minipage}{0.45\linewidth}
\centering
\fbox{
\includegraphics[width=60mm, page=4]{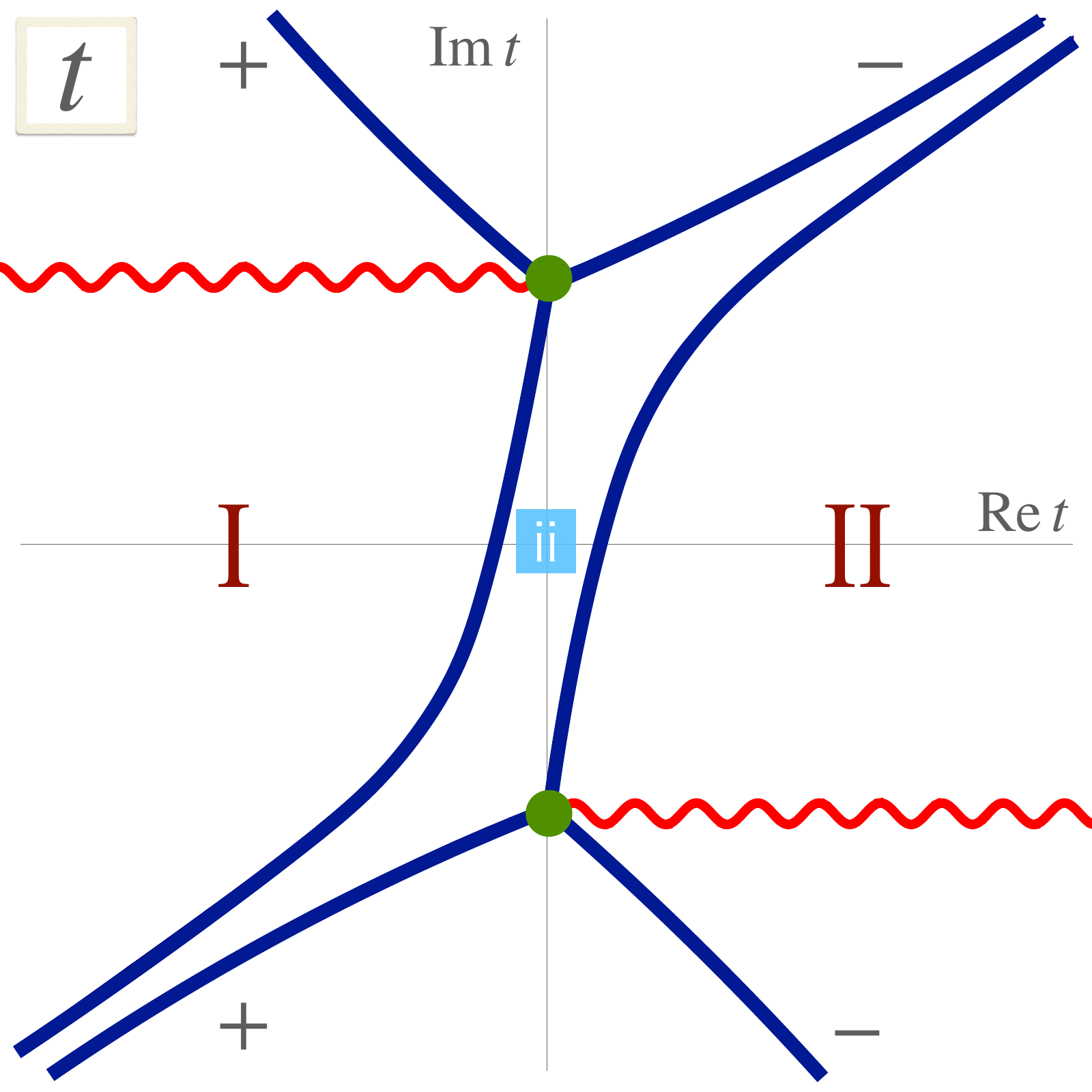}}
\caption{Stokes graph for $\arg \hbar < 0$}
\label{fig:LZ_graph_complex_m}
\end{minipage}
\end{figure}

By using the mapping \eqref{eq:linear_combination} with $c_1 = c_2 = 1/\sqrt{2}$, the two-level system~\eqref{eq:LZ_Hamiltonian} can be mapped to the Weber equation, i.e., a second-order differential equation of the form Eq.~\eqref{eq:Sch_eq_WKB} with $Q(t) = Q_0(t) + \hbar Q_1(t) = v^2 t^2+\Delta^2 + i \hbar v$:
\begin{align}
\Big[ - \hbar^2 \partial_t^2 - (v^2 t^2+\Delta^2 + i \hbar v) \, \Big] \psi = 0, \hspace{10mm} \psi = \frac{1}{\sqrt{2}} (-\psi_1 + \psi_2).
\end{align} 
The Stokes graph of this system can be determined by solving the condition Eq.~\eqref{eq:def_curve}
with $Q_0(t) = -v^2 t^2 -\Delta^2$. 
As illustrated in Fig.~\ref{fig:LZ_graph}, the Stokes graph of this system has a Stokes curve connecting two turning points located at $t=\pm i \Delta/v \equiv t_{\pm}$. 
As shown in Figs.~\ref{fig:LZ_graph_complex_p} and \ref{fig:LZ_graph_complex_m}, this Stokes curve splits into two distinct Stokes curves, each emanating from one of the turning points, when the Planck constant $\hbar$ takes a complex value. 
This behavior is a general feature of Stokes curves connecting pairs of turning points. 
Therefore, we refer to such overlapping Stokes curves as \textit{degenerate Stokes curves}.  
Importantly, the structure of the resulting Stokes graph changes depending on the phase of the Planck constant $\arg \hbar$, as illustrated in Figs.~\ref{fig:LZ_graph_complex_p} and \ref{fig:LZ_graph_complex_m}. 
As a result, the connection formula also depends on $\arg \hbar$, as will be discussed below.

We now apply the exact-WKB analysis to analyze this system.
The formal power-series WKB solutions are expressed as
\begin{align}
\psi_{\rm P}^\pm(t,t_0) = \left( \frac{-1}{v^2 t^2+\Delta^2} \right)^\frac{1}{4} \exp \left[ \pm \int_{t_0}^t \left( \frac{i}{\hbar} \sqrt{v^2t^2+\Delta^2} + \frac{v}{2\sqrt{v^2t^2+\Delta^2}} \right) \right] \sum_{n=0}^\infty \psi_{n}^\pm(t,t_0) \, \hbar^n.
\end{align}
Although obtaining the explicit forms of the Borel transform $B^\pm(s,t,t_0)=\sum \psi_{n}^\pm(t,t_0) \, s^n/n!$ is difficult in this case, it can be shown that they have a singularity on the Borel plane at\footnote{The location of the fixed singularity can be determined from an integral expression for the solutions of the Weber equation that can be formally rewritten in a form of Borel resummation.}
\begin{align}
s = \pi \Delta^2/v.
\label{eq:LZ_fixed_singularity}
\end{align}

This singularity prevents the Laplace transform in Eq.~\eqref{eq:Borel_psi}, required for the Borel resummation, from being performed. As a result, the formal power series of the WKB solutions is non-Borel summable. 
Such singularities are referred to as \textit{fixed singularities}, distinguishing them from {\it movable singularities}, which depend on $t$ and make the series non-Borel summable only when $t$ lies on a Stokes curve.

\begin{figure}[t]
\centering
\fbox{
\includegraphics[width=100mm, bb=0 200 1040 560]{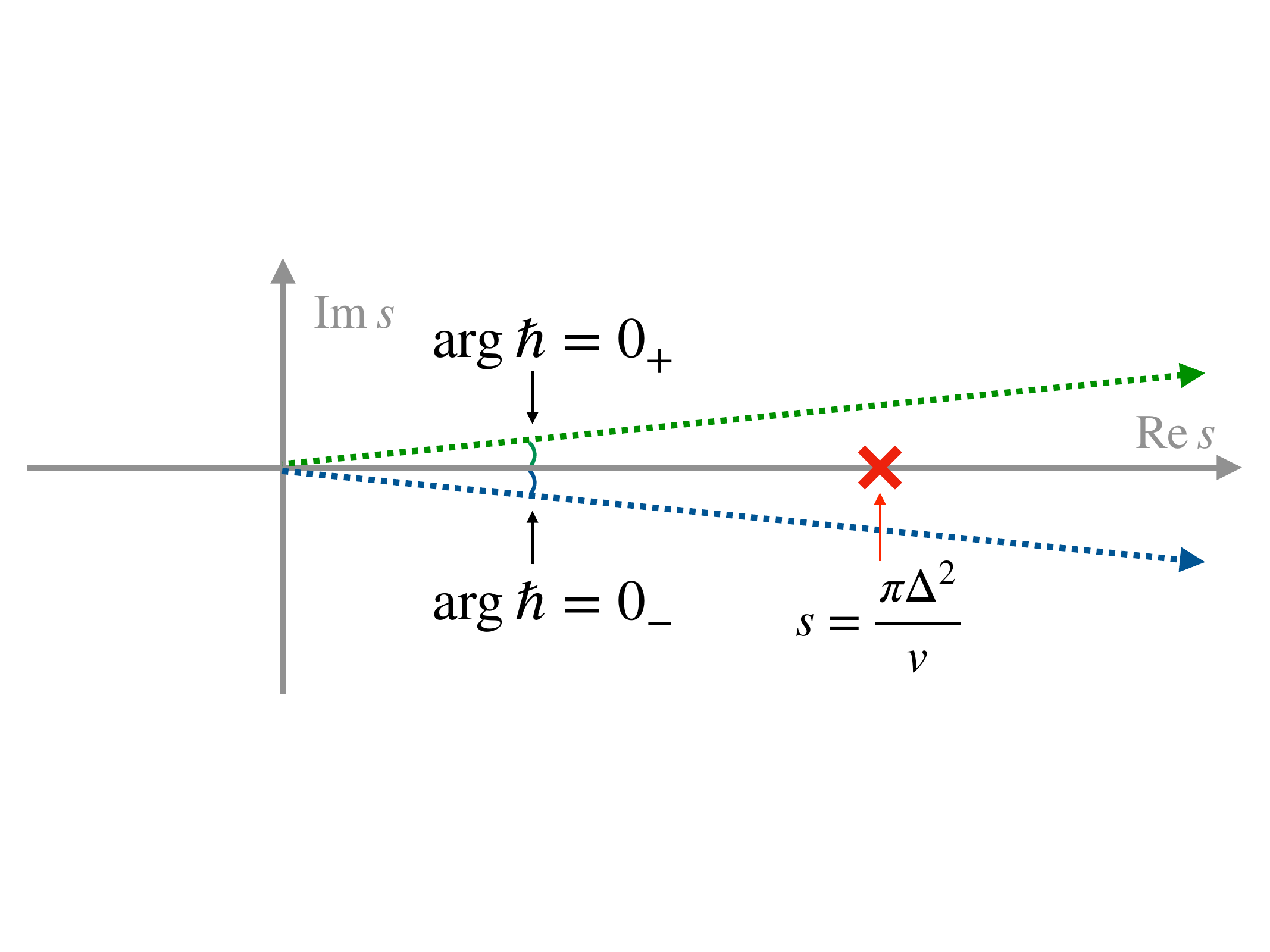}}
\caption{Rotated integration paths on the Borel plane.}
\label{fig:small_arghbar}
\end{figure}

To avoid the fixed singularity, we introduce a small phase to the Planck constant:
\begin{align}
\hbar \in \mathbb{R} \quad \mapsto \quad \hbar \in \mathbb{C} \qquad (\arg h \not = 0).
\end{align}
Correspondingly, we rotate the integration path on the Borel plane and define the Borel resummation along the rotated path as
\begin{align}
\mathcal{S}_{\arg \hbar}[\psi_{\rm P}^\pm(t, t_0)] = 
\frac{1}{\hbar} \frac{1}{Q_0(t)^{\frac{1}{4}}} 
\exp \qty(\pm \frac{1}{\hbar} W_0(t,t_0)) 
\int_0^{\infty \, e^{i \arg \hbar}} \! e^{-\frac{s}{\hbar}} B^\pm(s, t) \, ds.
\label{eq:Borel_psi_complex}
\end{align}
As shown in Fig.~\ref{fig:small_arghbar}, the rotated path avoids the fixed singularity.
Consequently, the series becomes Borel summable. 
However, due to the fixed singularity Eq.~\eqref{eq:LZ_fixed_singularity}, the Borel-resummed form $\mathcal{S}_{\arg \hbar}[\psi_{\rm P}^\pm(t, t_0)]$ depends on the sign of $\arg \hbar$ and exhibits a discontinuous jump when $\hbar$ crosses the real axis. 

First, we examine the case $\arg \hbar > 0$, where the degeneracy of the Stokes curve is lifted as illustrated in Fig.~\ref{fig:LZ_graph_complex_p}. 
This resolution allows us to apply the Airy-type connection formula given in Eq.~\eqref{eq:connection_nondegenerate}. 
After deriving the connection matrix, we take the limit $\arg \hbar \to 0_+$ (quantities in this limit are denoted by the subscript $0_+$). By repeatedly applying the Airy-type connection formula, we find that the WKB solutions in Regions I and II are related to each other as 
\begin{align}
(\psi^+_{\rm I,0_+}(t, t_+), \psi^-_{\rm I,0_+}(t, t_+)) 
&= (\psi^+_{\rm i,0_+}(t, t_+), \psi^-_{\rm i,0_+}(t, t_+)) \, T_- \\
&= (\psi^+_{\rm i,0_+}(t, t_-), \psi^-_{\rm i,0_+}(t, t_-)) \, \mathcal{N}_{t_-, t_+} T_- \notag \\
&= (\psi^+_{\rm II,0_+}(t, t_-), \psi^-_{\rm II,0_+}(t, t_-)) \, T_+^{-1} \mathcal{N}_{t_-, t_+} T_-,
\label{eq:connection_plus}
\end{align}
where $T_\pm$ are the connection matrices for non-degenerate Stokes curves defined in Eqs.~\eqref{eq:Tplus} and \eqref{eq:Tminus}. 
The matrix $\mathcal{N}_{t_-, t_+}$ is the normalization matrix defined in Eq.~\eqref{eq:normt0t1}, 
and is given explicitly by:
\begin{align}
\mathcal{N}_{t_-, t_+} = 
\begin{pmatrix}
B^{\frac{1}{2}} & 0 \\ 
0 & B^{-\frac{1}{2}}
\end{pmatrix},
\end{align}
where $B$ is defined by the integral along the cycle $\mathcal{B}$ in Fig.~\ref{fig:LZ_graph} that encloses the pair of turning points $t = t_\pm$:
\begin{align}
B = \exp \left( \oint_{\mathcal{B}} S_{\rm odd} \, dt \right) = -\exp \left(-\frac{\pi \Delta^2}{v \hbar} \right).
\end{align}
The integral is evaluated by enlarging the cycle ${\cal B}$ to an infinitely large circle and using the asymptotic behavior of $S_{\rm odd}=\partial_t W/\hbar$ for large $t$, which is determined from Eq.~\eqref{eq:Riccati}. 
Adjusting the normalization point $t_0$ to the origin $t = 0$ in Eq.~\eqref{eq:connection_plus} as
\begin{align}
(\psi^+_{\rm I,0_+}(t, t_\pm), \psi^-_{\rm I,0_+}(t, t_\pm)) = (\psi^+_{\rm I,0_+}(t, 0), \psi^-_{\rm I,0_+}(t, 0)) \, {\cal N}_{t_-,t_+}^{\pm \frac{1}{2}}, 
\end{align}
we obtain the connection formula for $\arg \hbar > 0$:
\begin{align}
(\psi^+_{\rm I,0_+}(t, 0), \psi^-_{\rm I,0_+}(t, 0)) 
= (\psi^+_{\rm II,0_+}(t, 0), \psi^-_{\rm II,0_+}(t, 0)) \, T_{0_+},
\end{align}
where the connection matrix is
\begin{align}
T_{0_+} = 
\mathcal{N}_{t_-, t_+}^{-\frac{1}{2}} 
\qty(T_+^{-1} \mathcal{N}_{t_-, t_+} T_-) 
\mathcal{N}_{t_-, t_+}^{-\frac{1}{2}} 
= 
\begin{pmatrix}
1 & -i B^{\frac{1}{2}} \\ 
i B^{\frac{1}{2}} & 1 + B
\end{pmatrix}.
\label{eq:T_plus}
\end{align}
Applying a similar procedure to the case of $\arg \hbar < 0$, we obtain the connection formula for $\arg \hbar < 0$:
\begin{align}
(\psi^+_{\rm I,0_-}(t, 0), \psi^-_{\rm I,0_-}(t, 0)) 
= (\psi^+_{\rm II,0_-}(t, 0), \psi^-_{\rm II,0_-}(t, 0)) \, T_{0_-},
\end{align}
with the connection matrix $T_{0_-}$ given by
\begin{align}
T_{0_-} = 
\mathcal{N}_{t_-, t_+}^{\frac{1}{2}} 
\left(T_- \mathcal{N}_{t_-, t_+} T_+^{-1}\right) 
\mathcal{N}_{t_-, t_+}^{\frac{1}{2}} 
= 
\begin{pmatrix}
1 + B & -i B^{\frac{1}{2}} \\ 
i B^{\frac{1}{2}} & 1
\end{pmatrix}.
\label{eq:T_minus}
\end{align}

Eqs.~\eqref{eq:T_plus} and \eqref{eq:T_minus} show that the connection matrix depends on the sign of $\arg \hbar$. This dependence originates from the fixed singularity, which leads to discontinuous jumps of the Borel-resummed solutions. The Borel-resummed solutions for $\arg \hbar > 0$ and $\arg \hbar < 0$ are related to each other as
\begin{align}
(\psi^+_{\rm \, I,\,0_+}(t, 0), \psi^-_{\rm \,I,\,0_+}(t, 0)) 
&= (\psi^+_{\rm \, I,\,0_-}(t, 0), \psi^-_{\rm \,I,\,0_-}(t, 0)) \, \Sigma(B), \label{eq:Borel_rel_1} \\
(\psi^+_{\rm II,0_+}(t, 0), \psi^-_{\rm II,0_+}(t, 0)) 
&= (\psi^+_{\rm II,0_-}(t, 0), \psi^-_{\rm II,0_-}(t, 0)) \, \Sigma(B)^{-1}, \label{eq:Borel_rel_2}
\end{align}
where the matrix $\Sigma(B)$ is defined as
\begin{align}
\Sigma(B) = 
\begin{pmatrix}
(1 + B)^{-\frac{1}{2}} & 0 \\ 
0 & (1 + B)^{\frac{1}{2}}
\end{pmatrix}. \label{eq:def_Sigma}
\end{align}
This map, relating the Borel-resummed solutions for $\arg \hbar > 0$ and $\arg \hbar < 0$, is an example of the \textit{Stokes automorphism}, which will be introduced in the next subsection.  

To write down the connection formula for the degenerate Stokes curve, it is convenient to introduce solutions that are intermediate between the Borel-resummed solutions for $\arg \hbar > 0$ and $\arg \hbar < 0$ defined as 
\begin{align}
(\psi^+_{\rm I}(t, 0), \psi^-_{\rm I}(t, 0)) 
&= (\psi^+_{\rm \, I,\,0_\pm}(t, 0), \psi^-_{\rm \, I , \,0_\pm}(t, 0)) \, \Sigma(B)^{\pm \frac{1}{2}}, \\
(\psi^+_{\rm II}(t, 0), \psi^-_{\rm II}(t, 0)) 
&= (\psi^+_{\rm II,0_\pm}(t, 0), \psi^-_{\rm II,0_\pm}(t, 0)) \, \Sigma(B)^{\mp \frac{1}{2}}.
\end{align}
The procedure to obtain these intermediate solutions is called \textit{median resummation}, which will be introduced in the next subsection. 
Using the median-resummed WKB solutions, the connection formula for the degenerate Stokes curve is expressed as
\begin{align}
(\psi^+_{\rm I}(t, 0), \psi^-_{\rm I}(t, 0)) 
= (\psi^+_{\rm II}(t, 0), \psi^-_{\rm II}(t, 0)) \, R,
\label{eq:LZ_connection}
\end{align}
where the connection matrix $R$ for the degenerate Stokes curve is given by
\begin{align}
R = 
\Sigma(B)^{\mp \frac{1}{2}} T_{0_\pm} \Sigma(B)^{\mp \frac{1}{2}} 
= 
\begin{pmatrix}
\sqrt{1 + B} & -i \sqrt{B} \\ 
i \sqrt{B} & \sqrt{1 + B}
\end{pmatrix}.
\label{eq:weber_connection}
\end{align}
From a technical standpoint, the connection matrices associated with degenerate Stokes curves have not been thoroughly investigated~(\ref{eq:weber_connection}) and, along with its generalization presented later in the paper (e.g., \ref{eq:Mex}), the first explicit formulas for such connection matrices will be obtained.

This connection formula determines the transition rate in the system described by Eq.~\eqref{eq:LZ_Hamiltonian}. 
To see this, we construct a pair of solutions to the Schr\"odinger equation, Eq.~\eqref{eq:LZ_Hamiltonian}, from the median-resummed WKB solutions in each Stokes region. Using the mapping given in Eq.~\eqref{eq:1st_to_2nd} with $c_1 = c_2 = 1/\sqrt{2}$, we obtain the following $2\times2$ matrix $G_I(t)$ whose columns are linearly independent solutions to the Schr\"odinger equation:
\begin{align}
(\psi^+_{\rm I}(t, 0), \psi^-_{\rm I}(t, 0)) 
\rightarrow G_{\rm I}(t) = \frac{-1}{\sqrt{2} v \Delta} 
\begin{pmatrix}
(i \hbar \partial_t + v t + \Delta) \psi_I^+(t, 0) & (i \hbar \partial_t + v t + \Delta) \psi_I^-(t, 0) \\
(i \hbar \partial_t + v t - \Delta) \psi_I^+(t, 0) & (i \hbar \partial_t + v t - \Delta) \psi_I^-(t, 0)
\end{pmatrix},
\end{align}
and similarly, we can construct $G_{\rm II}(t)$ from $(\psi^+_{\rm II}(t, 0), \psi^-_{\rm II}(t, 0))$. 
The connection formula Eq.~\eqref{eq:LZ_connection} implies that these solutions are related to each other as
\begin{align}
G_{\rm I}(t) = G_{\rm II}(t) R = G_{\rm II}(t) \begin{pmatrix}
\sqrt{1 - e^{-\frac{\pi \Delta^2}{v \hbar}}} & -e^{-\frac{\pi \Delta^2}{2v \hbar}} \\
e^{-\frac{\pi \Delta^2}{2v \hbar}} & \sqrt{1 - e^{-\frac{\pi \Delta^2}{v \hbar}}}
\end{pmatrix}.
\end{align}
Using this relation, we can determine the behavior of the solutions for $t \to \pm \infty$. 
The column vectors of the solutions $G_{\rm I}(t)$ and $G_{\rm II}(t)$ asymptotically approach the instantaneous eigenvectors of the Hamiltonian in the limit $t \rightarrow \mp \infty$:
\begin{align}
G_{\rm I}(t) \overset{t \to -\infty}{\longrightarrow}
\begin{pmatrix}
\xi_1^+(t) & \xi_1^-(t) \\
\xi_2^+(t) & \xi_2^-(t)
\end{pmatrix} 
e^{i \theta_{\rm I}(t) \sigma_3}, 
\quad
G_{\rm II}(t) \overset{t \to +\infty}{\longrightarrow}
\begin{pmatrix}
\xi_1^+(t) & \xi_1^-(t) \\
\xi_2^+(t) & \xi_2^-(t)
\end{pmatrix} 
e^{i \theta_{\rm II}(t) \sigma_3},
\end{align}
where $\xi^\pm(t) = (\xi_1^\pm(t), \xi_2^\pm(t))^{\rm T}$ are the instantaneous eigenvectors of $H(t)$ and $\theta_{\rm I,II}(t)$ represent the dynamical phases, which include additive constants. 
Using the connection formula, 
we can determine the behavior of the solution $G_{\rm I}(t)$ in the limit $t \to +\infty$: 
\begin{align}
G_{\rm I}(t) \overset{t \to +\infty}{\longrightarrow} 
\begin{pmatrix}
\xi_1^+(t) & \xi_1^-(t) \\
\xi_2^+(t) & \xi_2^-(t)
\end{pmatrix} 
e^{i \theta_{\rm II}(t) \sigma_3} 
\begin{pmatrix}
\sqrt{1 - e^{-\frac{\pi \Delta^2}{v \hbar}}} & -e^{-\frac{\pi \Delta^2}{2v \hbar}} \\
e^{-\frac{\pi \Delta^2}{2v \hbar}} & \sqrt{1 - e^{-\frac{\pi \Delta^2}{v \hbar}}}
\end{pmatrix}.
\end{align}
This implies that the off-diagonal elements of the matrix $R$ give the transition rate between the instantaneous eigenstates. This result is equivalent to the well-known Landau-Zener formula, which states that the transition rate between the ground and excited states in this system is
\begin{align}
|B| = \exp\left(-\frac{\pi \Delta^2}{v \hbar}\right).
\end{align}
This is a typical non-perturbative (non-adiabatic) effect that can be effectively analyzed using the exact-WKB analysis.

\subsection{Preparation for the application to Floquet systems}
\label{subsec:degenerate_Stokes}
\begin{figure}[htbp]
\centering
\includegraphics[clip, width=80mm]{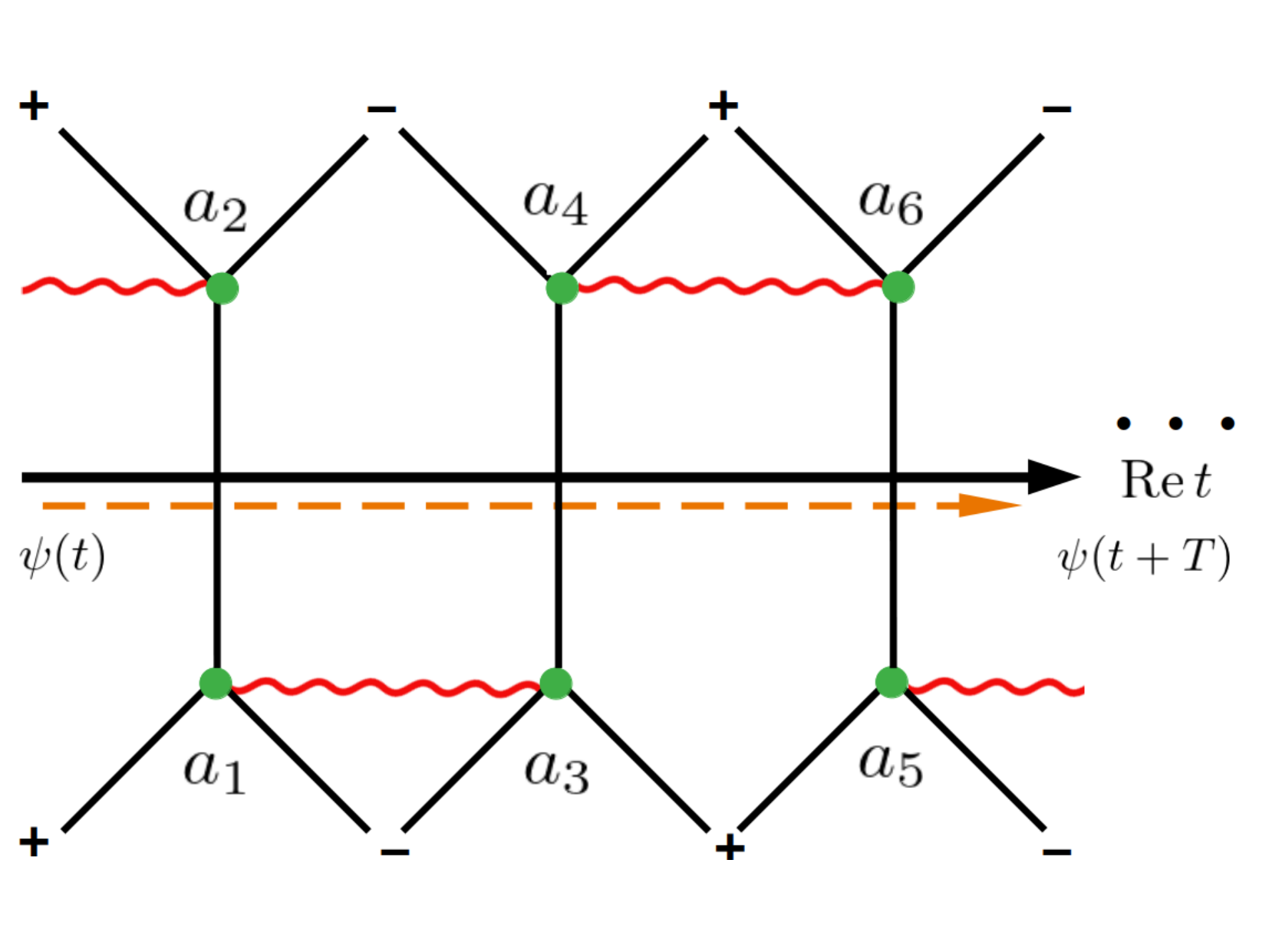}
\vspace{-8mm} \\
\caption{Typical Stokes graph in a Floquet system.}
\label{fig:degenerate_stokes}
\end{figure}

The application of exact-WKB analysis to Floquet systems inevitably involves degenerate Stokes curves.
Since the functions $f_i(t)~(i=1,2,3)$ are real when $t$ is on the real axis of the complex $t$-plane $(t \in \mathbb R)$, 
the function $Q_0(t) =- \sum_{i=1}^3 f_i^2(t)$ satisfies $\overline{Q_0(t)} = Q_0(\bar t)$.
Therefore, all turning points appear in complex conjugate pairs in Floquet systems.  These degenerate Stokes curves are composed of two individual Stokes curves emanating from each turning point in the pair. A typical Stokes graph in a Floquet system is illustrated in Fig.~\ref{fig:degenerate_stokes}.

\begin{figure}[tp]
  \begin{center}
    \begin{tabular}{cc}
      \begin{minipage}{0.5\hsize}
        \begin{center} 
            \includegraphics[clip, width=80mm]{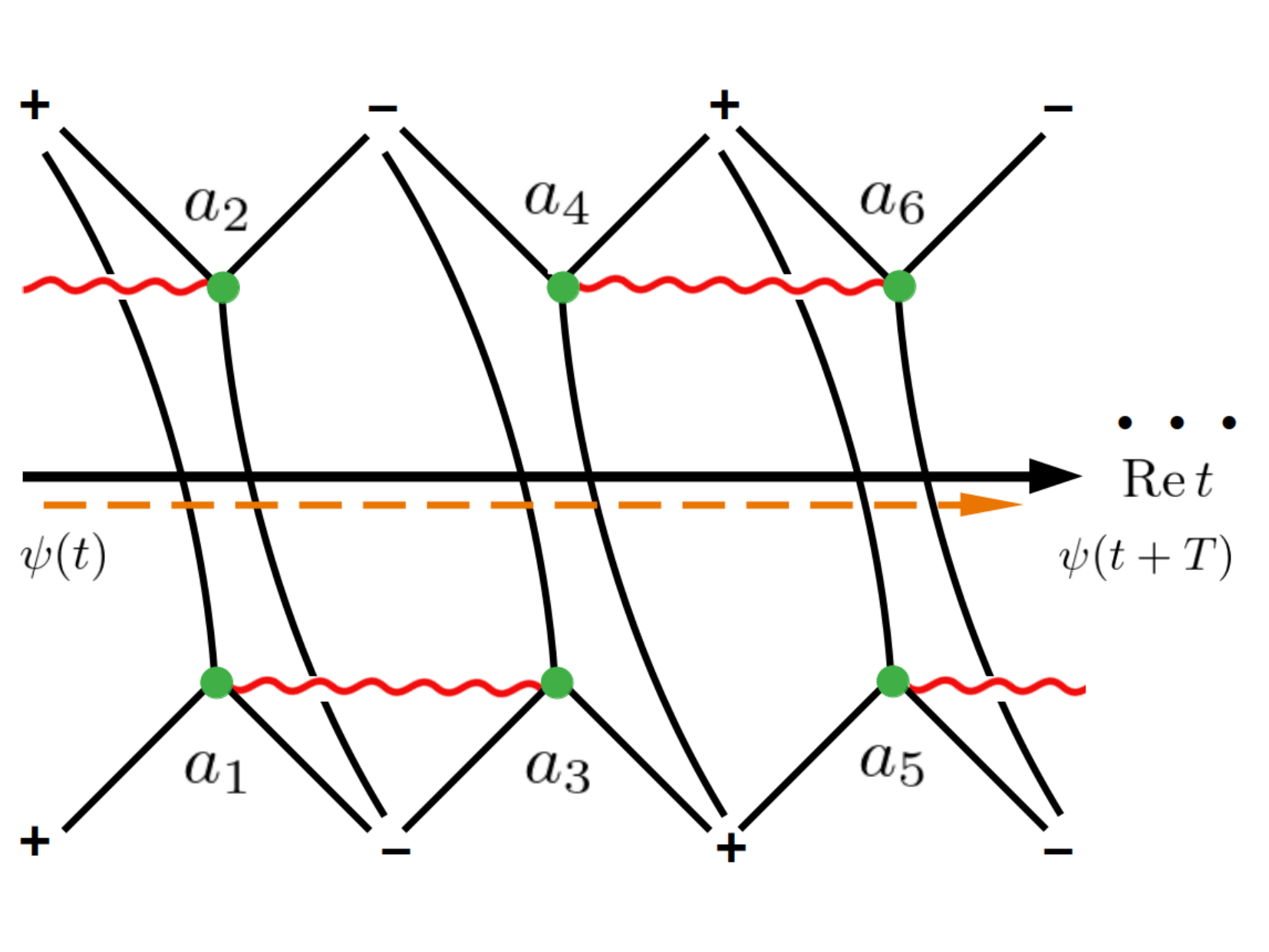} \vspace{-5mm} \\
(a) $\arg \hbar > 0$  
        \end{center}
      \end{minipage}
      \begin{minipage}{0.5\hsize}
        \begin{center} 
            \includegraphics[clip, width=80mm]{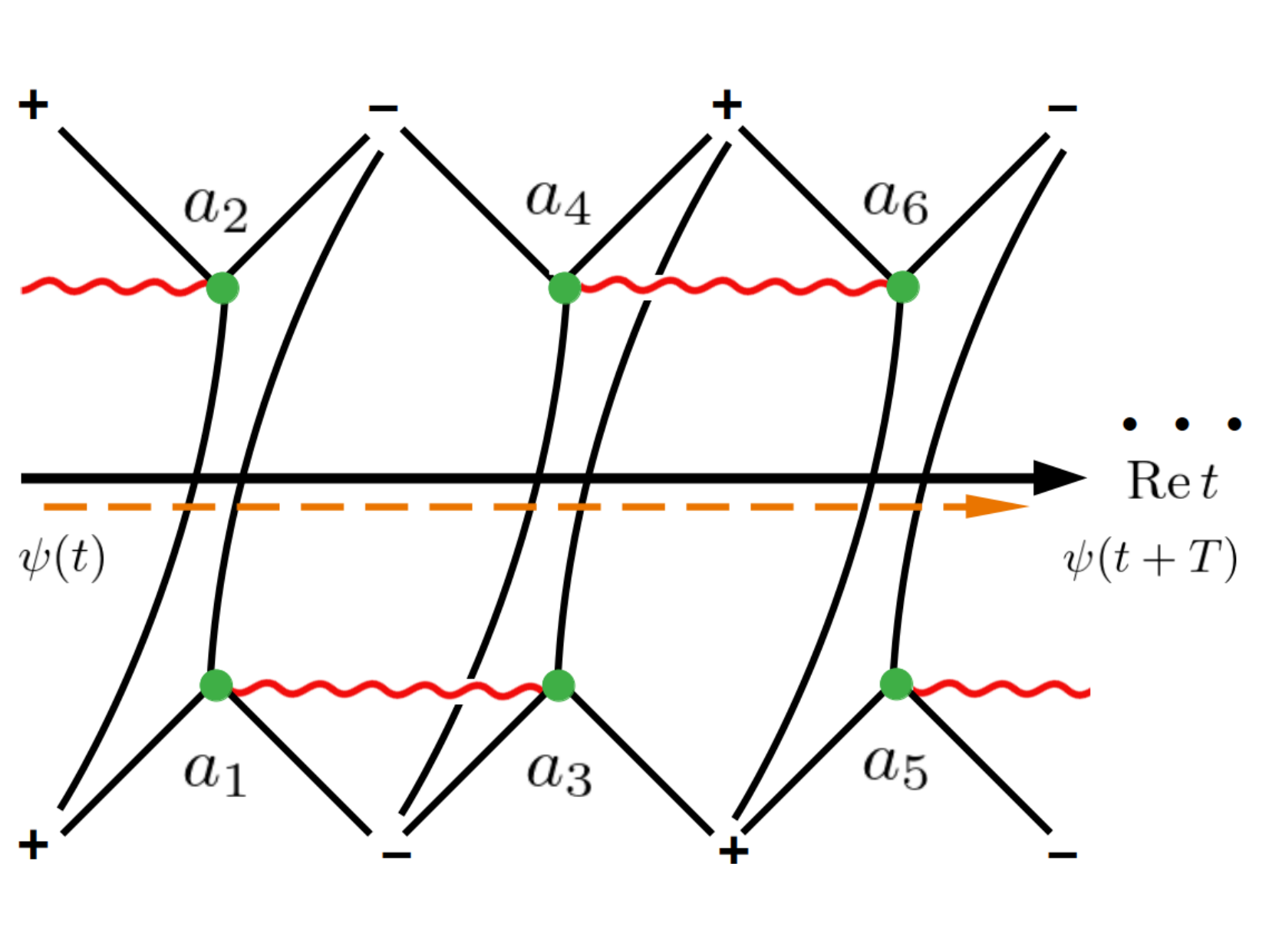} \vspace{-5mm} \\
(b) $\arg \hbar < 0$  
        \end{center}
      \end{minipage}      
    \end{tabular} 
    \caption{Resolved (lifted) versions of typical Stokes graphs in a Floquet system. To avoid the degeneracy of the Stokes curves, an infinitesimally small complex phase is introduced to $\hbar$, as illustrated in panels (a) and (b). Analytic continuation is performed along the real axis (orange dashed line) to compute the monodromy matrix under periodic boundary conditions.}
    \label{fig:stokes_graph graph1}
  \end{center}
\end{figure}

As in the previous case, 
we need to deform the system to resolve the degeneracy of the Stokes curves. This is achieved by introducing a small phase to $\hbar$.
Schematic diagrams of typical Stokes graphs for $\arg \hbar>0$ and $\arg \hbar<0$ are shown in Fig.~\ref{fig:stokes_graph graph1}. 
Each degenerate Stokes curve connecting a pair of turning points splits into two distinct curves.
This allows us to safely apply the Airy-type connection formula mentioned in the previous section \eqref{eq:connection_matrix2} to obtain the connection matrices.
However, since the structures of the Stokes graphs differ for $\arg \hbar > 0$ and $\arg \hbar <0$, there is a mismatch between the connection matrices obtained by taking the limits $\arg \hbar \rightarrow \pm 0$. 

To obtain an unambiguous result, we need to consider an additional discontinuity at $\arg \hbar = 0$ (see Appendix~\ref{sec:Quant_DDP_Stokes}). 
This discontinuity arises from a fixed singularity on the Borel plane. Through the analysis detailed in Sec.~\ref{subsec:3-2}, we can show that for each degenerate Stokes curve, a connection matrix of the form given by Eq.~\eqref{eq:weber_connection} can be used to relate the WKB solutions resummed in neighboring Stokes regions. Using these connection formulas, we can construct a monodromy matrix for each Floquet system.

\subsubsection{Median resummation and monodromy matrix for Floquet systems} \label{subsec:3-2}

Here, we outline the construction of the monodromy matrix using the exact-WKB analysis in a typical setup of Floquet systems
(see Appendix~\ref{sec:Quant_DDP_Stokes} and Ref.~\cite{Kamata:2021jrs} for more details). 
In generic contexts of the resurgence theory, this type of problem can be solved by constructing \textit{Stokes automorphism} and \textit{median resummation}. 

Suppose that a Stokes graph has degenerate Stokes curves when $\arg \hbar=0$. 
In such a case, we need to introduce an infinitesimally small complex phase to $\hbar$ to deform the Stokes graph as shown in Figs.~\ref{fig:stokes_graph graph1} (a) and (b).
Since the structure of the Stokes graph depends on the sign of $\arg \hbar$ in the limit $\arg \hbar \rightarrow \pm 0$, the resulting monodromy matrix $M_{0_\pm}$ also depends on the direction from which the limit is taken.  Nevertheless, the characteristic equations of $M_{0_\pm}$, from which the quasi-energies are determined, become identical after applying the Borel resummation:
\begin{align}
{\cal S}_{0_+} \left[ \det(M_{0_+} -e^{-i\epsilon T/\hbar}) \right] = {\cal S}_{0_-} \left[ \det(M_{0_-} - e^{-i\epsilon T/\hbar}) \right],
\label{eq:detMpm}
\end{align}
where ${\cal S}_{0_\pm}$ denote the Borel resummation along the rotated paths ${\cal S}_{\arg \hbar}$, defined in Eq.~\eqref{eq:Borel_psi_complex}, in the limit $t \rightarrow \pm 0$, respectively.

\begin{figure}[tp]
  \begin{center}
        \includegraphics[page=5, clip, width=105mm]{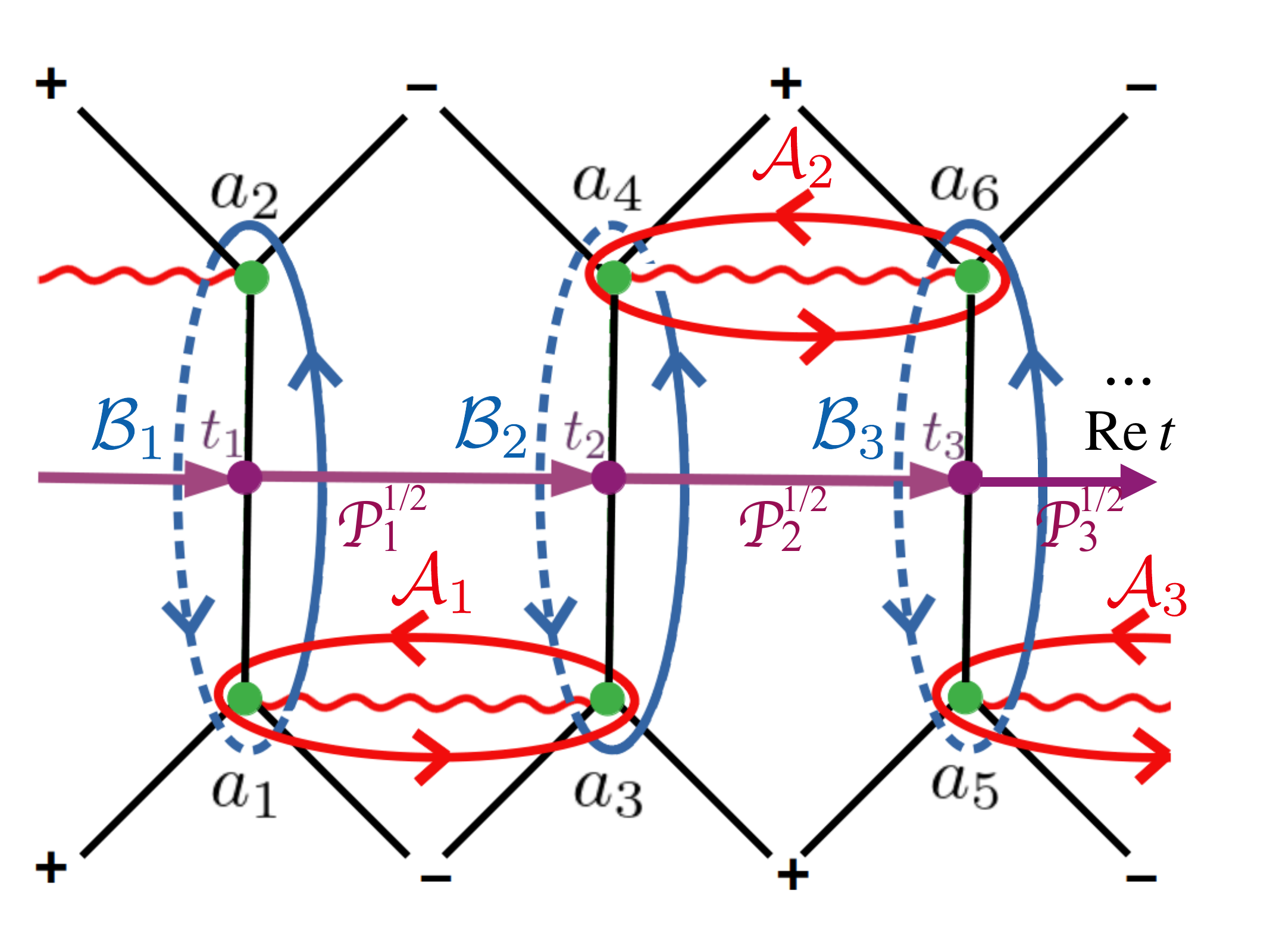}
    \caption{Cycles on the typical Stokes graph in Fig.~\ref{fig:degenerate_stokes}.   The colored cycles show the $A$- and $B$-cycles defined in Eq.~(\ref{eq:AB_def}).
    }
    \label{fig:cycles1}
  \end{center}
\end{figure}

As we will see, the monodromy matrices $M_{0_\pm}$ can be expressed as combinations of the connection matrices given in Eqs.~\eqref{eq:Tplus} and \eqref{eq:Tminus}, and the normalization matrices~\eqref{eq:normt0t1}.  The normalization matrix ${\mathcal N}_{t_1,t_0}$ between turning points $t_0$ and $t_1$ can be expressed by the integral along the cycle enclosing the turning points $t_0$ and $t_1$. 
Specifically, to construct a monodromy matrix, we need cycle integrals (the Voros symbols) along the ${\cal A}$- and ${\cal B}$-cycles going around turning points as shown in Fig.~\ref{fig:cycles1}.
The point is that the ${\cal A}$-cycle integral, which will be referred to just as $A$-cycle or $A$ below, and the wave function is not Borel summable when $\hbar$ is real positive.
In general, when a function of $\hbar$ gives a well-defined Borel resummation and is Borel non-summable in the limit $\arg(h) \rightarrow \pm 0$, its Borel-resummed form
can be obtained by using the median summation defined as
\begin{align}
{\cal S}_{\rm med} = {\cal S}_{0_+} \circ {\mathfrak S}^{-1/2} = {\cal S}_{0_-} \circ {\mathfrak S}^{+1/2},
\label{eq:S_med}
\end{align}
where ${\mathfrak S}^{\nu \in {\mathbb R}}$ is the (one parametrized) Stokes automorphism, which relates objects\footnote{More precisely, the Stokes automorphism relates trans-series for $\arg \hbar > 0$ and $\arg \hbar <0 $.} for $\arg \hbar > 0$ and $\arg \hbar <0$ [see Eq.~(\ref{eq:Stokes_aut}) in Appendix~\ref{sec:Quant_DDP_Stokes_story}].
The action of ${\mathfrak S}^{\nu}$ to $A$'s is known as the Delabaere-Dillinger-Pham (DDP) formula~\cite{DDP2,DP1}.
According to Appendices~\ref{sec:DDP_cycle} and \ref{sec:DDP_wave}, the DDP formula is given by
\begin{align}
&&   {\mathfrak S}^{\nu}[A^{\pm 1}_{2n+1}] = A^{\pm 1}_{2n + 1}\prod_{j=1}^2 (1+B_{2n+j})^{\mp \nu}, \qquad {\mathfrak S}^{\nu}[A^{\pm 1}_{2n+2}] = A^{\pm 1}_{2n+2}\prod_{j=1}^2 (1+B_{2n+1+j})^{\pm \nu}, \label{eq:St_auto_A_main} \\
&&   {\mathfrak S}^{\nu}[B_n] = B_n.
\end{align}
Furthermore, the DDP formula for a wave function normalized at $a_1$ is given by
\begin{align}
 {\mathfrak S}^{\nu}[\psi] = \psi [\Sigma(B_1)]^{\nu}, \qquad  \Sigma(B_1) = 
\begin{pmatrix}
(1 + B_1)^{-1/2} && 0 \\
0 && (1 + B_1)^{+1/2}  \\
\end{pmatrix}.
\end{align}
Notice that for a general function $f(A_1,A_2,\cdots, B_1,B_2,\cdots)$, the Stokes automorphism satisfies
\begin{align}
   {\mathfrak S}[f(A_1,A_2,\cdots, B_1,B_2,\cdots)] = f({\mathfrak S}[A_1],{\mathfrak S}[A_2],\cdots,{\mathfrak S}[B_1],{\mathfrak S}[B_2],\cdots).
\end{align}
As shown in Appendices~\ref{app:exact_monodromy} and \ref{app:2N_turning_points}, the monodromy matrix for the median-resummed wave functions can be obtained by applying the median resummation defined in Eq.~\eqref{eq:S_med} to $M_{\rm med}$,
\begin{align}
M_{\rm med} = \Sigma(B_1)^{\pm 1/2} {\mathfrak S}^{\pm 1/2}[ M_{0_\pm} ] \Sigma(B_1)^{\mp 1/2}.
\label{eq:Mmedsss}
\end{align}
Solving $\det (M_{\rm med} - e^{ i\epsilon T/\hbar})$, we can obtain a (trans-series form of) the quasi-energy without a discontinuity at $\arg(\hbar)=0$.

In a case that $Q_0(t)$ has $N$ complex conjugate pairs of turning points and that there is a degenerate Stokes curve intersecting the real axis between each pair of the turning points, 
the explicit form of $M_{\rm med}$ is given by (see Appendix~\ref{app:2N_turning_points})
\begin{align}
M_{\rm med} = 
\mathcal N_{t_1,t_0}^{-1} 
\left( \mathcal N_{t_1+T,\,t_N} \, R_N \cdots \mathcal N_{t_3,\,t_2} \, R_2 \ \mathcal N_{t_2,t_1} \, R_1 \right) 
\mathcal N_{t_1,t_0}, 
\label{eq:formula_Mmed}
\end{align}
where $t_n~(n=1,2,\cdots,N)$ are the intersections between the degenerate Stokes curves and the real axis, 
$\mathcal N_{t_{n},t_{n-1}}~(n=1,\cdots,N,N+1)$ with $t_{1}+T \equiv t_{N+1}$ the normalization matrices \eqref{eq:normt0t1}, and $R_n~(n=1,2,\cdots,N)$ the connection matrices for the degenerate Stokes curves:
\begin{align}
\mathcal N_{t_{n+1},t_n} = 
\left(
\begin{array}{cc}
P_n^{+\frac{1}{2}} & 0 \\
0 & P_n^{-\frac{1}{2}}
\end{array}
\right), \hspace{5mm}
R_n = 
\left(
\begin{array}{cc}
\sqrt{1+B_n} & -i \sqrt{B_n} \\
+ i \sqrt{B_n} & \sqrt{1+B_n} 
\end{array}
\right).
\end{align}
The quantities $P_{n}$ are given in terms of $A_n$ and $B_n$ as
\begin{align}
P_{n} = \left[ A_n B_{n}^{1/2} B_{n+1}^{-1/2} \right]^{(-1)^n},
\end{align}
which correspond to ${\cal P}_n$ in Fig.~\ref{fig:cycles1}.
In the next sections, we apply these formulas to a simple example of a Floquet system.

\section{Application of exact-WKB analysis to a Floquet system} \label{sec:4}

\subsection{A simple example}

In this section, we consider a simple model Hamiltonian to demonstrate how the exact-WKB analysis works in analyzing the non-perturbative aspects of a Floquet system. Our Hamiltonian is given by
\begin{align}
H = \sum_{i=1}^3 \sigma_i f_i(t), \hspace{5mm} \text{with} \hspace{5mm}
f_1(t) = v \sin(\omega t), \hspace{3mm} f_2(t) = 0, \hspace{3mm} f_3(t) = \Delta,
\label{eq:example1}
\end{align}
where $v$ and $\Delta$ are constants. For this Hamiltonian, the functions $\lambda(t)$ and $Q(t)$ in Eq.~\eqref{eq:2nd_order_diffeq} are given by
\begin{align}
\lambda = \frac{1}{2} \ln \frac{2 c_1 c_2 \Delta - (c_1^2 - c_2^2) v \sin(\omega t)}{i \hbar}, \hspace{10mm}
Q = Q_0(t) + Q_1(t) \hbar + Q_2(t) \hbar^2,
\end{align}
with
\begin{align}
Q_0 = -v^2 \sin^2(\omega t) - \Delta^2, \hspace{5mm}
Q_1 = -2 i \Delta \frac{c_1^2 + c_2^2}{c_1^2 - c_2^2} \partial_t \lambda(t), \hspace{5mm}
Q_2 = (\partial_t \lambda(t))^2 - \partial_t^2 \lambda(t).
\label{eq:Q_0etc}
\end{align}
For simplicity, we have set the coefficients of the linear combination $C = (c_1, c_2)^T$ in Eq.~\eqref{eq:linear_combination} to be time-independent. We will confirm that the results of the exact-WKB analysis are independent of these coefficients.

\begin{figure}[t]
\centering
\fbox{
\includegraphics[width=90mm, page=6]{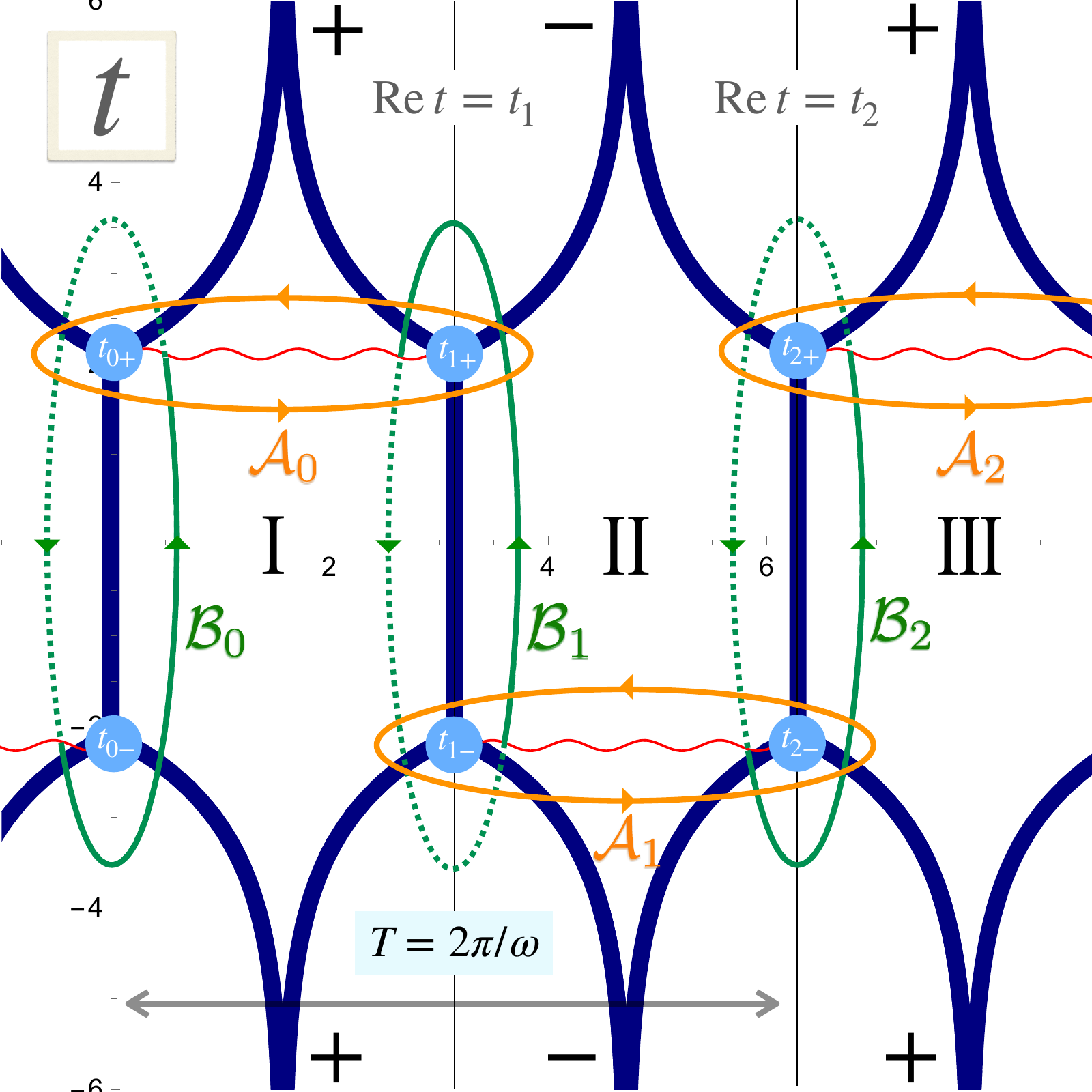}}
\caption{Stokes graph and integration cycles for $Q_0(t) = - v^2 \sin^2(t) -\Delta^2$ with parameters $\omega=1$, $v=0.25$ and $\Delta = 1$. The dots indicate the turning points, while the thick lines represent the Stokes curves. These Stokes curves partition the complex 
$t$-plane into distinct Stokes regions, labeled as I, II, III, etc. The period in this case is $T = 2\pi$.}
\label{fig:stokes_undeformed}
\end{figure}

The turning points, i.e., the zeros of $Q_0(t)$, are located at
\begin{align}
t_{n\pm} = \frac{1}{\omega} \left[ n \pi \pm \arcsin\left(\frac{i \Delta}{v}\right) \right], \quad n \in \mathbb{Z}.
\end{align}
The Stokes graph for $Q_0(t)$ is shown in Fig.~\ref{fig:stokes_undeformed}. The Stokes curves divide the complex $t$-plane into several Stokes regions. Regions overlapping with the real axis are labeled as Region I, II, III, and so on, numbered sequentially from $t = 0$.

The WKB solutions resummed in Regions I and III are related to each other as
\begin{align}
\mqty(\psi_{\rm I}^+(t, t_0) & \psi_{\rm I}^-(t, t_0)) 
&= \mqty(\psi_{\rm II}^+(t, t_0) & \psi_{\rm II}^-(t, t_0)) T_{\rm II, I}(t_0) \notag \\
&= \mqty(\psi_{\rm III}^+(t, t_0) & \psi_{\rm III}^-(t, t_0)) T_{\rm III, II}(t_0) T_{\rm II, I}(t_0),
\label{eq:connection_I_III}
\end{align}
where $T_{\rm II, I}(t_0)$ and $T_{\rm III, II}(t_0)$ are the connection matrices from Region I to II and from Region II to III, respectively.
The periodicity of the Hamiltonian $H(t+T)=H(t)~(T = 2\pi/\omega)$ implies the following relation between $\psi_{\rm I}^\pm$ and $\psi_{\rm III}^\pm$:\footnote{To derive this relation, note that the periodicity of the system allows the quantities in the Borel-resummation formula \eqref{eq:Borel_psi} to be shifted as 
\begin{align}
W_0(t, t_0) = W_0(t - T, t_0 - T) \quad \text{and} \quad B(s, t) = B(s, t - T). \notag
\end{align}
Since $t - T$ lies in Region I when $t$ is in Region III, the Borel resummation gives the relation 
\begin{align}
\psi_{\rm III}^\pm(t, t_0) = \psi_{\rm I}^\pm(t - T, t_0 - T). \notag
\end{align}
Through analytic continuation, this relation extends to all regions of the complex $t$-plane.
}
\begin{align}
\psi_{\rm III}^\pm(t, t_0) = \psi_{\rm I}^\pm(t - T, t_0 - T).
\label{eq:I_III}
\end{align}

Combining Eqs.~\eqref{eq:connection_I_III} and \eqref{eq:I_III}, we find that a pair of the WKB solutions $(\psi_{\rm I}^+(t, t_0), \psi_{\rm I}^-(t, t_0))$ transforms over one period as 
\begin{align}
\mqty(\psi_{\rm I}^+(t, t_0) & \psi_{\rm I}^-(t, t_0)) 
&= \mqty(\psi_{\rm I}^+(t - T, t_0 - T) & \psi_{\rm I}^-(t - T, t_0 - T)) T_{\rm III, II}(t_0) T_{\rm II, I}(t_0) \notag \\
&= \mqty(\psi_{\rm I}^+(t - T, t_0) & \psi_{\rm I}^-(t - T, t_0)) \mathcal{N}_{t_0, t_0 - T} T_{\rm III, II}(t_0) T_{\rm II, I}(t_0),
\label{eq:monodromy0}
\end{align}
where $\mathcal{N}_{t_0, t_0 - T}$ is the normalization matrix, relating the WKB solutions normalized at $t_0$ and $t_0 - T$ [see Eq.~\eqref{eq:normalization_matrix}].

Shifting $t \to t + T$ and using the periodicity $\mathcal{N}_{t_0, t_0 - T} = \mathcal{N}_{t_0 + T, t_0}$ in Eq.~\eqref{eq:monodromy0}, we find that the monodromy matrix $M(t_0)$ is given by:
\begin{align}
M(t_0) = \mathcal{N}_{t_0 + T, t_0} T_{\rm III, II}(t_0) T_{\rm II, I}(t_0).
\label{eq:monodromy_matrix}
\end{align}

Thus, the monodromy matrix $M(t_0)$ can be determined by calculating the connection matrices $T_{\rm II, I}(t_0)$ and $T_{\rm III, II}(t_0)$ through the connection formula. Once $M(t_0)$ is determined, the time-evolution unitary matrix $U(t)$ can be computed via Eq.~\eqref{eq:unitary_monodromy}, allowing us to derive physical quantities such as the Floquet Hamiltonian and quasi-energies. For this purpose, the matrix $M_{\rm med}$ given in Eq.~\eqref{eq:Mmedsss} is required.

As discussed in the previous section, the monodromy matrix $M_{\rm med}$ can be expressed in terms of the Voros symbols. Let us define the quantities $A_n$ and $B_n$ $(n \in \mathbb{Z})$ as follows:
\begin{align}
A_n = \exp \left( \oint_{\mathcal{A}_n} S_{\rm odd}(t') \, dt' \right), \hspace{10mm}
B_n = \exp \left( \oint_{\mathcal{B}_n} S_{\rm odd}(t') \, dt' \right),
\label{eq:AB_def}
\end{align}
where $\mathcal{A}_n$ and $\mathcal{B}_n$ are the cycles illustrated in Fig.~\ref{fig:stokes_undeformed}. 
The cycle $\mathcal{A}_n$ encircles the pairs of turning points $(t_{n+}, t_{(n+1)+})$ for even $n$ and $(t_{n-}, t_{(n+1)-})$ for odd $n$. The cycle $\mathcal{B}_n$, on the other hand, encircles the pair of turning points $(t_{n-}, t_{n+})$.
These quantities are necessary for constructing the component matrices that constitute the matrix 
$M_{\rm med}$ given in Eq.~\eqref{eq:formula_Mmed}:
\begin{align}
{\mathcal N}_{t_{n+1},t_n} = 
\left(
\begin{array}{cc}
P_n^{+\frac{1}{2}} & 0 \\
0 & P_n^{-\frac{1}{2}}  
\end{array}
\right),
\hspace{7mm}
R_n = 
\left(
\begin{array}{cc}
\sqrt{1+B_n} & -i \sqrt{B_n} \\
+ i \sqrt{B_n} & \sqrt{1+B_n} 
\end{array}
\right),
\label{eq:RinB}
\end{align}
where $P_n$ is defined as\footnote{
$P_n^{1/2}$ in Fig.~\ref{fig:cycles1} is an \textit{intuitive} picture, in the sense that it cannot be naively expressed by a line integral from $t_n$ to $t_{n+1}$ in general. 
This picture is valid only when ${\rm Im}[t_{n+}]$ are the same for all $n$, otherwise one should use the contour integral using $A$- and $B$-cycles, as is expressed in Eq.~(\ref{eq:AB_def}).
}
\begin{align}    
P_{n} = \left[ A_n B_{n}^{1/2} B_{n+1}^{-1/2} \right]^{(-1)^n}.
\label{eq:P_def}
\end{align}

Applying the formula for $M_{\rm med}$ given in Eq.~\eqref{eq:formula_Mmed} (see also Appendix \ref{app:exact_monodromy} for the derivation of the monodromy-matrix formula), we find that the monodromy matrix~\eqref{eq:monodromy_matrix} can be expressed as
\begin{align}
M_{\rm med}(t_0) = 
\mathcal N_{t_1,t_0}^{-1} 
\left( \mathcal N_{t_1+T,\,t_2} \, R_2 \ \mathcal N_{t_2,t_1} \, R_1 \right) 
\mathcal N_{t_1,t_0}.
\label{eq:Mex}
\end{align}
Note that this formula is equivalent to Eq.~\eqref{eq:monodromy_matrix}, with the connection matrices written in the following forms:
\begin{align}
T_{\rm II,I}(t_0) = 
{\mathcal N}_{t_1,t_0}^{-1}
R_1 \,
{\mathcal N}_{t_1,t_0}, \hspace{10mm}
T_{\rm III,II}(t_0) = 
{\mathcal N}_{t_2,t_0}^{-1}
R_2 \,
{\mathcal N}_{t_2,t_0}.
\end{align}

It is worth noting that the monodromy matrix Eq.~\eqref{eq:Mmed} derived in Appendix~\ref{app:exact_monodromy} has a distinct form $M_{\rm med}(t_1) = {\mathcal N}_{t_1+T,t_2} R_2 \, {\mathcal N}_{t_2,t_1} R_1$. This difference arises from the choice of the normalization points: $t_0$ in the main text and $t_1$ in the appendix. However, the difference in the normalization points does not affect the equation for the quasi-energy $\det(M_{\rm med} - e^{-i\epsilon T/\hbar}) = 0$. 

In the following, we calculate the quasi-energies and effective Hamiltonian using the monodromy matrix~\eqref{eq:Mex} and compare these results with direct numerical computations.

\subsubsection{Quasi-energy}
In the previous part, we have derived the formal expression~\eqref{eq:Mex} for the monodromy matrix $M_{\rm med}$ in terms of the quantities $B_n$ and $P_n$ defined in Eqs.~\eqref{eq:AB_def} and \eqref{eq:P_def}. Here, we explicitly write down the equation for the quasi-energy in terms of $ P_n $ and $ B_n $. 

As shown in Eq.~\eqref{eq:detM}, the quasi-energy $\epsilon$ is determined by solving the characteristic equation:
\begin{align}
D(\theta) = \det (M_{\rm med} - e^{i \theta} ) = 0, 
\hspace{10mm}
\theta := -\frac{\epsilon T}{\hbar} = -\frac{2 \pi \epsilon}{\hbar \omega}.
\end{align}
From the monodromy matrix \eqref{eq:Mex}, we find that $D(\theta)$ can be expressed as
\begin{align}
D(\theta) &= e^{i\theta} \bigg[ 2 \cos \theta 
- \sqrt{1+B_1} \sqrt{1+B_2} \left( \sqrt{P_1 P_2} + \frac{1}{\sqrt{P_1 P_2}} \right)  - \sqrt{B_1 B_2} \left( \sqrt{\frac{P_1}{P_2}} + \sqrt{\frac{P_2}{P_1}} \right) \bigg].
\label{eq:D_theta}
\end{align}
By solving this equation for $\theta$, 
we can obtain expressions for the quasi-energies in terms of $B_n$ and $P_n$.
For simplicity, we will refer to the dimensionless quasi-energy $\theta$ just as ``quasi-energy" in the remainder of this paper.

\begin{figure}[htbp]
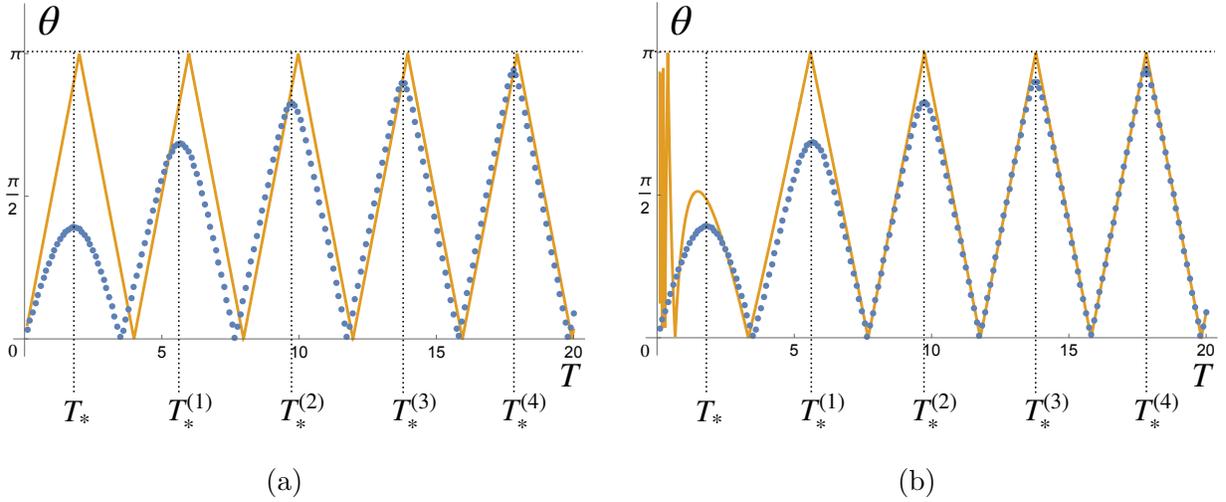

\begin{minipage}[b]{0.48\linewidth}
\centering
\includegraphics[width=80mm, page=7, bb=0 50 768 700, clip]{floquet_fig4.pdf} \vspace{-8mm} \\
(a)
\end{minipage}
~~
\begin{minipage}[b]{0.48\linewidth}
\centering
\includegraphics[width=80mm, page=8, bb=0 50 768 700, clip]{floquet_fig4.pdf} \vspace{-8mm} \\
(b)
\end{minipage}
\caption{The numerical results (dots) and (a) the leading-order perturbative result (solid line), (b) the next-leading-order perturbative result (solid line) of the quasi-energy $\theta = \epsilon T/\hbar$. Both figures show $\theta$ as a function of the period $T=2\pi/\omega$ with $\Delta=1, v = 1.8, \hbar=1$, $2\pi \hbar/\Delta \approx 6.28$. Since the quasi-energies appear in pairs of positive and negative values, only the positive values are plotted.
}
\label{fig:theta_leading}
\end{figure}

To validate our exact-WKB approach, 
let us evaluate $B_n$ and $P_n$ to compare our results for the quasi-energy 
with numerical results obtained by directly solving the Schr\"odinger equation~(\ref{eq:Sch_eq}). 
The quantities $A_n$ and $B_n$ are expressed as the cycle integrals of $S_{\rm odd}$ in Eq.~\eqref{eq:AB_def}. 
The small-$\hbar$ expansion of $S_{\rm odd} = \partial_t W/\hbar$ can be obtained by solving Eq.~\eqref{eq:Riccati} perturbatively. The first three terms are given by
\begin{align}
S_{\rm odd} = \frac{1}{\hbar} \sqrt{Q_0} 
+ \frac{Q_1}{2\sqrt{Q_0}} 
+ \frac{\hbar}{2\sqrt{Q_0}} \left( Q_2 - \frac{1}{4} \frac{Q_1^2}{Q_0} 
+ \frac{1}{4} \frac{\partial_t^2 Q_0}{Q_0} 
- \frac{5}{16} \left( \frac{\partial_t Q_0}{Q_0} \right)^2 \right) +  \mathcal{O}(\hbar^2).
\label{eq:Soddhbar}
\end{align}
Using the explicit forms of $Q_0(t)$, $Q_1(t)$, and $Q_2(t)$ given in Eq.~\eqref{eq:Q_0etc}, 
evaluating the integrals in Eq.~\eqref{eq:AB_def} and 
using the relation Eq.~\eqref{eq:P_def}, 
we find that $\sqrt{P_n}$ and $\sqrt{B_n}$ behave in the small $\hbar$ limit as 
\begin{align}
\sqrt{P_n} &= \exp \left( + \frac{i}{2} \theta_0 \right) \left[ 1 - \frac{i\hbar \omega}{12\Delta} \mathcal{F} \left( \frac{\pi}{2} \right)  + \mathcal{O}(\hbar^2) \right], \\
\sqrt{B_n} &= \exp \left( - \frac{1}{2} \beta \right) \, \left[ 1 - \frac{i\hbar \omega}{12\Delta} \mathcal{F} \left( \arcsin \frac{i\Delta}{v} \right) + \mathcal{O}(\hbar^2) \right] \exp \left( (-1)^{n+1} \frac{\pi i}{2} \right),
\end{align}
where $\theta_0$ is the dynamical phase and $\beta$ is a positive real constant that determines the non-perturbative exponent: 
\begin{align}
\theta_0 \ &= \frac{1}{\hbar} \int_0^T \sqrt{-Q_0} \, dt = + \frac{4\Delta}{\hbar \omega} E \left( - \frac{v^2}{\Delta^2} \right), \\
\beta \ &= \frac{1}{\hbar} \ \oint_{\mathcal{B}} \sqrt{-Q_0} \, dt = - \frac{4i\Delta}{\hbar \omega} E\left( \arcsin \frac{i\Delta}{v}, - \frac{v^2}{\Delta^2} \right).
\end{align}
The function $\mathcal{F}(x)$ is defined as
\begin{align}
\mathcal{F}(x) = F\left(x, - \frac{v^2}{\Delta^2} \right) - \left( 1 +\frac{v^2}{\Delta^2+v^2} \right) E\left(x, - \frac{v^2}{\Delta^2} \right),
\end{align}
where $F(x, m)$ and $E(x, m)$ are the elliptic integrals of the first and second kinds, respectively:
\begin{align}
F(x,m) &= \int_0^x \frac{1}{\sqrt{1-m \sin^2 z}} \, dz, \hspace{10mm}
K(m) = F \left( \frac{\pi}{2} , m \right), \\
E(x,m) &= \int_0^x \sqrt{1-m \sin^2 z} \, dz, \hspace{10mm}
E(m) = E \left( \frac{\pi}{2} , m \right).
\end{align}
Note that $P_n$ and $B_n$ are independent of the coefficients of the linear combination $(c_1, c_2)$, even though $Q(t)$ depends on $(c_1, c_2)$. 
Using these asymptotic forms of $P_n$ and $B_n$ in the small $\hbar \omega$ limit, 
$D(\theta)$ in Eq.~\eqref{eq:D_theta} can be expanded as 
\begin{align}
D(\theta) = 2 e^{i \theta} \Big[ \cos \theta - \left\{ \cos \theta_0 + \cdots \right\} - e^{-\beta} \left\{ 1 - \cos \theta_0 + \cdots \right\} + \mathcal{O}(e^{-2\beta}) \Big],
\label{eq:D_theta_asymptotic}
\end{align}
where $\cdots$ denotes terms with higher powers of $\hbar$. 
Since $e^{-\beta}$ is exponentially suppressed when $\hbar \omega \ll \Delta$, 
Eq.~\eqref{eq:D_theta_asymptotic} implies that the leading-order asymptotic solutions to $D(\theta) = 0$ are given by
\begin{align}
\theta = \pm \theta_0 + \cdots ~~~({\rm mod} \, 2\pi),  \label{eq:LOtheta}
\end{align}
where the corrections $\cdots$ involve higher-order terms in $\hbar$. 
This leading-order result provides a good approximation in the slow-frequency limit $\hbar \omega \ll \Delta$ ($T \gg 2\pi \hbar / \Delta$), as shown in Fig.~\ref{fig:theta_leading}-(a). 
This result is equivalent to that obtained by applying the adiabatic approximation to the differential equation for the time-evolution unitary matrix~\eqref{eq:1st_order_diffeq} and using the formula for the dynamical phase in Eq.~\eqref{eq:dynamical_phase}.

As shown in Fig.~\ref{fig:theta_leading}-(a), for $\hbar \omega \gtrsim \Delta$ (or equivalently $T \lesssim 2\pi \hbar/\Delta$), the discrepancy between the leading-order and numerical results becomes significant, indicating that higher-order $\hbar$ corrections become non-negligible. Thus, it is necessary to incorporate the next-to-leading-order $\hbar$ correction in $S_{\rm odd}$~\eqref{eq:Soddhbar}. 

As illustrated in Fig.~\ref{fig:theta_leading}-(b), the next-to-leading-order correction improves the agreement with the numerical results. However, a notable discrepancy remains around $T = T_\ast^{(n)}$ at which $\sqrt{P_1} = \sqrt{P_2} = \pm i$, i.e., 
\begin{align}
\int_0^{T_\ast^{(n)}} \!\! S_{\rm odd} \, dt = (2n-1) \frac{\pi i}{2} \quad (n=1,2,\cdots).
\label{eq:Tasti}
\end{align}
At these periods, the perturbative quasi-energies degenerate ($\theta = \pm \pi$), whereas the numerical quasi-energies do not. 

These discrepancies can be lifted by including non-perturbative terms in Eq.~\eqref{eq:D_theta}:
\begin{align}
\cos \theta = \left( \cos \theta_0 + \cdots \right) + e^{-\beta} \left( 1 - \cos \theta_0 + \cdots \right),
\label{eq:theta1}
\end{align}
where $\cdots$ denotes power-series corrections of order $\mathcal{O}(\hbar)$. The second term represents a non-perturbative correction, as it exhibits exponential dependence of the form $e^{-\beta} \propto e^{-c/(\hbar \omega)}$, where $c$ is a constant. Consequently, $\theta$ acquires non-perturbative dependence on $\hbar$. 

\begin{figure}[htbp]
\centering
\includegraphics[width=100mm, page=9, bb=0 50 768 700, clip]{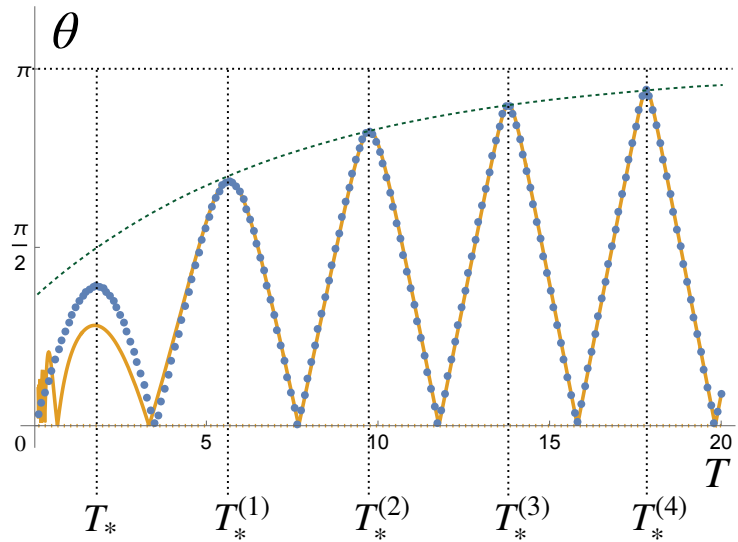} \vspace{-8mm} \\
\caption{Comparison of numerical results (dots) and exact-WKB results (solid line) incorporating next-to-leading-order perturbative and non-perturbative corrections for the quasi-energy $\theta = \epsilon T/\hbar$. The parameters are set to $\Delta = 1$, $v = 1.8$, $\hbar = 1$, $2\pi \hbar/\Delta \approx 6.28$. The inclusion of non-perturbative corrections significantly improves the accuracy, particularly near $T = T_\ast^{(n)}$, except at $T=T_\ast^{(1)}$. The dashed line represents $\theta = \arccos(-1+2e^{-\beta})$ which correctly capture the quasi-energy gaps at $T=T_\ast^{(n)}$.}
\label{fig:theta_npcorrected}
\end{figure}

The quasi-energy $\theta$ with the non-perturbative correction obtained from Eq.~\eqref{eq:theta1} is compared with numerical results in Fig.~\ref{fig:theta_npcorrected}. This non-perturbative correction significantly improves the agreement with numerical results. 
In particular, the degeneracies of the perturbative quasi-energy at $T=T_\ast^{(n)}$ are lifted and 
the gaps are accurately captured. 

Note that although there is a peak of the quasi-energy at $T=T_\ast$, this point is not in the series Eq.~\eqref{eq:Tasti},
and the degeneracy is already lifted at the next-leading order [see Fig.~\ref{fig:theta_leading}-(a) and (b)]. Since a large perturbative correction exists at this point, the exact-WKB result containing only up to the next-to-leading order in the  perturbative sector does not provide an accurate approximation. 

To evaluate the quasi-energy gaps, let us focus on $T = T_\ast^{(n)}$, where perturbative quasi-energies degenerate. 
The asymptotic form of $D(\theta)$ in Eq.~\eqref{eq:D_theta_asymptotic} implies that 
\begin{align}
\cos \theta(T_\ast^{(n)}) = -1 + 2 e^{-\beta(T_\ast^{(n)})} + \cdots .
\end{align}
Therefore, the gap at $T = T_\ast^{(n)}$ can be estimated as
\begin{align}
\theta(T_\ast^{(n)}) = \pm \left(\pi-2e^{-\frac{1}{2} \beta(T_\ast^{(n)})} + \cdots \right).
\label{eq:theta_omega_ast}
\end{align}
This estimate of the gap is represented by the dashed line in Fig.~\ref{fig:theta_npcorrected} corresponding to the function $\theta = \arccos(-1+2e^{-\beta})$, 
which shows a good agreement with numerical results. 
In the next section, we will demonstrate how these non-perturbative corrections provide insight 
into resonant transitions between two levels, a phenomenon that cannot be captured by the adiabatic analysis. 
The resonance behavior is reminiscent of the Rabi oscillation.  
In fact, our Hamiltonian can be continuously deformed by introducing a new parameter $\chi$ as  
\begin{align}
f_1 = v \cos \chi \sin \omega t , \hspace{5mm}
f_2 = -v \sin \chi \cos \omega t , \hspace{5mm}
f_3 = \Delta.
\label{eq:elliptic_oscillation}
\end{align}
In the limit of $\chi \to 0$, the Hamiltonian reduces to the original form given in Eq.~\eqref{eq:example1}. 
On the other hand, when $\chi = \pm \pi /4$, the system becomes exactly solvable, 
and the lowest resonance frequency $\omega_\ast = 2\pi/T_\ast$ coincides with that observed in the standard Rabi oscillation.
This relation can also be analyzed by applying the rotating-wave approximation to the system~\eqref{eq:example1}. 
For further details on the system~\eqref{eq:elliptic_oscillation} and its connection to the Rabi problem, see Appendix~\ref{appendix:elliptic}.  

Finally, note that the degeneracy of the quasi-energies at $\theta = 0$ is a special property of this specific system. In a more general system, this degeneracy would be lifted, leading to the emergence of a non-perturbative gap in the quasi-energy spectrum around $\theta = 0$.

\subsubsection{Effective Hamiltonian}
In this subsection, we calculate the effective Hamiltonian. As shown in Eq.~\eqref{eq:unitary_monodromy}, the time-evolution matrix $U(T)$ is related to the monodromy matrix $M_{\rm med}$ as follows:
\begin{align}
U(T) = G(0) M_{\rm med} G(0)^{-1},
\end{align}
where $G(0)$ represents the initial value of the WKB solutions obtained using Eq.~\eqref{eq:unitary_g}. Up to the next-to-leading order, the explicit form of $G(0)$ is given by
\begin{align}
G(0) = \left[  
\begin{pmatrix}
1 & 0 \\ 
0 & 1
\end{pmatrix} 
- \frac{iv \hbar \omega}{\Delta^2}
\begin{pmatrix}
0 & 1 \\ 
1 & 0
\end{pmatrix} 
+ \mathcal{O}(\hbar^2) \right] R_0^{-\frac{1}{2}},
\label{eq:G0_R0}
\end{align}
where $R_0^{-\frac{1}{2}}~(=R_2^{-\frac{1}{2}})$ is the matrix satisfying $R_0^{-\frac{1}{2}} R_0^{-\frac{1}{2}} = R_0^{-1}$. The explicit form of $R_0^{-\frac{1}{2}}$ is
\begin{align}
R_0^{-\frac{1}{2}} = \frac{1}{2}
\begin{pmatrix}
\sqrt{1+i\sqrt{B_0}}+\sqrt{1-i\sqrt{B_0}} & \sqrt{1-i\sqrt{B_0}}-\sqrt{1+i\sqrt{B_0}} \\ 
\sqrt{1+i\sqrt{B_0}}-\sqrt{1-i\sqrt{B_0}} & \sqrt{1+i\sqrt{B_0}}+\sqrt{1-i\sqrt{B_0}}
\end{pmatrix}.
\label{eq:Rhalf}
\end{align}
This matrix $R_0^{-\frac{1}{2}}$ plays a crucial role in handling discontinuities arising from the movable singularity at $t=0$, 
which must be properly taken into account when using the explicit forms of the WKB wave functions.  
The discontinuities associated with movable singularities can be treated in a similar manner to those arising from fixed singularities by introducing a Stokes automorphism associated with $t=t_0$. For further details, see Appendix~\ref{appendix:alien_movable}.

\begin{figure}[htbp]
\begin{minipage}[b]{0.45\linewidth}
\centering
\includegraphics[width=70mm, page=10]{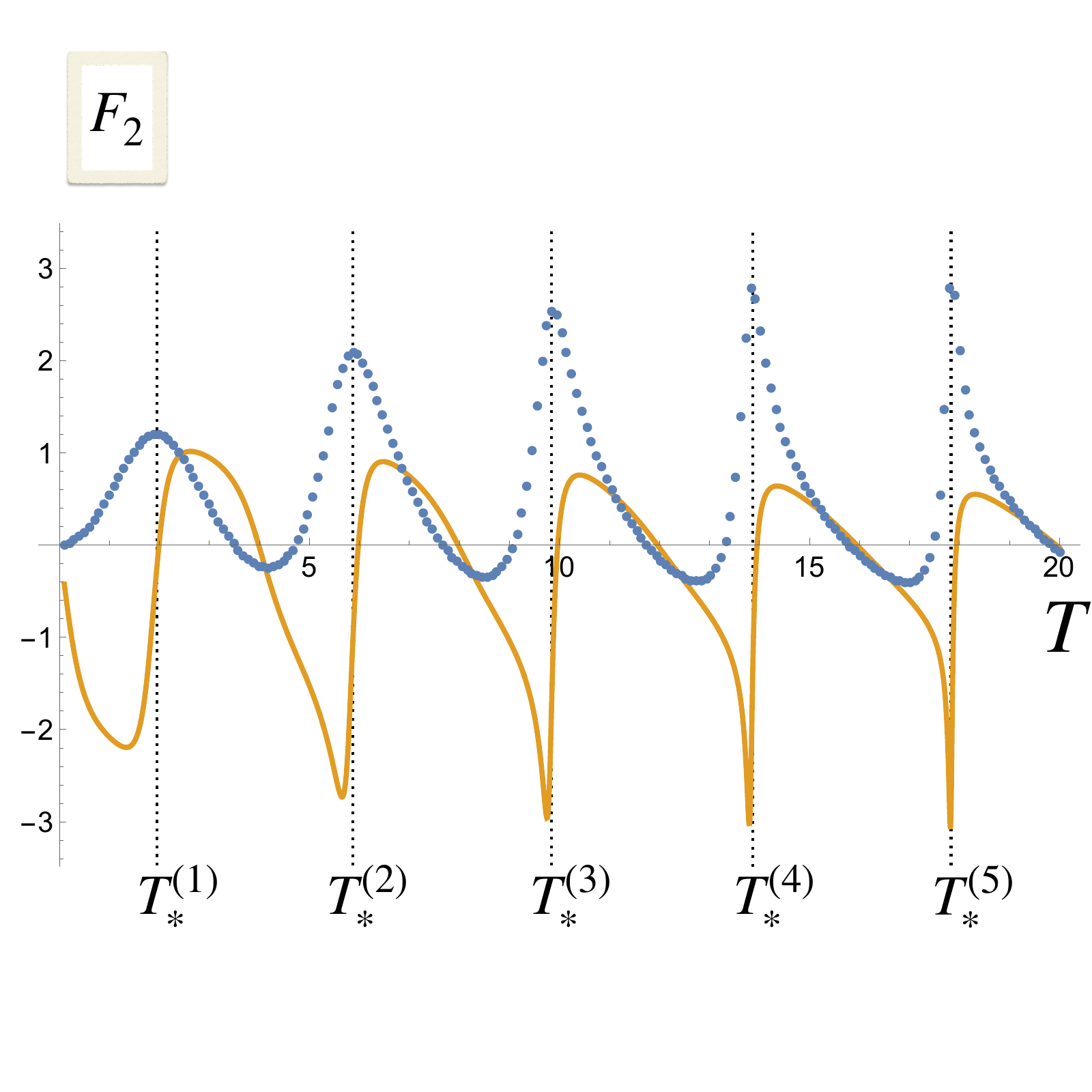} 
\end{minipage}
\begin{minipage}[b]{0.45\linewidth}
\centering
\includegraphics[width=70mm, page=11]{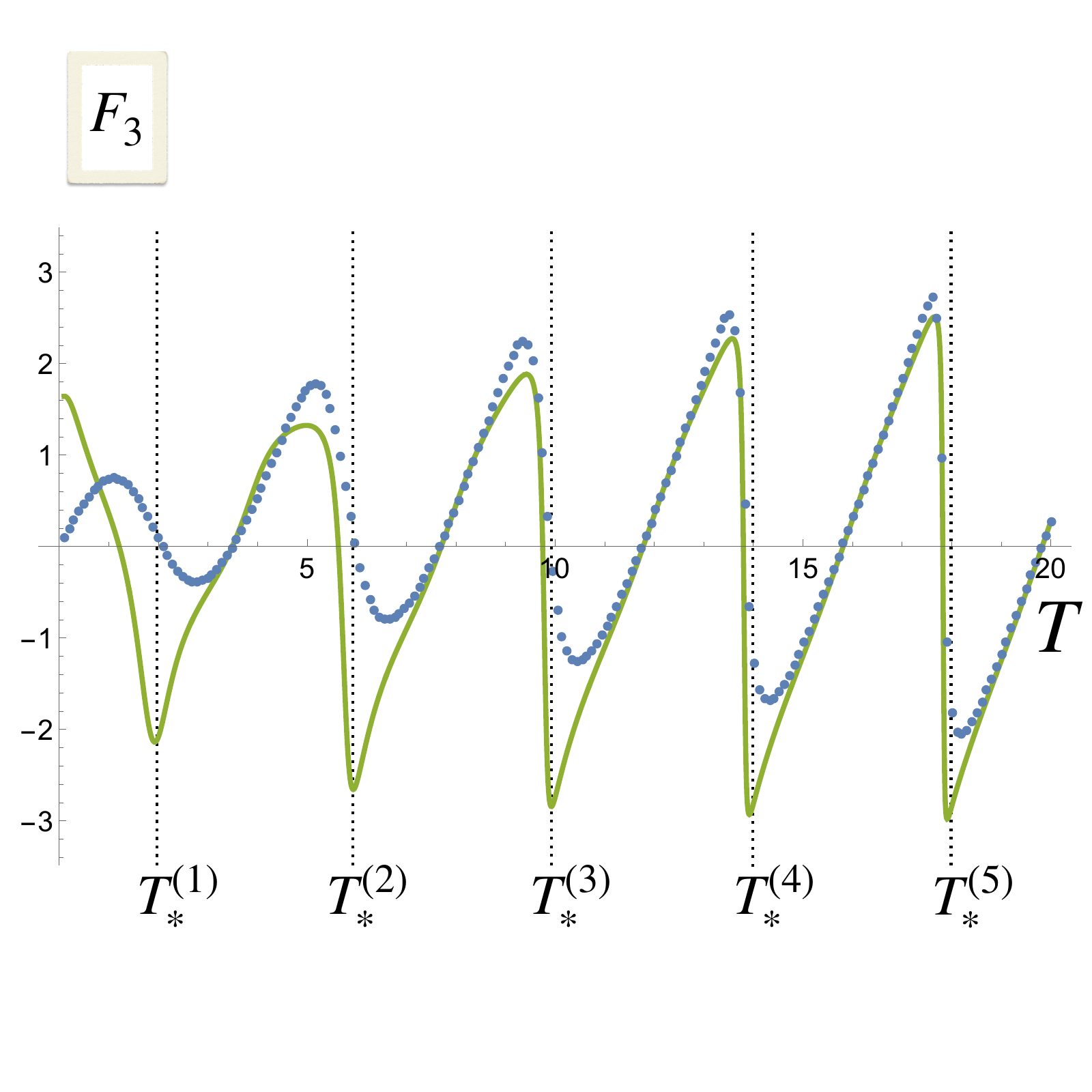} 
\end{minipage}
\caption{The components $F_{2,3}$ (corresponding to $\sigma_{2,3}$ components) of the leading-order effective Hamiltonian with $\Delta=1, v = 1.8, \hbar=1$, and $2\pi \hbar/\Delta \approx 6.28$. The dots indicate the numerical results, and the solid lines are the analytical result truncated at the first order in $\hbar \omega = 2\pi\hbar/T$.
} 
\label{fig:Heff_leading} 
\end{figure}

\begin{figure}[htbp]
\begin{minipage}[b]{0.45\linewidth}
\centering
\includegraphics[width=70mm, page=12]{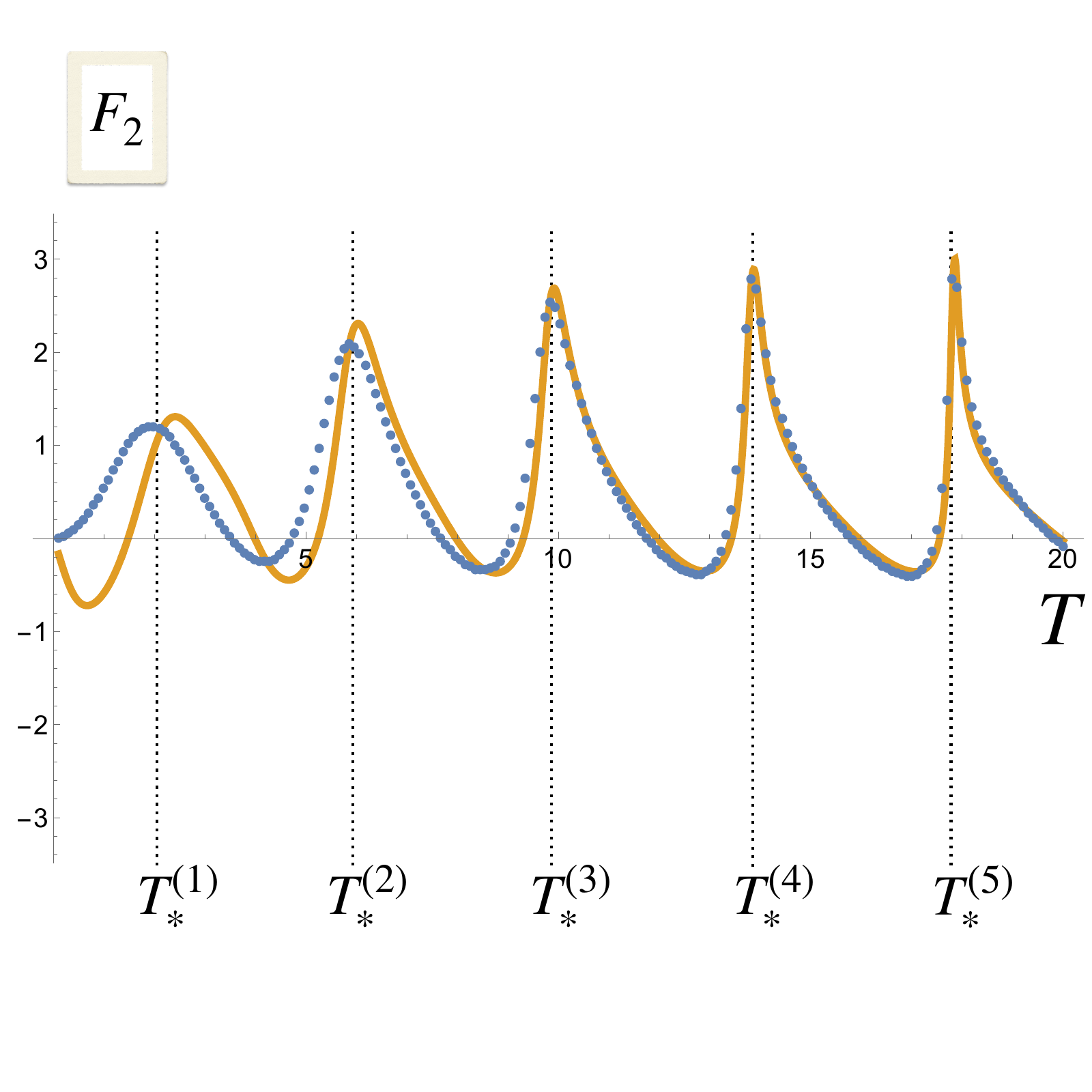} 
\end{minipage}
\begin{minipage}[b]{0.45\linewidth}
\centering
\includegraphics[width=70mm, page=13]{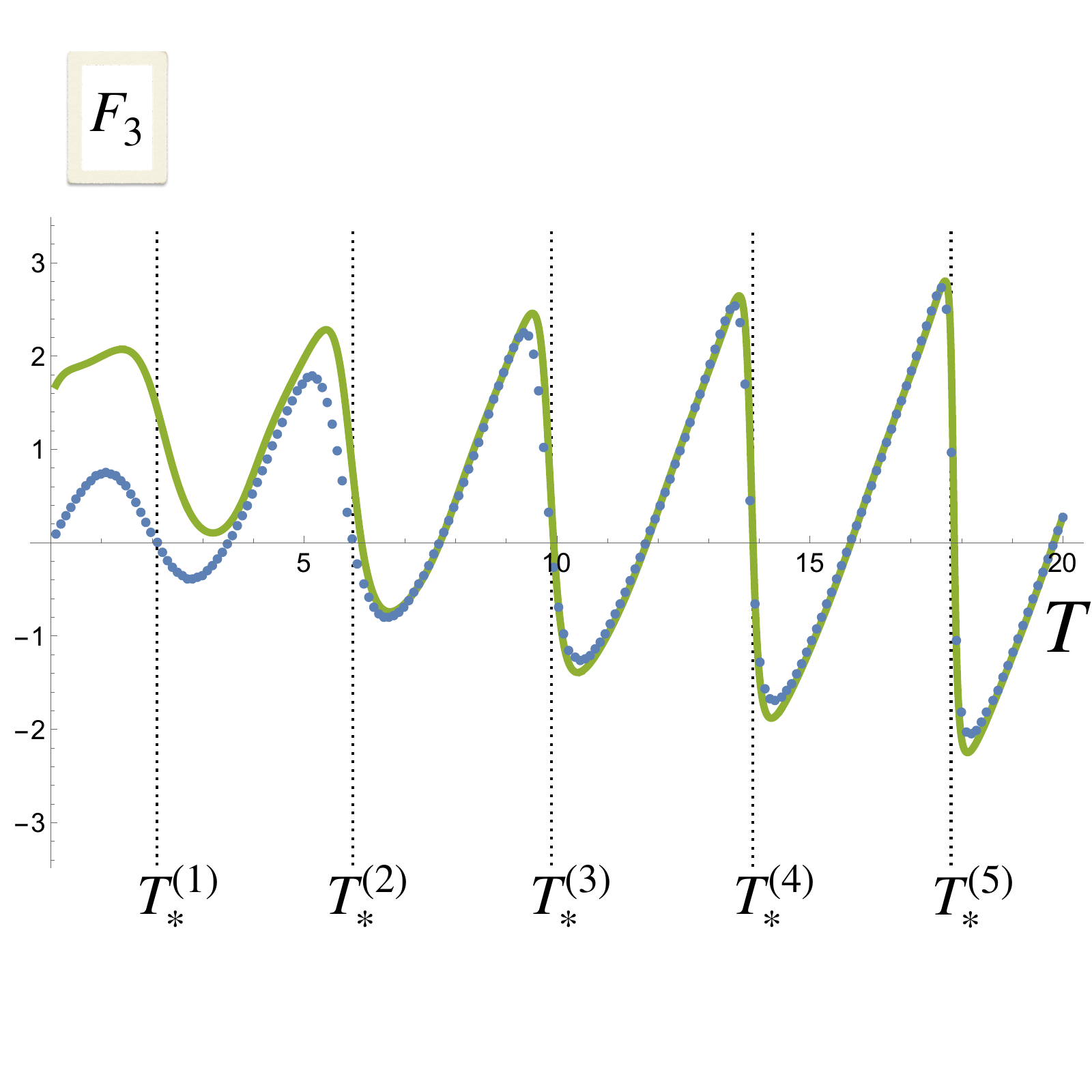} 
\end{minipage}
\caption{The components $F_{2,3}$ of the effective Hamiltonian obtained through the exact-WKB analysis with $\Delta=1, v = 1.8, \hbar=1$, and $2\pi \hbar/\Delta \approx 6.28$. The dots and solid lines, respectively, represent the numerical and exact-WKB results whose perturbative and non-perturbative parts are truncated at the first order in $\hbar \omega = 2\pi\hbar/T$. }
\label{fig:Heff_eWKB} 
\end{figure}

Using the matrix $G(0)$ and the monodromy matrix $M_{\rm med}$ obtained through the exact-WKB analysis, we can compute the time-evolution matrix $U(T)$ and the effective Hamiltonian. 
For small frequencies $\hbar \omega \ll \Delta$ ($T \gg 2\pi \hbar/\Delta$), 
the unitary matrix takes the form
\begin{align}
U(T) &= U_{\rm P}(T) + U_{\rm NP}(T) \notag \\
&= \Big[ \cos \theta_0 \mathbf 1 + i \sin \theta_0 \sigma_3 + \mathcal O(\hbar \omega) \Big] + e^{-\frac{\beta}{2}} \Big[ (1-\cos \theta_0)\sigma_2 + \mathcal O(\hbar \omega) \Big] + \mathcal O(e^{-\beta}).
\label{eq:unitary_result}
\end{align}
The terms in the first square brackets
are the leading perturbative (adiabatic) part $U_{\rm P}(T)$, which gives the leading-order quasi-energy in the slow-frequency limit. 
The terms in the second square brackets are the non-perturbative part $U_{\rm NP}(T)$, obtained through the exact-WKB analysis. 
Notably, this unitary matrix $U(T)$ has no $\sigma_1$ component,
which is consistent with the property of the time-evolution unitary matrix:
\begin{align}
\sigma_3 U(T)^{\rm T} \sigma_3 = U(T),
\label{eq:propertyU}
\end{align}
for systems described by the Hamiltonian with the time-reversal symmetry characterized by the property $\sigma_3 H_{\rm eff}^\ast \sigma_3 = H_{\rm eff}$ [see Appendix \ref{appendix:symH} for a proof of the property~\eqref{eq:propertyU} in systems with the time-reversal symmetry].
Due to the time-reversal symmetry, 
the effective Hamiltonian takes the form:
\begin{align}
H_{\rm eff}(T) = \frac{i\hbar}{T} \log U(T) = \sum_{i=1}^3 F_i(T) \sigma_i ~~~\mbox{with}~~~ F_1(T) = 0.
\end{align}
Comparisons between the exact-WKB and numerical results for the $\sigma_2$ and $\sigma_3$ components of the effective Hamiltonian, $F_2(T)$ and $F_3(T)$, are shown in Figs.~\ref{fig:Heff_leading} and \ref{fig:Heff_eWKB}. 

Firstly, Fig.~\ref{fig:Heff_leading} illustrates the components of the perturbative effective Hamiltonian truncated at the first order in $\hbar \omega = 2\pi \hbar/T$. 
The perturbative results exhibit limited accuracy, particularly near $T=T_\ast^{(n)}$, 
at which they show discontinuities that are absent in the numerical result. 
These discrepancies can be lifted by incorporating the non-perturbative contributions, 
which have been obtained through the exact-WKB analysis. 

Fig.~\ref{fig:Heff_eWKB} 
shows the results including both perturbative and non-perturbative contributions, 
demonstrating an improved agreement with the numerical results. 
Note that for large $T$, the exact-WKB results closely match the numerical results, 
whereas for small $T$, some discrepancies remain, 
since both perturbative and non-perturbative expansions are truncated at the first order in $\hbar \omega = 2\pi\hbar/T$.  

As demonstrated in this example, the exact-WKB analysis provides non-perturbative information on the behavior of the effective Hamiltonian near the points $ T = T_\ast^{(n)} $.
Let us now discuss the physical implications of these non-perturbative corrections obtained through the exact-WKB analysis.

At $T = T_\ast^{(n)}$, the perturbative part of the time-evolution unitary matrix reduces to $U_{\rm P}(T) = -\mathbf{1}$, which gives a degenerate leading-order effective Hamiltonian:
\begin{align}
H_{\rm eff} = \frac{\pi \hbar}{T_\ast^{(n)}} \mathbf{1} = \frac{1}{2} \hbar \omega_\ast^{(n)} \mathbf{1}.
\end{align}
This suggests that, in the absence of non-perturbative corrections, the state vector would return to its original state with an overall sign flip. However, due to the presence of non-perturbative terms $U_{\rm NP}(T)$, the effective Hamiltonian acquires small off-diagonal elements, lifting the degeneracy.
As a result, the true eigenstates of the effective Hamiltonian become superpositions of the perturbative eigenvectors.

In other words, non-perturbative transition amplitudes emerge between the perturbative eigenstates, 
leading to an oscillation of the state vector between these two states. 
Similarly to the case of the Landau-Zener problem discussed in Sec.~\ref{sec:3-2}, 
the exact-WKB analysis gives information on the rate of the non-perturbative transition. 
In particular, the period of the oscillation of the state vector can be estimated from the non-perturbative contribution $U_{\rm NP}(T)$ as
\begin{align}
T_{\rm t}^{(n)} = \frac{\pi}{2} T e^{\frac{1}{2} \beta} \Big|_{T=T_\ast^{(n)}} \approx 
\frac{\pi^2}{\omega} \exp \left[ - \frac{2i\Delta}{\hbar \omega} E\left( \operatorname{arcsin} \frac{i\Delta}{v}, - \frac{v^2}{\Delta^2} \right) \right] \Bigg|_{\omega = 2\pi/T_\ast^{(n)}}.
\end{align}
Since the transition amplitude is purely non-perturbative, the oscillation period is exponentially long for large $T$. This highlights the significance of non-perturbative effects, which we have successfully captured through the exact-WKB analysis.

\begin{figure}[htbp]
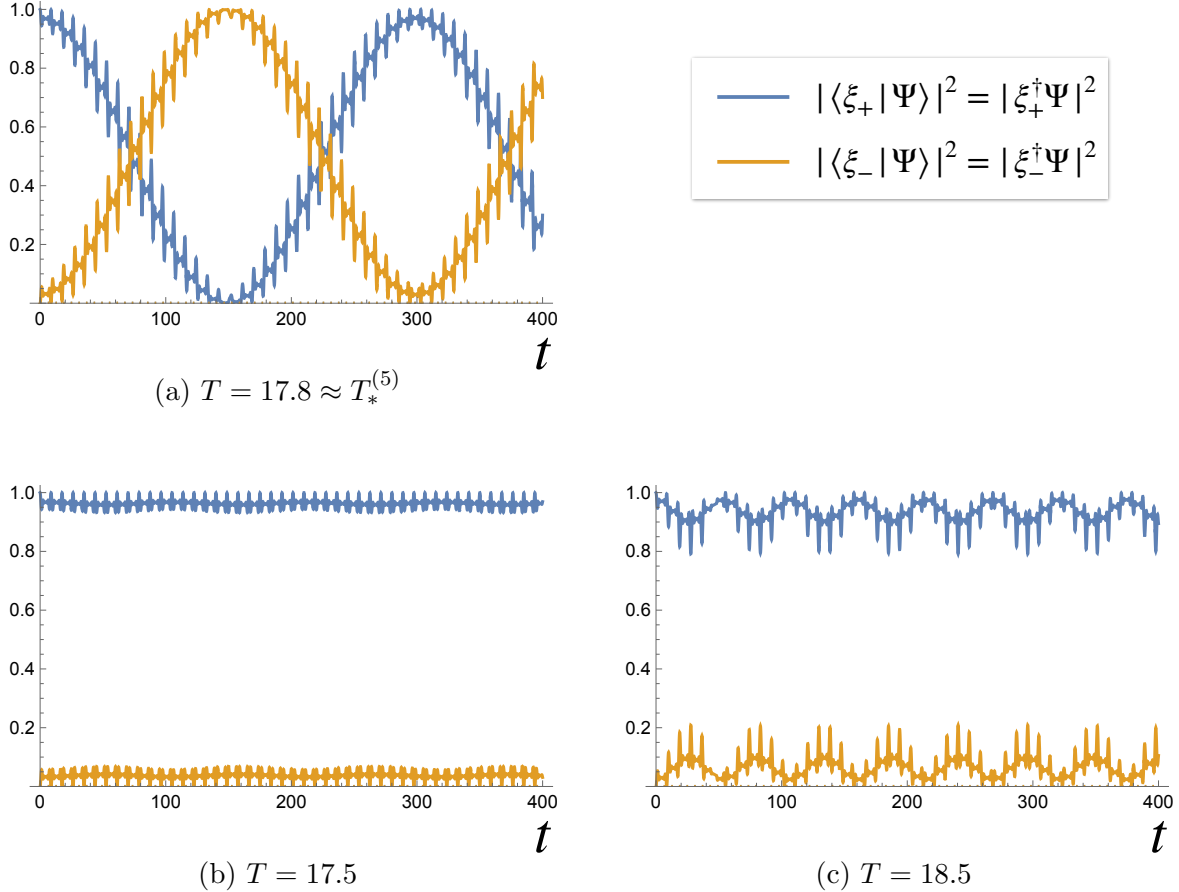

\begin{minipage}[b]{0.49\linewidth}
\centering
\includegraphics[width=75mm, page=14, bb= 0 0 750 512, clip]{floquet_fig4.pdf} \\ \vspace{-3mm}
(a) $T = 17.8 \approx T_\ast^{(5)}$
\end{minipage} 
\begin{minipage}[b]{0.49\linewidth}
\centering
\includegraphics[width=75mm, page=15, bb= 0 0 750 512, clip]{floquet_fig4.pdf} \\ 
\end{minipage} 

\vspace{10mm}
\begin{minipage}[b]{0.49\linewidth}
\centering
\includegraphics[width=75mm, page=16, bb= 0 0 750 512, clip]{floquet_fig4.pdf} \\ \vspace{-3mm}
(b) $T = 17.5$
\end{minipage}
\begin{minipage}[b]{0.49\linewidth}
\centering
\includegraphics[width=75mm, page=17, bb= 0 0 750 512, clip]{floquet_fig4.pdf} \\ \vspace{-3mm}
(c) $T = 18.5$
\end{minipage}
\caption{Inner products of the instantaneous eigenvectors $\xi_{\pm}$ and the numerical solutions for $\Psi(t)$ with the initial condition $\Psi(0)=\xi_+$ ($\Delta=1, v=1.8, \hbar=1$). When the period is close to $T=T_\ast^{(5)}$, the state vector oscillates between the instantaneous eigenvectors, $\xi_+$ and $\xi_-$, with a period $T_{\rm t}^{(5)} \approx 300$ [see (a)]. 
For general values of the period $T$, the state vector $\Psi$ remains nearly proportional to the instantaneous eigenvector $\xi_+$, as shown in (b) and (c).}
\label{fig:transition}
\end{figure}

Fig.~\ref{fig:transition} shows the inner products of instantaneous eigenvectors $\xi_{\pm}$ 
with the numerical solutions $\Psi(t)$ initialized as $\Psi(0)=\xi_+$. 
For general values of $T$, the inner products remain nearly constant with small fluctuations, as shown in Figs.~\ref{fig:transition}-(b) and (c). 
On the other hand, at $T = T_\ast^{(5)}$, as depicted in Fig.~\ref{fig:transition}-(a), they exhibit slow oscillations, 
indicating tunneling between the instantaneous eigenvectors $\xi_{\pm}$. 
This is a clear example of an explicit non-perturbative phenomenon that cannot be captured by the perturbative analysis alone.

\subsection{Generalization} 
\label{sec:5}

\subsubsection{Non-perturbative corrections}
The analysis in the simple example discussed in Sec.~\ref{sec:4} can be straightforwardly extended to general two-level Hamiltonians. 
As mentioned in Sec.~\ref{subsec:degenerate_Stokes}, 
if $t=t_{n+}$ is a turning point $Q_0(t_0)=0$, its complex conjugate $t=t_{n-}=\bar t_{n+}$ is also a turning point. 
Assuming that $Q_0(t)$ has $N$ complex conjugate pairs of turning points, 
let us first consider the case in which there is a degenerate Stokes curve intersecting the real axis between each pair of the turning points (see Fig.~\ref{fig:cycles1}). We will briefly comment on the case in which there is a pair of turning points without a degenerate Stokes curve in the next subsection.
In Appendix~\ref{app:2N_turning_points}, we derive the expression of the median-resummed monodromy matrix \eqref{eq:MMex_PB} for the case with $N$ complex conjugate pairs of turning points.
Taking account of the fact that we take $t_1$ as the normalization point in this section, the expression means that the unambiguous monodromy matrix for this case is
\begin{align}
M_{\rm med} = 
\mathcal N_{t_1,t_0}^{-1} 
\left( \mathcal N_{t_1+T,\,t_N} \, R_N \cdots \mathcal N_{t_3,\,t_2} \, R_2 \ \mathcal N_{t_2,t_1} \, R_1 \right) 
\mathcal N_{t_1,t_0}, 
\label{eq:Mex_general}
\end{align}
where $t_n~(n=1,2,\cdots,N)$ are the intersections between the degenerate Stokes curves and the real axis, 
$\mathcal N_{t_{n},t_{n-1}}~(n=1,\cdots,N,N+1)$ the normalization matrices~\eqref{eq:normt0t1}, 
and $R_n~(n=1,2,\cdots,N)$ the connection matrices for the degenerate Stokes curves:
\begin{align}
\mathcal N_{t_{n+1},t_n} = 
\left(
\begin{array}{cc}
P_n^{+\frac{1}{2}} & 0 \\
0 & P_n^{-\frac{1}{2}}
\end{array}
\right), \hspace{5mm}
R_n = 
\left(
\begin{array}{cc}
\sqrt{1+B_n} & -i \sqrt{B_n} \\
+ i \sqrt{B_n} & \sqrt{1+B_n} 
\end{array}
\right).
\end{align}
This monodromy matrix~\eqref{eq:Mex_general} has a different expression from Eq.~\eqref{eq:MMex_PB} in Appendix~\ref{app:2N_turning_points} just by $N_{t_1,t_0}$ due to the choice of the normalization point. 
All elements of the monodromy matrix $M_{\rm med}$ can be expressed with the quantities $P_n$ and $B_n$, which are given in terms of $S_{\rm odd}$ as in Eq.~\eqref{eq:AB_def}.  We can show that 
$B_n$ becomes exponentially small 
in the small $\hbar$ limit
\begin{align}
B_n \approx \exp(-\beta_n) ~~~ \mbox{with} ~~~ 0 < \beta_n = \mathcal O(1/\hbar \omega). 
\end{align}
Therefore, in the small $\hbar$ limit, the matrix $R_n$ can be expanded as
\begin{align}
R_n = 
\begin{pmatrix}
1 & 0 \\ 0 & 1
\end{pmatrix}
+ i \sqrt{B_n}
\begin{pmatrix}
0 & -1 \\ 1 & 0
\end{pmatrix}
+ \frac{1}{2} B_n
\begin{pmatrix}
1 & 0 \\ 0 & 1
\end{pmatrix} + \cdots.
\label{eq:R_expansion}
\end{align}
Substituting this expression into Eq.~\eqref{eq:Mex_general}, we find that the expanded monodromy matrix takes the form:
\begin{align}
M_{\rm med} = \mathcal N_{t_1,t_0}^{-1} 
\left( 
\begin{array}{cc}
e^{+ i \theta_0} & 0 \\ 0 & e^{-i \theta_0} 
\end{array}
\right)
\bigg[ 
 1 + X^{(1)} + X^{(2)} + \cdots \bigg] \mathcal N_{t_1,t_0} , 
\end{align}
where $\theta_0$ is given by 
\begin{align}
e^{2 i \theta_0} = P_1 P_2 \cdots P_N ~~~~ \mbox{or equivalently} ~~~~ \theta_0 = \int_0^T S_{\rm odd} \, dt.
\end{align}
The first two non-perturbative corrections to the monodromy matrix are given by
\begin{align}
X^{(1)} = 
\sum_{n=1}^N i \sqrt{B_n}
\left( 
{\renewcommand{\arraystretch}{0.9}
{\setlength{\arraycolsep}{0.8mm}
\begin{array}{cc}
0 & -e^{- i \chi_{1n}} \\  e^{i \chi_{1n}} & 0 
\end{array}}}
\right), ~~~~
X^{(2)} = \frac{1}{2}
\sum_{n=1}^N \sum_{m=1}^N 
\sqrt{B_m B_n}
\left( 
{\renewcommand{\arraystretch}{0.9}
{\setlength{\arraycolsep}{0.8mm}
\begin{array}{cc}
 e^{-i \chi_{mn}} & 0 \\ 0 & e^{i \chi_{mn}}
\end{array}}}
\right),
\end{align}
where $\chi_{mn}$ is given by
\begin{align}
\exp \left( i \chi_{mn} \right) = 
P_m P_{m+1} \cdots P_{n-1} ~~ (\mbox{for $m<n$}), ~~~~ 
\chi_{mn} = \chi_{nm} ~~ (\mbox{for $m>n$}), ~~~ \chi_{mm} = 0.
\end{align}

Although $X^{(1)}$ gives the leading-order non-perturbative correction to the monodromy matrix, the leading-order non-perturbative correction to the quasi-energy is of order $X^{(2)}$. 
The function $D(\theta)$ is given by 
\begin{align} 
D(\theta) = 2 e^{i \theta} \left[ \cos \theta - \cos \theta_0 - \frac{1}{2} \sum_{n=1}^N \sum_{m=1}^N \sqrt{B_m B_n} \cos (\theta_0-\chi_{mn}) + \cdots \right].
\end{align}
Then, the condition $D(\theta) = 0$ gives the expression of
the quasi-energy $\epsilon = \hbar \omega \theta / 2\pi$ as 
\begin{align}
\epsilon = \pm \frac{\hbar \omega}{2\pi} \arccos \left( \cos \theta_0 + \frac{1}{2} \sum_{n=1}^N \sum_{m=1}^N \sqrt{B_mB_n} \cos (\theta_0-\chi_{mn}) + \cdots \right),
\end{align}
where $\cdots$ denotes the higher-order non-perturbative corrections. 
This exact-WKB result implies that the 
non-perturbative effects become important around the point in the parameter space where the perturbative quasi-energies degenerate ($\theta_0 = k \pi$ with $k \in \mathbb Z$). 
When the oscillation frequency of the Hamiltonian is at a point $\omega = \omega_\ast^{(k)}$ ($T=2\pi/\omega_\ast^{(k)}$) at which $\theta_0 = k \pi$, we find a resonant behavior with the resonance frequency:
\begin{align}
\Omega^{(k)} = \frac{\Delta \epsilon}{2\hbar} = \frac{\omega_\ast^{(k)}}{2\pi} \left| \sum_{n=1}^N \sum_{m=1}^N \sqrt{B_mB_n} \cos \chi_{mn} \right|^\frac{1}{2} + \cdots.
\end{align}

The leading-order non-perturbative correction to the effective Hamiltonian is obtained in a parallel manner. As with the case of $N=2$, the matrix $G(0)$ is decomposed into the perturbative factor $G_p(0)$ and $R_0^{-\frac{1}{2}}$ as
\begin{align}
G(0) = G_P(0) R_0^{-\frac{1}{2}}, 
\end{align}
where $R_0^{-\frac{1}{2}}$ is a matrix such that $R_0^{-\frac{1}{2}} R_0^{-\frac{1}{2}} = R_0^{-1}$ [see Eq.~\eqref{eq:Rhalf}]. 
Using this matrix and the expansion of $R_n$ given in Eq.~\eqref{eq:R_expansion}, we obtain the time-evolution unitary matrix $U(T)$ as
\begin{align}
U(T) &= G(0) M_{\rm med} G(0)^{-1} = G_P(0) {\cal N}_{t_1 t_0}^{-1}  ( Y^{(0)} + Y^{(1)} + \cdots ) {\cal N}_{t_1 t_0} G_P(0)^{-1}, 
\end{align}
where
\begin{align}
Y^{(0)} &=  \left( 
{\renewcommand{\arraystretch}{0.9}
{\setlength{\arraycolsep}{0.8mm}
\begin{array}{cc}
e^{+ i \theta_0} & 0 \\ 0 & e^{-i \theta_0} 
\end{array}}}
\right), \\
Y^{(1)} &= Y^{(0)} \left[ \sum_{n=1}^N i \sqrt{B_n}
\left( 
{\renewcommand{\arraystretch}{0.9}
{\setlength{\arraycolsep}{0.8mm}
\begin{array}{cc}
0 & -e^{- i \chi_{1n}} \\  e^{i \chi_{1n}} & 0 
\end{array}}}
\right) - \sqrt{B_N} \sin \theta_0
\left( 
{\renewcommand{\arraystretch}{0.9}
{\setlength{\arraycolsep}{0.8mm}
\begin{array}{cc}
0 & e^{- i \theta_0} \\ e^{i \theta_0} & 0 
\end{array}}}
\right) \right].
\end{align}
For a general value of $\theta_0$, 
the effective Hamiltonian is given by
\begin{align}
H_{\rm eff} = \frac{i \hbar}{T} \log U(T) = \frac{\hbar}{T} G_P(0) {\cal N}_{t_1 t_0}^{-1} \left(\theta_0 \sigma_3 + \frac{\theta_0}{\sin \theta_0}  Y^{(1)} + \cdots \right) {\cal N}_{t_1 t_0} G_P(0)^{-1}. 
\end{align}
When $\theta_0 = k \pi ~ (k \in \mathbb Z)$, the leading part becomes $Y^{(0)} = (-1)^k {\bf 1}$, and hence the effective Hamiltonian takes the form:
\begin{align}
H_{\rm eff} = \frac{\hbar}{T} G_P(0) {\cal N}_{t_1 t_0}^{-1} \left[ \left( 
{\renewcommand{\arraystretch}{0.9}
{\setlength{\arraycolsep}{0.8mm}
\begin{array}{cc}
\theta' & 0 \\ 0 & \theta'
\end{array}}}
\right) +  \sum_{n=1}^N i \sqrt{B_n}
\left( 
{\renewcommand{\arraystretch}{0.9}
{\setlength{\arraycolsep}{0.8mm}
\begin{array}{cc}
0 & -e^{- i \chi_{1n}} \\   e^{i \chi_{1n}} & 0 
\end{array}}}
\right) + \cdots \right] {\cal N}_{t_1 t_0} G_P(0)^{-1},
\end{align}
where $\theta' = 0$ for even $k$ and $\theta' = \pi$ for odd $k$. 
This form of the effective Hamiltonian indicates that the perturbative degeneracy of the quasi-energies is lifted by the non-perturbative corrections of order $\sqrt{B_n}~(n=1,\cdots,N)$. 
This implies a resonant behavior at $\theta_0 = n \pi$.\footnote{In the case of the system with $N=2$ discussed in Sec.~\ref{sec:4}, there is no non-perturbative correction for $\theta_0 = k \pi$ with even $k$, since the degeneracies are not lifted.}

\subsubsection{A case with a pair of turning points without a degenerate Stokes curve}
So far, we have assumed that each pair of turning points is connected by a degenerate Stokes curve crossing the real axis on the complex $t$-plane. However, there can also be pairs of turning points that are not connected by such a Stokes curve.
In fact, a degenerate Stokes curve crossing the real axis can transform into Stokes curves that no longer intersect the real axis. This is due to a class of Stokes phenomena occurring when parameters in the Hamiltonian go across the boundary of different Stokes regions in the parameter space.

\begin{figure}[htbp]
\begin{minipage}[b]{0.49\linewidth}
\centering
\includegraphics[width=75mm, page=18, clip]{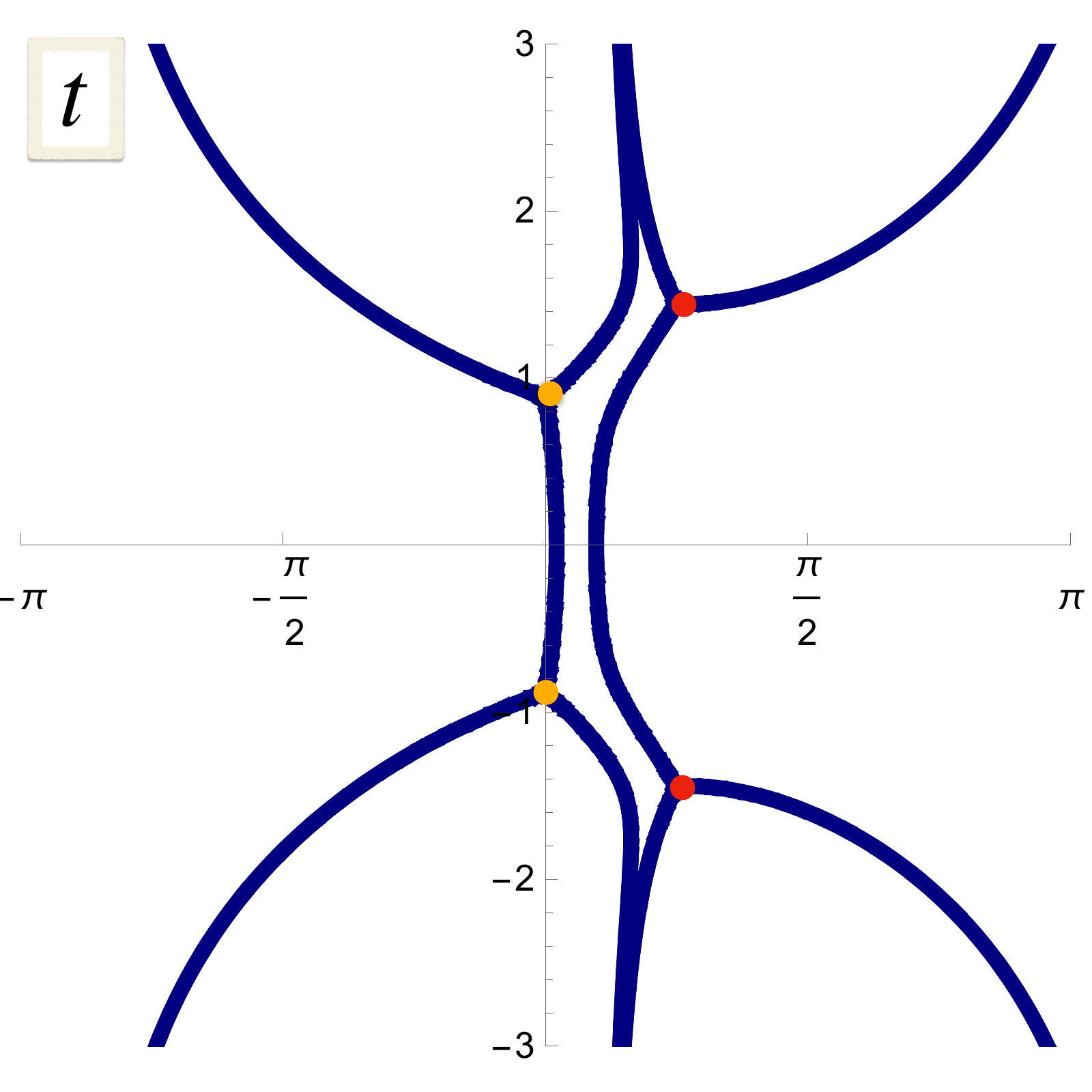} \\
(a) $(s_1,r_1)=(0, 0.9),~(s_2,r_2)=(\pi/4, 1.4)$
\end{minipage} 
\begin{minipage}[b]{0.49\linewidth}
\centering
\includegraphics[width=75mm, page=19, clip]{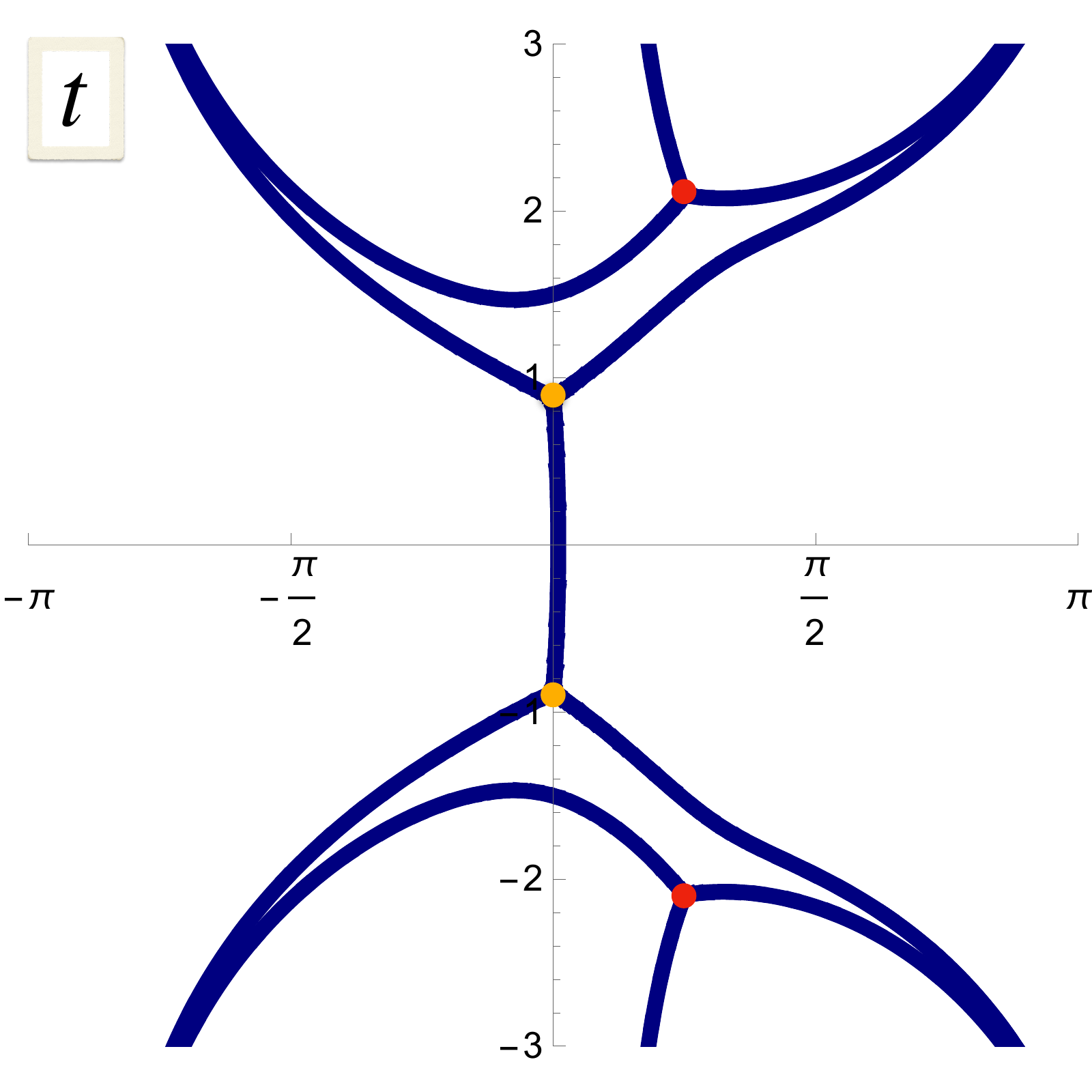} \\ 
(b) $(s_1,r_1)=(0, 0.9),~(s_2,r_2)=(\pi/4, 2.1)$
\end{minipage} 
\caption{Stokes graphs with two pairs of turning points. When the parameters in the Hamiltonian are varied across the Stokes curve in the parameter space, the shape of the graph changes from (a) to (b), or vice versa.}
\label{fig:Stokes_phenomenon}
\end{figure}

\begin{figure}[htbp]
\centering
\includegraphics[width=100mm, page=20]{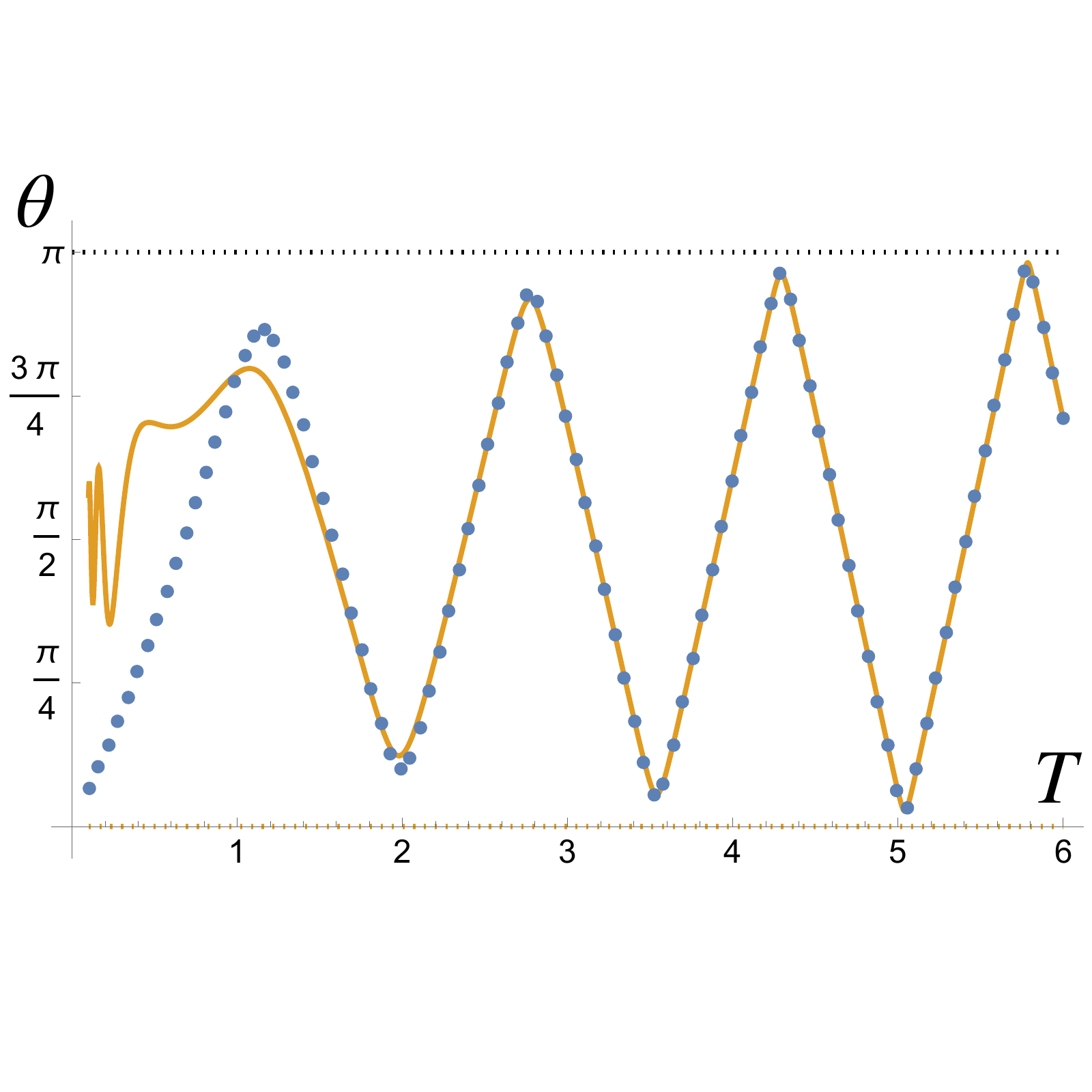} \\
\caption{Quasi-energy corresponding to the case in Fig.~\ref{fig:Stokes_phenomenon}-(b). The parameters are set to $\hbar=1$, $v=1.8$ and $(s_1,r_1)=(0, 0.9),~(s_2,r_2)=(\pi/4, 2.1)$. The solid lines are the exact-WKB results approximated by truncating $P_1$ and $B_1$ at the next leading order in the $\hbar$ expansion. When the period $T$ is large, the approximated exact-WKB results are consistent with the numerical results indicated by the dots.}
\label{fig:single_DS}
\end{figure}

As an example, let us consider the Hamiltonian $H=\sum_i f_i(t) \sigma_i$ where the components $f_i~(i=1,2,3)$ are given by
\begin{align}
f_1 = a_1 \sin(\omega t+\phi_0) + b_1, ~~~~~
f_2 = 0, ~~~~~
f_3 = a_3 \cos(\omega t+\phi_0) + b_3. \label{eq:fs_one_pair}
\end{align}
In this case, the function $Q_0(t)$ takes the form:
\begin{align} 
Q_0(t) = -\sum_{i=1}^3 f_i(t)^2 = - v \big( \cos (\omega t - s_1)-\cosh r_1 \big) \big( \cos (\omega t - s_2)-\cosh r_2 \big),
\end{align}
where we have changed the parameters from $(a_1,a_2,b_1,b_2,\phi_0)$ to $(s_1,r_1,s_2,r_2,v)$ by 
\begin{align}
a_1 = + v \sinh \frac{r_1+r_2}{2}, ~~~~~ 
& b_1 = v \sinh \frac{r_1-r_2}{2} \sin\frac{s_1- s_2}{2}, 
\\
a_3 = - v \cosh \frac{r_1+r_2}{2},~~~~~
& b_3 = v \cosh \frac{r_1-r_2}{2} \cos\frac{s_1-s_2}{2},
\end{align}
and $\phi_0 = (s_1+s_2)/2$.
The turning points, i.e., the zeros of $Q(t)$, are located at
\begin{align}
t = \frac{1}{\omega}(s_i \pm i r_i + 2 \pi n), ~~~~ i=1,2,~~~ n \in \mathbb Z. 
\end{align}
As shown in Fig.~\ref{fig:Stokes_phenomenon}, 
varying the positions of the turning points $(s_i,r_i)$ causes a Stokes phenomenon. 
For the case in Fig.~\ref{fig:Stokes_phenomenon}-(a), the structure of the Stokes graph with two degenerate Stokes curves intersecting the real axis is essentially the same as that of the system discussed in Sec.~\ref{sec:4}, and hence the function $D(\theta)$ takes the same form as Eq.~\eqref{eq:D_theta}. 
The Stokes graph in Fig.~\ref{fig:Stokes_phenomenon}-(b), however, has a different structure: only a single degenerate Stokes curve intersects the real axis. Therefore, the monodromy matrix takes the form: 
\begin{align}
M_{\rm med} = 
\begin{pmatrix}
P_1^{+\frac{1}{2}} & 0 \\ 0 & P_1^{-\frac{1}{2}}
\end{pmatrix}
\begin{pmatrix}
\sqrt{1+B_1} & -i \sqrt{B_1} \\ +i \sqrt{B_1} & \sqrt{1+B_1},
\end{pmatrix},
\end{align}
where $P_1$ and $B_1$ are given in Eqs.~\eqref{eq:AB_def} and \eqref{eq:P_def} with $t_1=0$, $t_2=2\pi/\omega$ and $t_{1\pm}=s_1 \pm i r_1$. 
For this monodromy matrix, the function $D(\theta)$ is
\begin{align}
D(\theta) = 2 e^{i\theta} \left[ \cos \theta - \sqrt{1+B_1} \, \frac{P_1^{+\frac{1}{2}}+P_1^{-\frac{1}{2}}}{2} \right].
\end{align}
Therefore, the quasi-energy is given by
\begin{align}
\epsilon = \frac{\hbar}{T} \theta =  \frac{\hbar \omega\theta}{2\pi} = \pm \frac{\hbar \omega}{2\pi} \arccos \left( \sqrt{1+B_1} \, \frac{P_1^{+\frac{1}{2}}+P_1^{-\frac{1}{2}}}{2} \right).
\end{align}
Fig.~\ref{fig:single_DS} shows the comparison of the numerical result and the exact-WKB result approximated by truncating $P_1$ and $B_1$ at the next leading order in the $\hbar$ expansion. 
As this example implies, we can ignore the effect of the degenerate Stokes curves that disappear due to the Stokes phenomena in the parameter space. 
Note that the physical quantities such as the quasi-energy must be smoothly connected even when there are Stokes phenomena in the parameter space. 
The discontinuous change of the function $D(\theta)$ merely means that the asymptotic form of the physical quantity changes, but the full expression is smoothly connected by certain connection formulas.

One can construct a monodromy matrix specific to the case that an additional Stokes phenomenon occurs in the parameter space while keeping $\arg(\hbar)=0$.
Here, we consider Figs~.\ref{fig:Stokes_phenomenon} (a) and (b) and discuss the Stokes phenomenon between the two cases.
The corresponding median-resummed monodromy matrices are given by
\begin{align}
M_{\rm med}^{(2)} = {\cal N}_{t_3,t_2} R_2 {\cal N}_{t_2,t_1} R_1, \qquad  M_{\rm med}^{(1)} = {\cal N}_{t_3,t_1} R_1, \label{eq:Mmed_2_1}
\end{align}
where $M_{\rm med}^{(2)}$ and $M_{\rm med}^{(1)}$ are the monodromy matrices corresponding to  Figs.~\ref{fig:Stokes_phenomenon} (a) and (b), respectively.
Then, we introduce another Borel resummation ${\cal S}_{\cal P}$, where ${\cal P}$ denotes a parameter set in $f_{i}(t)$ of the Hamiltonian in Eq.~\eqref{eq:H_fsig}, and suppose that the Stokes phenomenon happens at a certain parameter set, ${\cal P}={\cal P}_*$.
Thus, the two monodromy matrices in Eq.~\eqref{eq:Mmed_2_1} are connected to each other by the associated Borel resummation, ${\cal S}_{{\cal P}_*}$ as
\begin{align}
{\cal S}_{{\cal P}_* + {0}_+}[M_{\rm med}^{(2)}] = {\cal S}_{{\cal P}_* + {0}_-} [M_{\rm med}^{(1)}].
\end{align}
By introducing the Stokes automorphism, ${\mathfrak S}_{{\cal P}_*}^{\nu}$, defined as ${\cal S}_{{\cal P}_* + {0}_+} = {\cal S}_{{\cal P}_* + {0}_-} \circ {\mathfrak S}_{{\cal P}_*}^{\nu = 1}$, one can find
\begin{align}
 {\mathfrak S}^{\nu = 1}_{{\cal P}_*}[M_{\rm med}^{(2)}] = M_{\rm med}^{(1)} \quad 
    &\Rightarrow \quad {\mathfrak S}^{\nu}_{{\cal P}_*}[N_{t_3,t_1}]= (N_{t_3,t_2} R_2 N_{t_2,t_3} )^{-\nu} N_{t_3,t_1}  .
\end{align}
As a result, we end up with the monodromy matrix without the discontinuity as
\begin{align}
M_{\rm med}^{(2=1)} &:= {\mathfrak S}^{+\frac{1}{2}}_{(2 \rightarrow 1)}[M_{\rm med}^{(2)}] = {\mathfrak S}^{-\frac{1}{2}}_{(2 \rightarrow 1)}[M_{\rm med}^{(1)}] \nonumber\\
&= (N_{t_3,t_2} R_2 N_{t_2,t_3} )^{\frac{1}{2}} N_{t_3,t_1} R_1. 
\end{align}

It is noteworthy that this is a generalizations of the standard approach to Borel resummation involving  the variation of the complex phase of $\hbar$ (or a coupling constant).
In general, a Stokes phenomenon can occur not only through changes in $\arg(\hbar)$ but also via variations in other parameters of the theory or observables. The movable singularity depending on the coordinate in the wave function is an illustrative example of such a broader case.
This indicates that discontinuities of physical quantities are induced not exclusively by changes in $\arg(\hbar)$ but also by variations in various parameters.
From this viewpoint, both $\arg(\hbar)$ and the coordinate $t$ can be naturally incorporated into the set of parameters ${\cal P}$ in the above analysis.
In such situations, the discontinuities in the observable ${\cal O}_{{\cal P}_*+{0_\pm}}$, which originate from the variations of multiple parameters are eliminated by individually constructing ${\cal O}_{\rm med}$ for each parameters causing the Stokes phenomena.

\section{Summary} 
\label{sec:6}

We have applied the exact-WKB analysis to two-level Floquet systems and established a systematic procedure for calculating the quasi-energy and Floquet effective Hamiltonian.  In our exact-WKB approach, the time-evolution unitary matrix, and the resulting quasi-energy and Floquet effective Hamiltonian, are determined from the monodromy matrix, which is given by a product of the connection matrices for the degenerate Stokes curves appearing in the Stokes graph of Floquet systems.  The connection matrices are obtained by evaluating the cycle integrals (the Voros symbols), which can be done systematically using the low-frequency expansion.  
 
Considering a simple Floquet system, we have demonstrated that our approach is especially effective in the low-frequency regime and can be improved systematically by including higher-order $\hbar$ corrections. We have also shown that there exist non-perturbative modifications to the quasi-energy $\propto e^{-1/\hbar}$, which is significant for capturing the resonance-like behaviors that appear when the perturbative quasi-energies degenerate and are lifted by the non-perturbative effects.  

In addition, we have found that, unlike the quasi-energy, the effective Hamiltonian explicitly depends on the wave functions. This requires the removal of discontinuities of the wave functions from both fixed and movable singularities on the Borel plane. We have found that considering the Stokes automorphism for both movable and fixed singularities is necessary to obtain results that match the numerical calculations.

As a technical outcome of the exact-WKB analysis, we have established in this paper a method to eliminate discontinuities arising from degenerate Stokes curves. We have then derived the formulas for connection matrices for the degenerate Stokes curves.
This method can be applied to time-dependent Schr"odinger-type differential equations with real potentials and degenerate Stokes curves, which commonly arise in various contexts in physics such as the Schwinger effect~\cite{Taya:2020dco}.

A possible generalizations of the present work is its application to multi-level Floquet systems. 
For this purpose, we need to generalize the exact-WKB analysis to higher-order differential equations.
Such a generalization has long been discussed by mathematicians, but applications to physical problems remain limited (see e.g. \cite{Enomoto:2022nuj}).
A detailed exploration of this topic remains a subject for future analysis.

\begin{acknowledgments}
We would thank Muneto~Nitta and Norisuke~Sakai for enlightening discussions.
This work is supported by the Ministry of Education, Culture, 
Sports, Science, and Technology (MEXT)-Supported Program for the Strategic Research Foundation at Private Universities ``Topological Science" (Grant No. S1511006) 
and by the Japan Society for the Promotion of Science Grant-in-Aid for Scientific Research (JSPS KAKENHI) Grant No.~18H01217. 
T. F is supported by JSPS KAKENHI Grant No.~21K03558. 
S. K is supported by JSPS KAKENHI Grant No.~22H05118.
T. M. is supported by JSPS KAKENHI Grant Nos.~23K03425 and 22H05118.
H. T. is supported by JSPS KAKENHI Grant No.~24K17058 and the RIKEN TRIP initiative (RIKEN Quantum).
\end{acknowledgments}

\appendix
\section{WKB expansion} \label{appendix:WKB expansion}

\subsection{WKB expansion and formal WKB solutions} \label{appendix:recursion}

In this appendix, we briefly review the WKB expansion (see, for example, Ref.~\cite{Kawai1} for details). 
We start with the Schr\"{o}dinger equation (as in the main text, we denote the independent variable as $t$ instead of $x$),
\begin{align}
\left[ - \hbar^2 \frac{\partial^2}{\partial t^2} + Q(t, \hbar) \right] \psi(t,\hbar) = 0. \qquad \label{eq:app_schr}
\end{align}
We assume that the potential $Q(t,\hbar)=V(t,\hbar)-E$ has a (formal) expansion of $\hbar$ given by
\begin{align}
Q(t,\hbar) = \sum_{n=0}^\infty Q_n(t) \hbar^{n}. \qquad (Q_0(t) \ne 0)
\end{align}
It is convenient to change the dependent variable from the wave function $\psi(t,\hbar)$ to $\mathcal W(t,\hbar)$ defined by
\begin{align}
\psi(t,\hbar) = \exp \left( \frac{\mathcal W(t,\hbar)}{\hbar} \right).
\end{align}
Substituting it into the Schr\"odinger equation, we obtain 
the following differential equation for $\mathcal W$
\begin{align}
(\partial_t \mathcal W)^2 + \hbar (\partial_t^2 \mathcal W ) = Q.
\label{eq:Riccati_W}
\end{align}
Let $S(t,\hbar)$ be the formal power series for $\partial_t \mathcal W/\hbar$
\begin{align}
\frac{\partial_t \mathcal W}{\hbar} = S(t,\hbar) = \sum_{n=-1}^\infty S_{n}(t)\hbar^n .
\end{align}
Substituting it into Eq.~\eqref{eq:Riccati_W}, 
we obtain the following recursion relation, from which the expansion coefficients $S_n(t)$ can be determined order-by-order:
\begin{align}
&& S_{-1} = \pm \sqrt{Q_0}\,,                                                    \quad\quad
	2S_{-1}S_n+\sum_{j=0}^{n-1}S_jS_{n-1-j} + \frac{\partial S_{n-1}}{\partial t} = Q_{n+1}. \qquad  (n = 0,1,2,\cdots)\,
\end{align}
Depending on the sign of $S_{-1} = \pm \sqrt{Q_0}$, one finds two independent solutions, which we call $S^\pm(t,\hbar)$.
Then, we define $S_{\rm odd}(t,\hbar)$ and $S_{\rm even}(t,\hbar)$ as
\begin{align}
S_{\rm odd}(t,\hbar) := \frac{S^+(t,\hbar)-S^-(t,\hbar)}{2}, \qquad S_{\rm even}(t,\hbar) := \frac{S^+(t,\hbar) + S^-(t,\hbar)}{2}.
\end{align}
We can show that $S_{\rm even}(t,\hbar)$ is expressed using $S_{\rm odd}(t,\hbar)$ as
\begin{align}
S_{\rm even}(t,\hbar) = - \frac{1}{2} \frac{\partial \log S_{\rm odd}(t,\hbar)}{\partial t}.
\end{align}
Since $S^{\pm} = \pm S_{\rm odd} + S_{\rm even}$, the two independent wave functions, $\psi^{\pm}(t,\hbar)$, can be written as
\begin{align}
\psi^{\pm}(t,\hbar) = \frac{\exp \left[ \pm \int_{a}^t d t^\prime \,  S_{\rm odd}(t^\prime,\hbar)\right] }{\sqrt{S_{\rm odd}(t,\hbar)}} ,
\label{eq:WKB_sol_appendix}
\end{align}
where $a$ is an arbitrary normalization point on the complex $t$-plane, which determines the normalization factor of the wave function.
Using a function $ W(t,\hbar)$ satisfying $\partial_t W(t,\hbar) = \hbar S_{\rm odd}(t,\hbar)$,
the WKB solutions can also be written as 
\begin{align}
\psi^{\pm}(t,\hbar) =  C_\pm(\hbar;a) \frac{\exp \left[ \pm W(t,\hbar)/\hbar \right] }{\sqrt{ \partial_t  W(t,\hbar)}},
\label{eq:WKB_solutions}
\end{align}
where $C(\hbar;a)$ is a normalization constant depending on the normalization point $a$ and $\hbar$.

\subsection{WKB expansion of first-order linear differential equation}
\label{subsec:WKB_1st_order}
In this subsection, we summarize the WKB expansion of the first-order linear differential equation 
\begin{align}
i\hbar \pdv{t} \Psi(t) = H(t) \Psi(t), 
\hspace{1cm} \Psi(t) = \mqty( \psi_1(t) \\ \psi_2(t)),
\label{eq:1st_SchEq}
\end{align}
where $H(t)$ is the time-dependent Hamiltonian
\begin{align}
H(t) = \sum_{i=1}^3 \sigma_i f_i(t).
\end{align}
Let $\xi^\pm(t)$ be the following instantaneous eigenvectors of the Hamiltonian $H(t)$
\begin{align}
\xi^+ &= \mqty( \xi^+_1(t) \\ \xi^+_2(t)) = \frac{e^{-i\gamma(t)}}{\sqrt{\epsilon_0(t)(\epsilon_0(t)+f_3(t))}} \mqty( \epsilon_0(t)+f_3(t) \\ f_1(t)+i f_2(t)) , \\
\xi^- &= \mqty( \xi^-_1(t) \\ \xi^-_2(t)) = \frac{e^{+i\gamma(t)}}{\sqrt{\epsilon_0(t)(\epsilon_0(t)+f_3(t))}} \mqty( f_1(t)-i f_2(t) \\ -\epsilon_0(t)-f_3(t)), 
\end{align}
where $\pm \epsilon_0(t) = \pm \sqrt{f_1(t)^2+f_2(t)^2+f_3(t)^2}$ are the instantaneous eigenvalues of $H(t)$ and $\gamma(t)$ is the Berry phase determined from the condition $\xi^\pm \partial_t \xi^\pm = 0$, or equivalently
\begin{align}
\partial_t \gamma(t) = \frac{f_1(t) \partial_t f_2(t) - f_2(t) \partial_t f_1(t)}{\epsilon_0(t)(\epsilon_0(t)+f_3(t))}.
\end{align}
It is convenient to change the dependent variable from $\Psi(t)$ to $\Phi(t)$ by using the unitary matrix $U_0(t)$ that diagonalizes the Hamiltonian $H(t)$ at each $t$
\begin{align}
\Psi(t) = U_0(t) \Phi(t), 
\hspace{5mm}
U_0(t) = 
\begin{pmatrix}
\xi_1^{+}(t) & \xi_1^{-}(t) \\
\xi_2^{+}(t) & \xi_2^{-}(t)
\end{pmatrix},
\hspace{5mm}
\Phi(t)=
\begin{pmatrix}
\phi^{+}(t) \\
\phi^{-}(t)
\end{pmatrix},
\label{eq:linear_combi_eigenv}
\end{align} 
Substituting it into Eq.~\eqref{eq:Sch_eq}, 
we obtain a linear differential equation for $\Phi(t)$
\begin{align}
\partial_t \Phi(t) = \frac{i}{\hbar}
\begin{pmatrix}
-\epsilon_0(t) & 0 \\ 0 & + \epsilon_0(t)
\end{pmatrix}
\Phi(t) 
 - \begin{pmatrix}
0 & i \gamma_+(t) \\
-i \gamma_-(t)  & 0
\end{pmatrix} \Phi(t),
\label{eq:1st_Phi}
\end{align}
where we have defined 
\begin{align}
i \gamma_+(t) = \overline{-i \gamma_-(t)} = \xi^{+\dagger} \partial_t \xi^{-}. 
\end{align}
By expanding this equation in powers of $\hbar$, we can determine the WKB solutions for $\phi^+(t)$ and $\phi^-(t)$ order-by-order. 
We can show that both $\phi^+(t)$ and $\phi^-(t)$ takes the form in Eq.~\eqref{eq:WKB_solutions} by eliminating $\phi^-(t)$ and $\partial_t \phi^-(t)$ [or $\phi^+(t)$ and $\partial_t \phi^+(t)$] from Eq.~\eqref{eq:1st_Phi}
and recasting it into a second-order differential equation of the form in Eq.~\eqref{eq:app_schr}.  
There are two linearly independent solutions for $\Phi=(\phi^+(t),\phi^-(t))$
\begin{align}
\Phi^\pm = 
\begin{pmatrix} \displaystyle
\sqrt{\frac{i \hbar \, \gamma_+(t)}{\partial_t W^+(t,\hbar)}} \exp \left(\pm \frac{W^+(t,\hbar)}{\hbar} \right) \\ \displaystyle
\sqrt{\frac{-i\hbar  \, \gamma_-(t)}{\partial_t W^-(t,\hbar)}} \exp \left( \pm \frac{W^-(t,\hbar)}{\hbar} \right)
\end{pmatrix}, 
\end{align}
where $W^\pm(t,\hbar)$ are the functions whose asymptotic expansions are given by
\begin{align}
W^\pm(t,\hbar) &= \int_a^t dt' \, S_{\rm odd}^\pm(t',\hbar),  \\
S_{\rm odd}^\pm(t,\hbar) &= i \epsilon_0 \pm \frac{\hbar}{2} \partial_{t} \log \frac{\pm i \hbar \, \gamma_\pm}{\epsilon_0} + \frac{i\hbar^2}{2} \left[ \frac{\gamma_+ \gamma_-}{\epsilon_0} - \partial_t \left( \frac{1}{2\epsilon_0} \partial_{t'} \log \frac{\pm i \hbar \, \gamma_\pm}{\epsilon_0} \right) \right] + \mathcal O(\hbar^3).
\label{eq:Wpm_expansion}
\end{align}
From these asymptotic forms of $W^\pm(t,\hbar)$, 
we can show that the leading-order expressions of the linearly independent solutions $\Phi^\pm(t)$ are 
\begin{align}
\Phi^+ = \exp \left( + i \int_0^t dt' \epsilon_0(t') \right) \begin{pmatrix}
0 + \mathcal O(\hbar) \\
1 + \mathcal O(\hbar)
\end{pmatrix}, \hspace{5mm}
\Phi^- = 
\exp \left( - i \int_0^t dt' \epsilon_0(t') \right) \begin{pmatrix}
1 + \mathcal O(\hbar) \\
0 + \mathcal O(\hbar)
\end{pmatrix}. 
\label{eq:adiabatic_approx}
\end{align}
We can see from Eq.~\eqref{eq:linear_combi_eigenv} that the leading-order approximation of the solutions is given by the instantaneous eigenvectors of the time-dependent Hamiltonian. 
This is consistent with the adiabatic theorem.
By calculating the higher-order corrections to Eq.~\eqref{eq:Wpm_expansion}, 
we can determine the corrections to the adiabatic approximation \eqref{eq:adiabatic_approx}.  

\subsection{WKB expansion of second-order linear differential equation}
\label{subsec:WKB_2nd_order}
Next, we discuss the WKB expansion of the second-order linear differential equation that is equivalent to the Shcr\"odinger equation \eqref{eq:1st_SchEq}. 
Let us consider an arbitrary linear combination of $\psi_1(t)$ and $\psi_2(t)$ 
\begin{align}
\psi(t) = \langle C(t) , \Psi(t) \rangle = - c_2(t) \psi_1(t) + c_1(t) \psi_2(t),
\end{align}
where $\Psi=(\psi_1(t), \psi_2(t))^T$ are the components of the solution to the first-order linear differential equation~\eqref{eq:1st_SchEq} and $C = (c_1(t),c_2(t))^T$ are arbitrary functions. 
The brackets $\langle \cdot \,, \cdot \rangle$ denote the anti-symmetric bilinear form for two-component vectors
\begin{align}
\langle A, B \rangle = - a_2 b_1 + a_1 b_2, \hspace{10mm}
A = 
\begin{pmatrix}
a_1 \\ a_2
\end{pmatrix},~~~
B = 
\begin{pmatrix}
b_1 \\ b_2
\end{pmatrix}.
\end{align}
We can show that if $\Psi=(\psi_1(t), \psi_2(t))^T$ satisfies Eq.~\eqref{eq:1st_SchEq}, 
the linear combination $\psi(t)=\langle C(t) , \Psi(t) \rangle$ with arbitrary coefficients $C = (c_1(t),c_2(t))^T$ satisfies the second-order differential equation
\begin{align}
    \qty[-\hbar^2\qty(\pdv{t}-\pdv{t}\lambda(t,\hbar))^2+Q(t,\hbar)] \psi = 0,
\label{eq:2nd_order_diffeq_app}
\end{align}
with
\begin{align}
Q(t,\hbar) = - \hbar^2 \left[ \frac{\langle {\cal D}C, {\cal D}^2C \rangle}{\langle C, {\cal D} C \rangle} + \partial_t \lambda(t,\hbar) - (\partial_t \lambda(t,\hbar))^2 \right], \hspace{7mm}
\lambda(t,\hbar) = \frac{1}{2} \log \langle C, {\cal D} C \rangle
,
\end{align}
where we have defined the derivative operator $\cal D$ as
\begin{align}
{\cal D} C = \left[ \partial_t - \frac{1}{i \hbar} H(t) \right] C. 
\end{align}
Compared with the standard form of the Schr\"odinger equation \eqref{eq:app_schr},  Eq.~\eqref{eq:2nd_order_diffeq_app} has an additional term with $\lambda$, which shifts $S_{\rm even}$. 
Therefore, the WKB ansatz for the wave function $\psi(t)$ becomes 
\begin{align}
\psi(t) = \frac{1}{\sqrt{\partial_t W(t,\hbar)}} \exp \left(\lambda(t,\hbar) \pm \frac{W(t,\hbar)}{\hbar} \right).
\end{align}
Substituting it into Eq.~\eqref{eq:2nd_order_diffeq_app}, we obtain the differential equation for $W(t,\hbar)$ 
\begin{align}
&(\partial_t W(t,\hbar))^2-\frac{\hbar^2}{2} \qty{ W(t,\hbar), t} = Q(t,\hbar), 
\end{align}
where $\{W,t\}$ denotes the Schwarzian derivative, 
\begin{align}
\{W,t\} = \frac{\partial_t^3 W}{\partial_t W} - \frac{3}{2} \qty(\frac{\partial_t^2 W}{\partial_t W} )^2.
\end{align}
The asymptotic form of the function $W(t,\hbar)$ is given by
\begin{align}
\frac{W(t,\hbar)}{\hbar} = \int_0^t dt' S_{\rm odd}(t',\hbar),
\end{align}
with
\begin{align}
S_{\rm odd} = \frac{\sqrt{Q_0(t)}}{\hbar} + \frac{Q_1(t)}{2\sqrt{Q_0(t)}} 
+ \frac{\hbar}{2\sqrt{Q_0(t)}} \left[ Q_2(t) - \frac{1}{4} \frac{Q_1(t)^2}{Q_0(t)} + \frac{1}{4} \frac{Q_0''(t)}{Q_0(t)} - \frac{5}{16} \frac{Q_0'(t')^2}{Q_0(t)^2} \right] + \cdots .
\end{align}
Explicitly, the asymptotic series for $S_{\rm odd}$ can be written as
\begin{align}
S_{\rm odd}(t,\hbar) = \frac{i \epsilon_0(t)}{\hbar} + \left[ \frac{1}{2} \frac{\varphi \partial_t \bar \varphi - \bar \varphi \partial_t \varphi}{1+|\varphi|^2} +  \partial_t w_0(t) \right] + \hbar \left[ \frac{i}{2\epsilon_0} \frac{|\partial_t \varphi|^2}{(1+|\varphi|^2)^2} + \partial_t w_1(t) \right] + \cdots,
\label{eq:explicit_Sodd}
\end{align}
where we have defined 
\begin{align}
\varphi(t) = \frac{f_1+i f_2}{f_3+\epsilon_0}. 
\hspace{5mm}
w_0(t) = \frac{1}{2} \log \frac{c_1 \varphi-c_2}{c_2 \bar \varphi+c_1},
\hspace{5mm}
w_1(t) = - \frac{i}{2\epsilon_0} \left[  \frac{\epsilon_0'(t)}{\epsilon_0(t)} + \frac{\langle C, H'(t) C \rangle}{{\langle C, H(t) C \rangle}} \right].
\end{align}
We can see from Eq.~\eqref{eq:explicit_Sodd} that the dependence on the arbitrary chosen coefficients $C=(c_1(t),c_2(t))^T$ is contained in the total derivative terms $\partial_t w_0(t)$ and $\partial_t w_1(t)$. 
In particular, for different  coefficients $C=(c_1(t),c_2(t))^T$ and $\tilde C=(\tilde c_1(t),\tilde c_2(t))^T$, the corresponding $S_{\rm odd}(C)$ and $S_{\rm odd}(\tilde C)$ are related to each other as
\begin{align}
S_{\rm odd}(C') = S_{\rm odd}(C) - \frac{1}{2} \partial_t \log \frac{(\lambda-W/\hbar-\frac{1}{2} \log \partial_t W) \langle \tilde C, C \rangle - \langle \tilde C, {\cal D} C \rangle}{(\lambda+W/\hbar-\frac{1}{2} \log \partial_t W) \langle \tilde C, C \rangle - \langle \tilde C, {\cal D} C \rangle}. 
\end{align}
This implies that the cycle integrals of $S_{\rm odd}$ (such as $\mathcal B$) are independent of the choice of the coefficients of the linear combination $C=(c_1(t),c_2(t))^T$.  

\section{Removing discontinuity of Borel resummation} \label{sec:Quant_DDP_Stokes}

In this appendix, we demonstrate how to eliminate discontinuities that arise when the Borel resummation is applied to a Borel non-summable trans-series.
We also discuss the Borel resummed form of monodromy matrix.
Our strategy is outlined from a general viewpoint in Sec.~\ref{sec:Quant_DDP_Stokes_story}. We then introduce Delabaere-Dillinger-Pham (DDP) formula for the cycle integrals and wave functions in Secs.~\ref{sec:DDP_cycle} and \ref{sec:DDP_wave}, respectively. 
By using the insights from Secs.~\ref{sec:DDP_cycle} and \ref{sec:DDP_wave}, we derive the formula for the discontinuity-free monodromy matrix for the $N=2$ case in Sec.~\ref{app:exact_monodromy} and then generalize it to the $2N$ case in Sec.~\ref{app:2N_turning_points}.
In Sec.~\ref{appendix:alien_movable}, we discuss how to manage movable singularities in a trans-series of a wave function, which is a crucial ingredient for computing the effective Hamiltonian.

\subsection{Stokes automorphism and median resummation} 
\label{sec:Quant_DDP_Stokes_story}

Before addressing our specific problem, let us briefly outline our strategy for obtaining an exact Borel-resummed form and its discontinuity-free trans-series.
Let $f(\hbar)$ represent a function, such as a wave function, a monodromy matrix, or other quantities that we want to calculate.
Suppose that we have its formal power series $f_{\rm P}(\hbar)$ expanded in powers of $\hbar$ in the limit $\hbar \rightarrow 0_+$:
\begin{align}
f(\hbar) ~~\xrightarrow{\hbar \rightarrow 0_+}~~ f_{\rm P}(\hbar) = \sum_{n=0}^\infty c_n(t) \hbar^n,
\end{align}
where ``$\xrightarrow{\hbar \rightarrow 0_+}$" indicates the asymptotic expansion as $\hbar$ approaches zero from the positive side. 
If $c_n \sim n!$ for large $n$, 
the power series $f_{\rm P}(\hbar)$ is a divergent asymptotic series. Therefore, to determine $f(\hbar)$ from $f_{\rm P}(\hbar)$, we need to apply the Borel resummation to $f_{\rm P}(\hbar)$.

Let ${\cal B}$ denote the Borel transformation, that is, a linear operator that acts on the powers of $\hbar$ as 
\begin{align}
{\cal B} [\hbar^n]:= \frac{\xi^{n-1}}{\Gamma(n)}.
\end{align}
The Borel transformation maps a series of $\hbar$ into that of $\xi$ as ${\cal B} [f_{\rm P}(\hbar)] \equiv \tilde{f}(\xi)$. 
Let ${\cal L}_\theta$ be the Laplace transformation with angle $\theta$ on the complex $\xi$-plane (Borel plane):
\begin{align}
{\cal L}_\theta[{\tilde f}(\xi)] = \int^{\infty e^{i \theta}}_0 d \xi \, e^{- \xi/\hbar} {\tilde f}(\xi).
\end{align}
Using ${\cal L}_\theta$ and ${\cal B}$, 
the Borel resummation ${\cal S}$ is expressed as 
\begin{align}
{\cal S}:= {\cal L}_0 \circ {\cal B}.
\end{align} 
Roughly speaking, the Borel resummation is a method to reconstruct $f(\hbar)$ from its asymptotic series. 
If ${\cal B} [f_{\rm P}(\hbar)]$ has no singularities along the real axis on the Boral plane, the Laplace integration $\mathcal L_0$ can be performed without any problem. In such cases, $f_{\rm P}(\hbar)$ is referred to as Borel summable; otherwise, it is referred to as Borel non-summable. 

When $f_{\rm P}(\hbar)$ is Borel non-summable, the integration can often be performed by introducing a small complex phase in the Laplace integral, such as $\theta = 0_\pm$, to avoid the singularities on the real axis.
We denote the Borel resummations with $\theta = 0_{\pm}$ as ${\cal S}_{0_\pm}$: 
\begin{align}
{\cal S}_{0_\pm}:={\cal L}_{0_\pm} \circ {\cal B} .
\end{align}
Although the Borel resummation with angle $\theta$ yields a finite function, it has a discontinuity at $\theta=0$ due to the singularities on the Borel plane. 
Moreover, in general, neither ${\cal S}_{0_+}[f_{\rm P}(\hbar)]$ nor ${\cal S}_{0_-}[f_{\rm P}(\hbar)]$ coincides with $f(\hbar)$ :
\begin{align}
{\cal S}_{0_+}[f_{\rm P}(\hbar)] \not = {\cal S}_{0_-}[f_{\rm P}(\hbar)] \not = f(\hbar).
\end{align}
The difference between ${\cal S}_{0_\pm}[f_{\rm P}(\hbar)]$ is exponentially decreasing in the limit $\hbar \rightarrow 0_+$. 
Therefore, by adding exponentially small terms to the power series $f_{\rm P}(\hbar) = \sum_{n=0}^\infty c^{(0)\pm}_n(t) \hbar^n$, we can construct a series
$f_{0_+}(\hbar)$ and $f_{0_-}(\hbar)$ 
whose Borel resummations agree with $f(\hbar)$:
\begin{align}
{\cal S}_{0_+}[f_{0_+}(\hbar)] = {\cal S}_{0_-}[f_{0_-}(\hbar)] = f(\hbar) .
\label{eq:f0+_f0-}
\end{align}
In general, $f_{0_+}(\hbar)$ and $f_{0_-}(\hbar)$ take the form:
\begin{align}
f_{0_\pm} = \sum_{n=0}^\infty c^{(0)\pm}_n(t) \hbar^n + e^{-\frac{S^{(1)}}{\hbar}} \sum_{n=0}^\infty c^{(1)\pm}_n(t) \hbar^n + e^{-\frac{S^{(2)}}{\hbar}} \sum_{n=0}^\infty c^{(2)\pm}_n(t) \hbar^n +\cdots .
\label{eq:trans-series_f0pm}
\end{align}
The series of this form is called \textit{trans-series}. 

As mentioned above, ${\cal S}_{0_+}$ and ${\cal S}_{0_-}$ do not agree in general. 
The \textit{Stokes automorphism} ${\mathfrak S}$ is an operation that compensates the mismatch between ${\cal S}_{0_+}$ and ${\cal S}_{0_-}$:
\begin{align}
{\cal S}_{0_+} = {\cal S}_{0_-} \circ {\mathfrak S}. 
\label{eq:Stokes_aut}
\end{align}
Let us define the \textit{median resummation} ${\cal S}_{\rm med}$ and the trans-series $f_{\rm med}(\hbar)$ as 
\begin{align}
{\cal S}_{\rm med} := {\cal S}_{0_+} \circ {\mathfrak S}^{-1/2} = {\cal S}_{0_-} \circ {\mathfrak S}^{+1/2}, \label{eq:def_med}
\end{align}
where ${\mathfrak S}^{\nu}~(\nu \in \mathbb R)$ is the Stokes automorphism extended to a one-parameter group using the generators called \textit{alien derivative} $\bul{\Delta}$.
It is formally expressed as
\begin{align}
{\mathfrak S}^{\nu} = \exp \left[ \nu \sum_{w \in \Gamma} \bul{\Delta}_w \right] 
= 1 + \sum_{k=1}^{\infty} \sum_{\{n_1,\cdots,n_k \ge 1\}} \frac{\nu^k}{k!} \prod_{s=1}^k \bul{\Delta}_{w_{n_s}}, \label{eq:Stokes_nu}
\end{align}
where $\Gamma$ is a set of singularities along the real axis in the Borel plane.
By definition, ${\mathfrak S}^{0} = {\rm id}$., ${\mathfrak S}^{+1} =: {\mathfrak S}$, and ${\mathfrak S}^{\nu} \circ {\mathfrak S}^{\nu^\prime} = {\mathfrak S}^{\nu + \nu^\prime}$.
Corresponding to the median resummation ${\cal S}_{\rm med}$, the trans-series $f_{\rm med}(\hbar)$ is defined as
\begin{align}
f_{\rm med}(\hbar) := {\mathfrak S}^{+ 1/2}[f_{0_+}(\hbar)] = {\mathfrak S}^{- 1/2}[f_{0_-}(\hbar)]. \label{eq:fex_fpm}
\end{align}
One can show that the median resummation of $f_{\rm med}(\hbar)$ agrees with $f(\hbar)$: 
\begin{align}
{\cal S}_{\rm med}[f_{\rm med}(\hbar)] = f(\hbar).
\end{align}
An important property of the median resummation ${\cal S}_{\rm med}$ is that it commutes with the complex conjugation ${\cal C}$\footnote{This property can be proven by using the following relations: 
\begin{align}
{\cal C} \circ {\cal S}_{0_+} = {\cal S}_{0_-} \circ {\cal C}, \qquad {\cal C} \circ {\mathfrak S}^{\nu}  = {\mathfrak S}^{-\nu} \circ {\cal C} , \qquad {\cal C}^2 = {\rm id}., 
\end{align}}: 
\begin{align}
{\cal C} \circ {\cal S}_{\rm med} = {\cal S}_{\rm med} \circ {\cal C}.
\end{align}
Therefore, if $f(\hbar)$ is a real function, 
the trans-series $f_{\rm med}(\hbar)$ is also real:
\begin{align}
{\cal C}[f(\hbar)] = f(\hbar) \quad \Longrightarrow \quad {\cal C}[f_{\rm med}(\hbar)] = f_{\rm med}(\hbar). 
\end{align}
In general, $f_{0_{\pm}}$ is not real and plagued with an imaginary ambiguity. 
On the other hand $f_{\rm med}(\hbar)$ is real and hence it can be used to approximate $f(\hbar)$. 
To apply the median resummation, we need to know how the Stokes automorphism acts on a trans-series. 
In the next subsection, we briefly review the DDP formula that determines the actions of the Stokes automorphism on the cycle integrals appearing in the connection formula.

\subsection{DDP formula for cycle integrals} \label{sec:DDP_cycle}
Here, we briefly explain the DDP formula for the cycle integrals, i.e., the Voros symbols
(see, for example, Refs.~\cite{DDP2,DP1,Kamata:2021jrs,Iwaki1} for details).

We start with the $A$- and $B$-cycles in Fig.~\ref{fig:cycles1}.
For simplicity, we focus on the $N=2$ case.
The DDP formula, which relates ${\cal S}_{0_+}$ and ${\cal S}_{0_-}$ for the cycle integrals, is given by
\begin{gather}
{\cal S}_{0_+}[A^{\pm 1}_1] = {\cal S}_{0_-}[A^{\pm 1}_1] \prod_{j=1}^2 \left( 1 + {\cal S}[B_j]\right)^{\mp 1}, \quad {\cal S}_{0_+}[A^{\pm 1}_2] = {\cal S}_{0_-}[A^{\pm 1}_2] \prod_{j=1}^2 \left( 1 + {\cal S}[B_j]\right)^{\pm 1}, \\
{\cal S}_{0_+}[B_j] = {\cal S}_{0_-}[B_j]. \qquad (j=1,2)
\end{gather}
Thus, the corresponding Stokes automorphism acts on the cycle integrals as 
\begin{gather}
{\mathfrak S}[A^{\pm 1}_1] = A^{\pm 1}_1\prod_{j=1}^2 (1+B_j)^{\mp 1}, \qquad {\mathfrak S}[A^{\pm 1}_2] = A^{\pm 1}_2\prod_{j=1}^2 (1+B_j)^{\pm 1}, \label{eq:St_auto_A} \\
{\mathfrak S}[B_j] = B_j.
\end{gather}
These formulas imply that the $A$-cycles are Borel non-summable, but the $B$-cycles are summable. 
One can identify the location of singularities on the Borel plane, denoted as $\Gamma$ in Eq.~(\ref{eq:Stokes_nu}), from Eq.~(\ref{eq:St_auto_A}) as
\begin{align}
\Gamma = \Gamma_1 \cup \Gamma_2, \qquad \Gamma_j = \{ \, n S_{b_j} \ | \ n \in {\mathbb N} \, \},  
\end{align}
where $S_{b_j} \in {\mathbb R}_+$ is defined as $S_{b_j} = - \lim_{\hbar \rightarrow 0_+} \hbar \log B_j$.
Labeling the elements as $w_1 < w_2 < \cdots \in \Gamma$, one can write down the action of the alien derivative to the cycle integrals as
\begin{align}
\bul{\Delta}_{w_n} A^{\pm 1}_j &= \mp (-1)^j A^{\pm 1}_j \nonumber \\ 
&\times
\begin{cases}
\frac{(-1)^{n_1}}{n_1} B_1^{n_1} & \ \mbox{if $w_n = n_1 S_{b_1}$ and $w_n \notin \Gamma_2$ with $\exists n_1 \in {\mathbb N}$} \\
\frac{(-1)^{n_2}}{n_2} B_2^{n_2} & \ \mbox{if $w_n = n_2 S_{b_2}$ and $w_n \notin \Gamma_1$ with $\exists n_2 \in {\mathbb N}$} \\
\frac{(-1)^{n_1}}{n_1} B_1^{n_1} + \frac{(-1)^{n_2}}{n_2} B_2^{n_2} & \ \mbox{if $w_n = n_1 S_{b_1} = n_2 S_{b_2}$ with $\exists n_1, \exists n_2 \in {\mathbb N}$} 
\end{cases}, \label{eq:DelA} \\ \nonumber \\
\bul{\Delta}_{w_n} B_j &= 0. \label{eq:DelB}
\end{align}
Therefore, from Eq.~(\ref{eq:Stokes_nu}), one can determine the action of the one-parameter Stokes automorphism on the cycle integrals as
\begin{gather}
{\mathfrak S}^{\nu}[A^{\pm 1}_1] = A^{\pm 1}_1\prod_{j=1}^2 (1+B_j)^{\mp \nu}, \qquad {\mathfrak S}^{\nu}[A^{\pm 1}_2] = A^{\pm 1}_2\prod_{j=1}^2 (1+B_j)^{\pm \nu}, \label{eq:St_auto_A_nu} \\
{\mathfrak S}^{\nu}[B_j] = B_j.
\end{gather}
Note that $S_{b_{1}} = S_{b_{2}}$ for $N=2$ and $w_n = n S_{b_{1,2}}~(n \in {\mathbb N})$. However, these equalities do not necessarily hold for a general $N$.

\subsection{DDP formula for wave functions} \label{sec:DDP_wave}

\begin{figure}[tp]
  \begin{center}
    \begin{tabular}{cc}
      \begin{minipage}{0.8\hsize}
        \begin{center} 
            \includegraphics[clip, width=100mm]{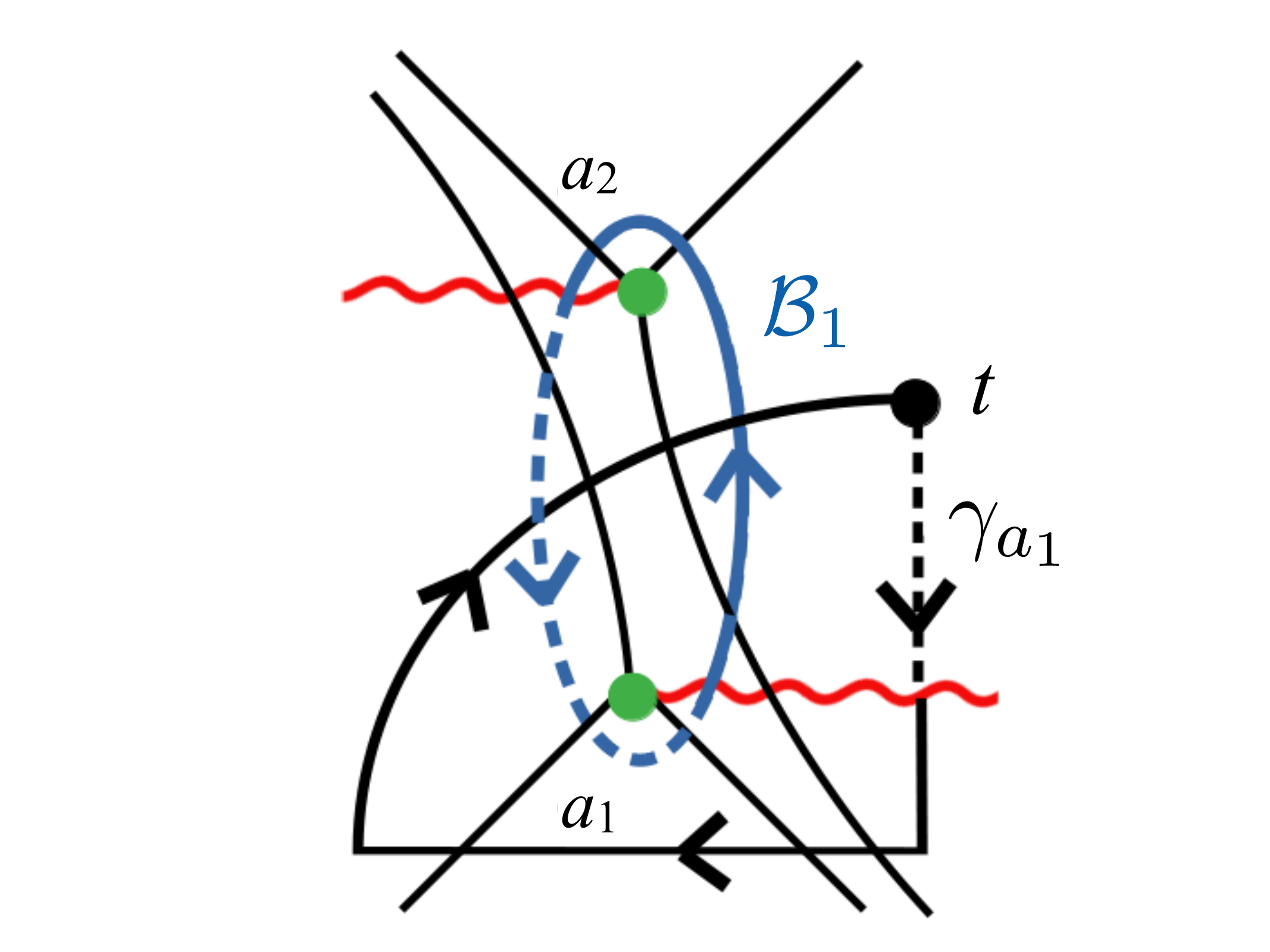}
        \end{center}
      \end{minipage}      
    \end{tabular} 
    \caption{Definition of the path of integration, $\gamma_a$.}
    \label{fig:path_gam}
  \end{center}
\end{figure}

When a degenerate Stokes curve emerges in a Stokes graph, the wave function typically becomes Borel non-summable, even at points away from the Stokes curve.
This phenomenon is due to the \textit{fixed singularity}.
In this part, we consider Stokes automorphism (i.e., DDP formula) of a wave function~\cite{Takei2,Iwaki1}.
This is also an important ingredient to construct an exact form of the monodromy matrix.

Consider the WKB solution~\eqref{eq:WKB_sol_appendix} normalized at a turning point $a_1$, and redefine the path of the integration as
\begin{align}
\pm \int_{a_1}^t dx \, S_{\rm odd}(x,\hbar) &\quad \rightarrow \quad \pm \frac{1}{2} \oint_{\gamma_{a_1}}^t dx \, S_{\rm odd}(x,\hbar), \label{eq:integ_gam}
\end{align}
where $\gamma_{a_1}$ is the path enclosing the points $t$ and $a_1$ (see Fig.~\ref{fig:path_gam}).
Since the wave function can be expressed as 
\begin{align}
\psi^{\pm}(t,a_1) = \frac{1}{\sqrt{S_{\rm odd}(t,\hbar)}} \exp \left[ \pm \frac{1}{2} \oint_{\gamma_{a_1}}^t dx \, S_{\rm odd}(x,\hbar) \right],
\end{align}
and the integration path has the intersection with the $B$-cycle, the DDP formula for the wave function $\psi(t,a_{1}) = (\psi^+(t,a_{1}),\psi^-(t,a_{1}))$ can be formulated in a similar way to the case of $A$-cycles:
\begin{align}
&& {\cal S}_{0_+}[\psi(t,a_{1})] =   {\cal S}_{0_-}[ \psi(t,a_{1})  \Sigma(B_1)] , \qquad {\mathfrak S}[\psi (t,a_{1})] = \psi(t,a_{1}) \Sigma(B_1),
\end{align}
where $\Sigma(B_1)$ is the matrix given by
\begin{align}
\Sigma(B_1) = 
\begin{pmatrix}
(1 + B_1)^{-1/2} && 0 \\
0 && (1 + B_1)^{+1/2}  \\
\end{pmatrix}.
\end{align}
Note that the exponent $\mp 1/2$ comes from the overall factor, $\pm 1/2$, in Eq.~(\ref{eq:integ_gam}).
The matrix $\Sigma(B)$ is Borel summable and invariant under the DDP formula, since it consists only of $B$-cycles, which is Borel summable as we discussed in Appendix~\ref{sec:DDP_cycle}.
The singularities on the Borel plane can be identified as
\begin{align}
\Gamma = \{ \, w_n = n S_{b_1} \in {\mathbb R}_+ \ | \ n \in {\mathbb N} \, \},
\end{align}
and one can find the alien derivative for the wave function as
\begin{align}
\bul{\Delta}_{w_n} \psi^{\pm}(t,a_1) = \mp \frac{(-1)^{n}}{2 n}  \psi^{\pm}(t,a_1) B_1^{n}.
\end{align}
Hence, the one-parameter Stokes automorphism is obtained as
\begin{align}
{\mathfrak S}^{\nu}[\psi(t,a_{1})] = \psi(t,a_{1}) [\Sigma(B_1)]^{\nu}. \label{Snu_psi}
\end{align}

As seen above, the DDP formula is determined by the turning point to normalize the wave function, so the resulting form in Eq.~(\ref{Snu_psi}) changes when a different turning point for the normalization is chosen.
However, the difference in the DDP formula among wave functions with different turning points can be compensated by taking into account the DDP formula of the normalization matrix, which is a part of $A$- and/or $B$-cycles.

\subsection{Monodromy matrix} 
\label{app:exact_monodromy}

We derive a formula for the monodromy matrix $M$. 
Here, we consider the case drawn in Fig.~\ref{fig:cycles1} and take $N=2$.
The generalization to general $N$ is straightforward and will be discussed in Appendix~\ref{app:2N_turning_points}.

The monodromy matrix $M$ depends on the choice of basis for the wave function. 
Here, we choose the pair of the median resummation of $\psi_{\rm P}^\pm(t,t_1)$ [see Eq.~\eqref{eq:powerseriessolution}] as the basis
\begin{align}
(\psi^{+}(t,t_1),\psi^{-}(t,t_1)) = ({\cal S}_{\rm med}[\psi_{\rm P}^{+}(t,t_1)],{\cal S}_{\rm med}[\psi_{\rm P}^{-}(t,t_1)]), 
\label{eq:basis_choice}
\end{align}
where $\psi_{\rm P}^\pm(t,t_1)$ are the power-series WKB solutions 
\begin{align}
\psi_{\rm P}^\pm(t,t_1) &= \frac{1}{Q_0(t)^{\frac{1}{4}}} \exp \qty( \pm \frac{1}{\hbar} \int_{t_1}^t \sqrt{Q_0(t')} \, dt' ) \sum_{n=0}^{\infty} \psi_{n}^\pm(t,t_1) \, \hbar^n .
\end{align}
In the following, we suppose that the wave functions are normalized at $t=t_1$.
Because of the choice of the basis in Eq.~\eqref{eq:basis_choice}, the power-series WKB solutions $\psi_{\rm P}^{\pm}(t,t_1)$ can be identified with $\psi_{\rm med}^{\pm}(t,t_1)$ defined by ${\cal S}_{\rm med}[\psi_{\rm med}^{\pm}(t,t_1)] = \psi^{\pm}(t,t_1)$. 

\begin{figure}[tp]
  \begin{center}
    \begin{tabular}{cc}
      \begin{minipage}{0.5\hsize}
        \begin{center} 
            \includegraphics[clip, width=80mm]{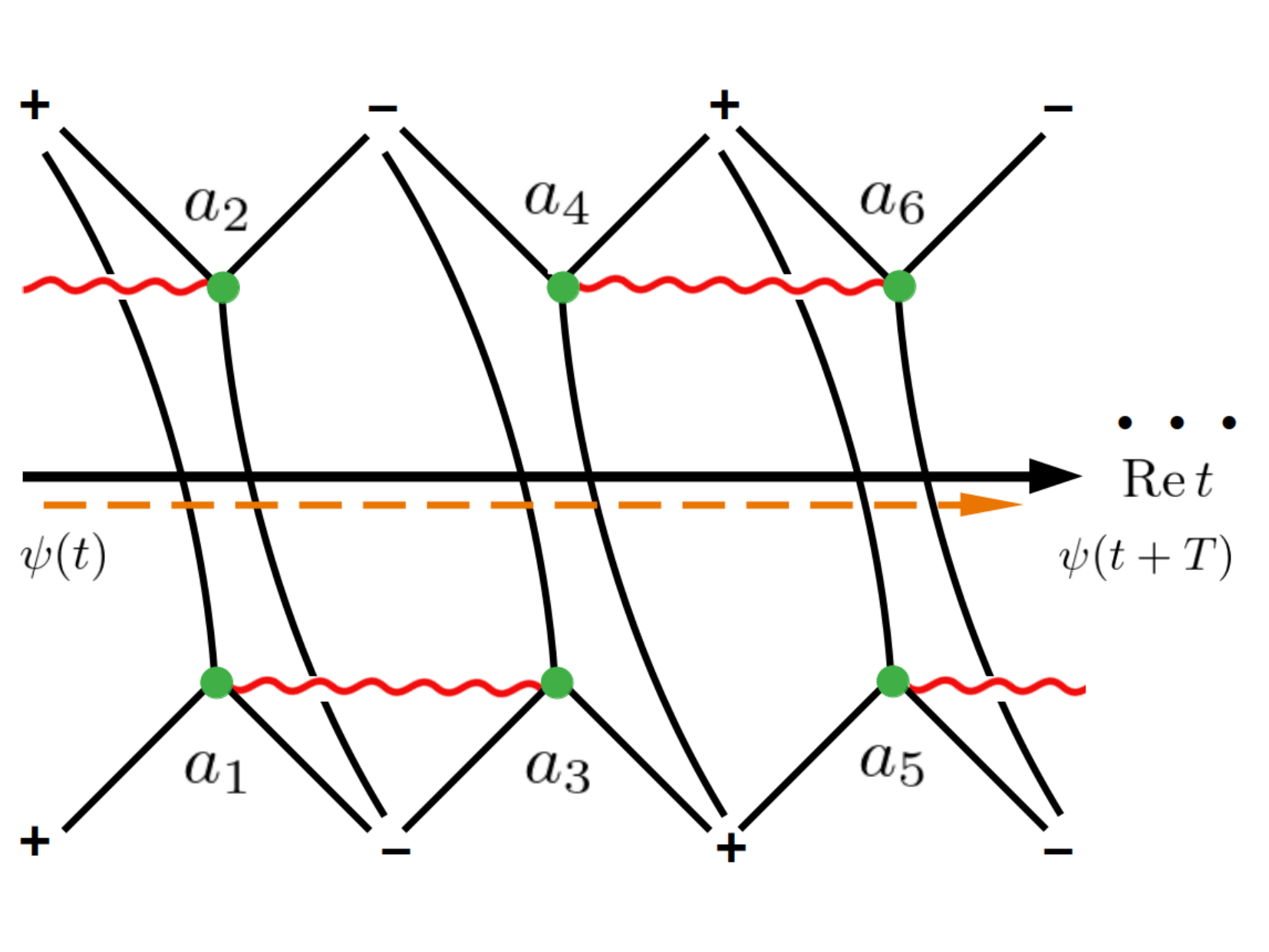} \vspace{-5mm} \\
(a) $\arg \hbar > 0$  
        \end{center}
      \end{minipage}
      \begin{minipage}{0.5\hsize}
        \begin{center} 
            \includegraphics[clip, width=80mm]{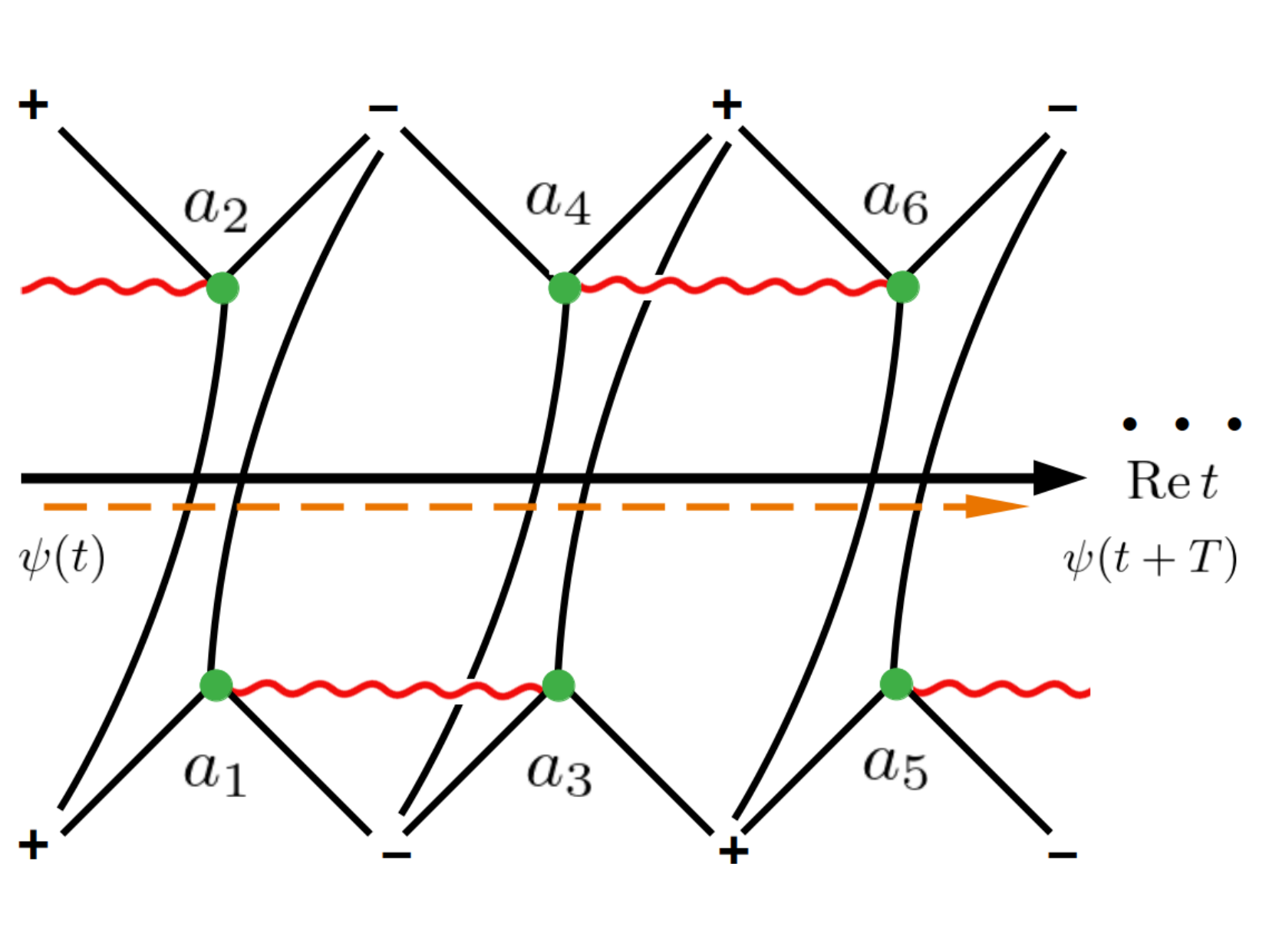} \vspace{-5mm} \\
(b) $\arg \hbar < 0$  
        \end{center}
      \end{minipage}      
    \end{tabular} 
    \caption{Resolved Stoke graphs in a Floquet system. 
    }
    \label{fig:stokes_graph graph}
  \end{center}
\end{figure}

Applying the connection formula for the non-degenerate Stokes curves in the graphs for $\arg \hbar > 0$ and $\arg \hbar < 0$ (see Fig.~\ref{fig:stokes_graph graph}), one can obtain monodromy matrices depending on the phase $\arg(\hbar) = 0_\pm$ as
\begin{align}
\bar{M}_{0_+} &= {\cal N}_{t_1 +T,a_4} ( T_{+}^{-1} {\cal N}_{a_4,a_3} T_{-} ) \, {\cal N}_{a_3,a_2} ( T_{+}^{-1} {\cal N}_{a_2,a_1} T_- ) \, {\cal N}_{a_1, t_1}, \label{eq:Monodromy_N2_p} \\
\bar{M}_{0_-} &= {\cal N}_{t_1+T,a_3} (T_{-} {\cal N}_{a_3,a_4} T_{+}^{-1}) \, {\cal N}_{a_4,a_1} (T_{-} {\cal N}_{a_1,a_2} T_{+}^{-1}) \, {\cal N}_{a_2, t_1},  \label{eq:Monodromy_N2_m}
\end{align}
where $T_{\pm}$ are the standard connection matrices associated with non-degenerate Stokes curves,
\begin{align}
T_{+} = 
\begin{pmatrix}
1 & 0 \\
-i & 1 
\end{pmatrix}, \qquad T_{-} = 
\begin{pmatrix}
1 & -i \\
0 & 1 
\end{pmatrix}, \label{eq:TpTm}
\end{align}
and ${\cal N}_{a_1,t_1}, \cdots$ are the normalization matrices, 
\begin{alignat}{3}
{\cal N}_{a_1,t_1} &= 
\begin{pmatrix} 
B_1^{-\frac{1}{4}} & 0 \\ 0 & B_1^{+\frac{1}{4}}
\end{pmatrix}, & \quad 
{\cal N}_{a_2,a_1} &= 
\begin{pmatrix} 
B_1^{+\frac{1}{2}} & 0 \\ 0 & B_1^{-\frac{1}{2}}
\end{pmatrix}, & \quad
{\cal N}_{a_3,a_1} &=
\begin{pmatrix}
A_1^{-\frac{1}{2}} & 0 \\ 0 & A_1^{+\frac{1}{2}}
\end{pmatrix}, \\
{\cal N}_{a_4,a_3} &= 
\begin{pmatrix} 
B_2^{+\frac{1}{2}} & 0 \\ 0 & B_2^{-\frac{1}{2}}
\end{pmatrix} , &\quad
{\cal N}_{a_6,a_4} &= 
\begin{pmatrix} 
A_2^{+\frac{1}{2}} & 0 \\ 0 & A_2^{-\frac{1}{2}}
\end{pmatrix}, &\quad
{\cal N}_{t_1+T,a_6} &= 
\begin{pmatrix} 
B_1^{-\frac{1}{4}} & 0 \\ 0 & B_1^{-\frac{1}{4}}
\end{pmatrix}.
\end{alignat}
The other normalization matrices can be constructed by using the relations
\begin{align}
{\cal N}_{t_1,t_3} = {\cal N}_{t_1,t_2} \, {\cal N}_{t_2,t_3}, \qquad {\cal N}_{t_1,t_2} = {\cal N}_{t_2,t_1}^{-1}. 
\end{align}
The matrices $\bar{M}_{0_\pm}$ are the monodromy matrices for the wave functions obtained by applying ${\cal S}_{0_\pm}$ to $\psi_{\rm P}(t,t_1) = (\psi_{\rm P}^+(t,t_1),\psi_{\rm P}^-(t,t_1))$: 
\begin{align}
{\cal S}_{0_\pm}[ \psi_{\rm P} (t+T,t_1)] = {\cal S}_{0_\pm} [\psi_{\rm P}(t,t_1) \bar{M}_{0_\pm}].
\end{align}
This relation can be rewritten in terms of the median resummation ${\mathcal S}_{\rm med}$ by using the relation ${\mathcal S}_{0_\pm} = {\mathcal S}_{\rm med} \circ {\mathfrak S}^{\pm \frac{1}{2}}$:
\begin{align}
{\cal S}_{\rm med} \Big[{\mathfrak S}^{\pm \frac{1}{2}} \big[ \psi_{\rm P} (t+T,t_1) \big] \Big] = {\cal S}_{\rm med} \Big[ {\mathfrak S}^{\pm \frac{1}{2}} \big[ \psi_{\rm P}(t,t_1)\bar{M}_{0_\pm} \big] \Big].
\end{align}
Using ${\mathfrak S}^{\pm \frac{1}{2}} [\psi_{\rm P}(t,t_1)\bar{M}_{0_\pm}] = {\mathfrak S}^{\pm \frac{1}{2}} [\psi_{\rm P}(t,t_1)] {\mathfrak S}^{\pm \frac{1}{2}} [\bar{M}_{0_\pm}]$ and ${\mathfrak S}^{\pm \frac{1}{2}}[ \psi_{\rm P}(t,t_1) ] = \psi_{\rm P}(t,t_1) \Sigma(B_1)^{\pm \frac{1}{2}}$, one can show that 
\begin{align}
{\cal S}_{\rm med}[ \psi_{\rm P} (t+T,t_1)] = {\cal S}_{\rm med} [ \psi_{\rm P}(t,t_1) M_{\rm med} ],
\end{align}
with 
\begin{align}
M_{\rm med} = \Sigma(B_1)^{\pm \frac{1}{2}} \,{\mathfrak S}^{\pm \frac{1}{2}} [\bar{M}_{0_\pm}] \, \Sigma(B_1)^{\mp \frac{1}{2}}. 
\end{align}
This is the matrix whose median resummation gives the 
monodromy matrix for $\psi(t,t_1)=(\psi^+(t,t_1),\psi^-(t,t_1))$:
\begin{align}
\psi(t+T,t_1) = \psi(t,t_1) M ~~~\mbox{with} ~~~ M = {\mathcal S}_{\rm med} [ M_{\rm med} ] . 
\end{align}
Using $P_i$ and $R_i$ defined as
\begin{align}
P_1 = A_1^{-1} B_1^{-\frac{1}{2}} B_1^{+\frac{1}{2}} , \qquad 
P_2 = A_2^{+1} B_1^{-\frac{1}{2}} B_1^{+\frac{1}{2}}, \qquad 
R_i = 
\begin{pmatrix}
\sqrt{1+B_i} & - i \sqrt{B_i} \\
+i \sqrt{B_i} & \sqrt{1+B_i}
\end{pmatrix},
\label{eq:def_PR}
\end{align}
the median-resummed monodromy matrix $M_{\rm med}$ can be explicitly written as
\begin{align}
M_{\rm med} = ( {\cal N}_{t_1+T,t_2} R_2 ) ( {\cal N}_{t_2,t_1} R_1 ) , 
\label{eq:Mmed}
\end{align}
where ${\cal N}_{t_2,t_1}$ and ${\cal N}_{t_1+T,t_2}$ are given by
\begin{align}
{\cal N}_{t_2,t_1} = 
\begin{pmatrix}
P_1^{+\frac{1}{2}} & 0 \\ 0 & P_1^{-\frac{1}{2}}
\end{pmatrix}, \qquad 
{\cal N}_{t_1+T,t_2} = 
\begin{pmatrix}
P_2^{+\frac{1}{2}} & 0 \\ 0 & P_2^{-\frac{1}{2}}
\end{pmatrix}. 
\end{align}
It is notable that the monodromy matrix~\eqref{eq:Mmed} has a different expression from Eq.~\eqref{eq:Mex}, since we take $t_0$ as the normalization point in the main text while it is $t_1$ here.
The formula~\eqref{eq:Mmed} means that the matrices $R_i~(i=1,2)$ are connection matrices for the degenerate Stokes curves. 
In the next subsection, we will generalize the formula \eqref{eq:Mmed} to more general cases.

\subsection{Generalization to 2{\it N} turning points}
\label{app:2N_turning_points}
\begin{figure}[tp]
  \begin{center}
    \begin{tabular}{cc}
      \begin{minipage}{1.\hsize}
        \begin{center} 
          \includegraphics[page=21, clip, width=110mm]{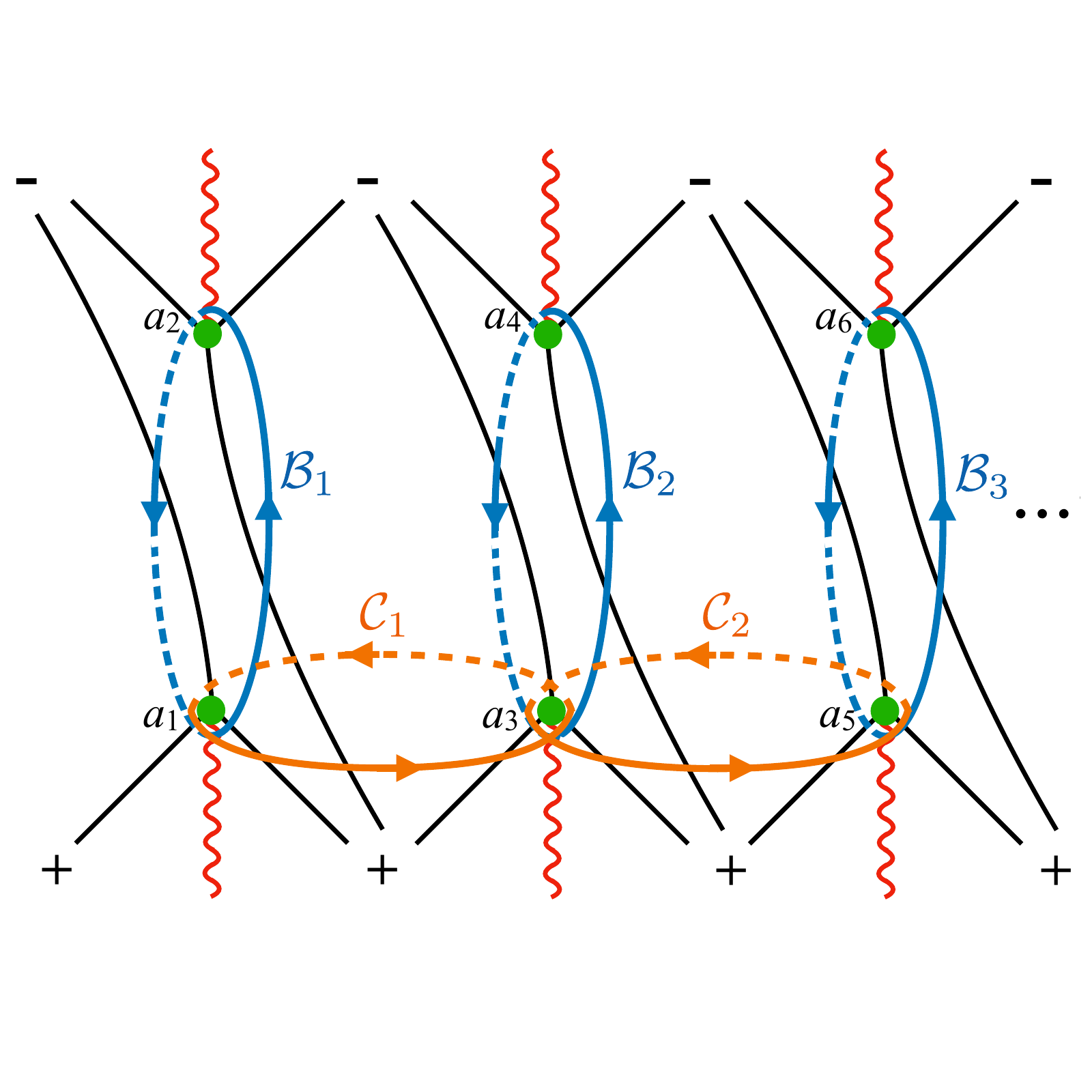}
        \end{center}
      \end{minipage}      
    \end{tabular} 
    \caption{The Stokes graph and the definition of cycles for the system with $2N$ turning points.}
    \label{fig:2N}
  \end{center}
\end{figure}

Here, we extend the formula for the monodromy matrix~\eqref{eq:Mmed} to the case of $2N$ turning points shown in Fig.~\ref{fig:2N}.
Following the same procedure outlined in Appendix~\ref{app:exact_monodromy}, the matrices $\bar M_{0_\pm}$ depending on the sign of $\arg h$ can be derived as 
\begin{align}
\bar M_{0_\pm} \ = \
\overset{\curvearrowleft}{\prod^{N}_{n=1}} ( {\cal N}_{t_{n+1},t_n} \bar R_{n,0_\pm}) \ = \ {\cal N}_{t_1+T,t_N} \bar R_{n,0_\pm} {\cal N}_{t_N,t_{N-1}} \bar R_{N-1,0_\pm} \cdots {\cal N}_{t_2,t_1} \bar R_{1,0_\pm} , 
\end{align}
where $\overset{\curvearrowleft}{\prod}$ denotes the left-product, ${\cal N}_{t_{n+1},t_n}$ are matrices given by
\begin{align}
{\cal N}_{t_{n+1},t_n} = \begin{pmatrix} 
P_n^{+\frac{1}{2}} & 0 \\ 0 & P_n^{-\frac{1}{2}}
\end{pmatrix} ~~~~ \mbox{with} ~~~~
P_{n} := C_n B_{n}^{-1/2} B_{n+1}^{+1/2}, \label{eq:Pn}
\end{align}
and $\bar R_{n,0_\pm}$ are the connection matrices for the degenerate Stokes curves resolved by introducing small positive and negative imaginary parts into the Planck constant: 
\begin{align}
\bar R_{n,0_+} \ &= \ {\cal N}_{t_n,a_{2n}} T_+^{-1} {\cal N}_{a_{2n},a_{2n-1}} T_- {\cal N}_{a_{2n-1},t_n}\,, \\
\bar R_{n,0_-} \ &= \ {\cal N}_{t_n,a_{2n-1}} T_{-} {\cal N}_{a_{2n-1},a_{2n}} T_{+}^{-1} {\cal N}_{a_{2n},t_{n}}\,.
\end{align}
Using the matrices $R_n$ given in Eq.~\eqref{eq:def_PR} 
and the explicit form of ${\cal N}$
\begin{align}
{\cal N}_{t_n,a_{2n}} = {\cal N}_{a_{2n-1},t_n} =
\begin{pmatrix}
B_n^{-\frac{1}{4}} & 0 \\ 0 & B_n^{+\frac{1}{4}}
\end{pmatrix}, 
 \qquad 
{\cal N}_{a_{2n},a_{2n-1}} = 
\begin{pmatrix} 
B_n^{+\frac{1}{2}} & 0 \\ 0 & B_n^{-\frac{1}{2}}
\end{pmatrix},
\end{align}
one can rewrite the matrices $\bar R_{n,0_\pm}$ as 
\begin{align}
\bar R_{n,0_\pm} = \Sigma(B_n)^{\pm \frac{1}{2}} \, R_n \, \Sigma(B_n)^{\pm \frac{1}{2}}.
\end{align}

As in the case of $N=2$, the matrix $M_{\rm med}$ can be obtained from $\bar M_{0_\pm}$ as
\begin{align}
M_{\rm med} &= \Sigma(B_1)^{\pm \frac{1}{2}} {\mathfrak S}^{\pm \frac{1}{2}}[ \bar M_{0_{\pm}} ] \Sigma(B_1)^{\mp \frac{1}{2}} \notag \\
& = \Sigma(B_1)^{\pm \frac{1}{2}} \overset{\curvearrowleft}{\prod^{N}_{n=1}} \left( {\mathfrak S}^{\pm \frac{1}{2}}[{\cal N}_{t_{n+1},t_n} ] \bar R_{n, 0_{\pm}} \right) \Sigma(B_1)^{\mp \frac{1}{2}}. 
\end{align}
From the DDP formula for $C_n$ 
\begin{align}
{\mathfrak S}^{\nu}[C_n] = C_n (1+B_{n})^{\nu}(1+B_{n+1})^{\nu},
\end{align}
we find 
\begin{align}
{\mathfrak S}^{\nu}[{\cal N}_{t_{n+1},t_n}] = \Sigma(B_{n+1})^{-\nu} {\cal N}_{t_{n+1},t_n} \Sigma(B_{n})^{-\nu} .
\end{align}
Thus, we obtain the following formula for the monodromy matrix $M_{\rm med}$:
\begin{align}
M_{\rm med} 
&= \overset{\curvearrowleft}{\prod^{N}_{n=1}} \left( {\cal N}_{t_{n+1},t_n} R_n \right), \label{eq:MMex_PB}
\end{align}
with 
\begin{align}
{\cal N}_{t_{n+1},t_n} R_n = 
\begin{pmatrix}
  P_n^{+1/2}  && 0  \\
  0 && P_n^{-1/2}
\end{pmatrix}
\begin{pmatrix}
  \sqrt{1+B_n} && -i \sqrt{B_n}  \\
  +i  \sqrt{B_n} && \sqrt{1+B_n}
\end{pmatrix}.
\end{align}
It is notable that the monodromy matrix~\eqref{eq:MMex_PB} has a different expression from Eq.~\eqref{eq:Mex_general} due to the choice of the normalization point.

\subsection{Median resummation of wave functions for a movable singularity} \label{appendix:alien_movable}

\begin{figure}[tp]
  \begin{center}
    \begin{tabular}{c}
      \begin{minipage}{1.\hsize}
        \begin{center} 
          \includegraphics[clip, width=130mm]{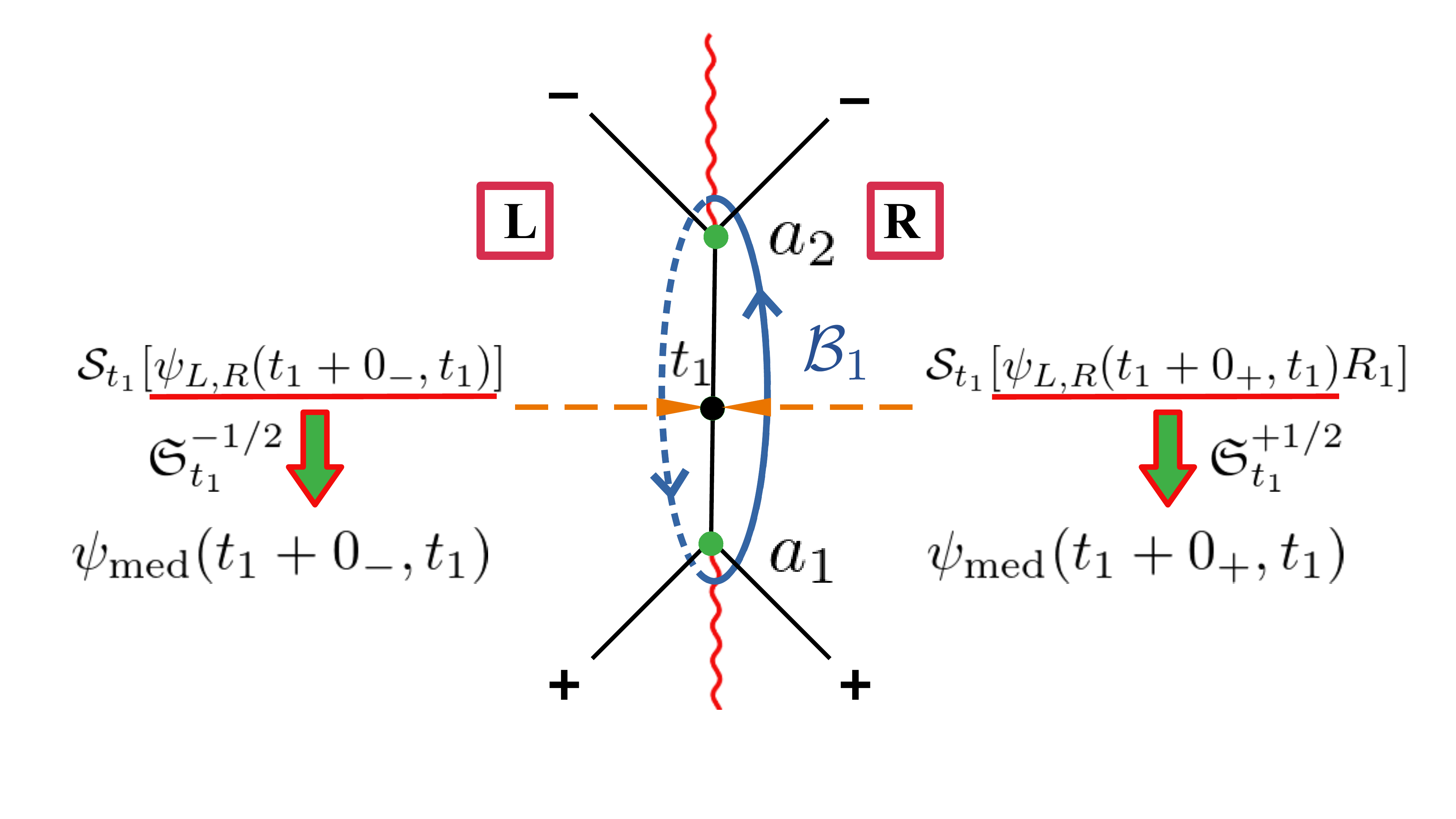}
        \end{center}
      \end{minipage}      
    \end{tabular} 
    \caption{The Stokes graph for a movable singularity.
    The trans-series of wave functions, denoted by $\psi_{R,L}(t_1 +0_\pm,t1)$, have a discontinuity at $t_1$, but the  Borel-resummed wave function can be continuous by taking into account the matrix, $R_1$ as Eq.~(\ref{eq:Stp_Stm}).    
    By appropriately operating the Stokes automorphism ${\mathfrak S}_{t_1}$ to the trans-series, the resulting trans-series wave function, $\psi_{\rm med}$, does not have a jump and is continuous at $t=t_1$.
    }
    \label{fig:med_movable}
  \end{center}
\end{figure}

In this part, we derive the median-resummed form of wave functions.
When an explicit form of a wave function is needed to calculate observables such as the effective Hamiltonian and the unitary matrix defined in Eq.~(\ref{eq:unitary_g}), one has to account for discontinuities arising not only from fixed singularities related to $\arg(\hbar)$ but also from movable singularities associated with the coordinate $t$.
The procedure for handling the latter case is almost parallel to the former, but it requires an additional Borel resummation.

Assuming that the discontinuities from fixed singularities in a trans-series of wave functions have been removed, we focus on the Stokes graph in Fig.~\ref{fig:med_movable}.
Let $\psi_{L}(t, t_1)=(\psi_{L}^+(t, t_1),\psi_{L}^-(t, t_1))$ and $\psi_{R}(t, t_1)=(\psi_{R}^+(t, t_1),\psi_{R}^-(t, t_1))$ be pairs of WKB wave functions normalized at $t = t_1$ that satisfy
\begin{align}
{\cal S}_{t_1 + 0_+}[\psi_{R}(t_1,t_1)] = {\cal S}_{t_1 + 0_-}[\psi_{L}(t_1,t_1)], \label{eq:Stp_Stm}
\end{align}
where ${\cal S}_{t_1 + 0_\pm}$ is the Borel resummations performed in the right and left domains relative to $t_1$.
Notice that the Borel resummations can also be expressed as ${\cal S}_{t_1 + 0_\pm}[\psi(t_1,t_1)] = {\cal S}_{t_1}[\psi(t_1+ 0_\pm,t_1)]$.

Our task is to construct a trans-series of $\psi(t,t_1)$ which does not have a discontinuity at $t=t_1$ from $\psi_{L,R}(t,t_1)$.
To do so, we define the Stokes automorphism and median summation associated with Eq.~(\ref{eq:Stp_Stm}) as
\begin{align}
& {\cal S}_{t_1 + 0_+} = {\cal S}_{t_1 + 0_-} \circ {\mathfrak S}_{t_1}^{+1}, \label{eq:Borel_t1} \\
& {\cal S}_{{\rm med},t_1}:= {\cal S}_{t_1 + 0_+} \circ {\mathfrak S}_{t_1}^{-1/2}= {\cal S}_{t_1 + 0_-} \circ {\mathfrak S}_{t_1}^{+1/2}. \label{eq:med_res_t1} 
\end{align}
The trans-series for the discontinuity-free wave function can be expressed using the Stokes automorphism as
\begin{align}
\psi_{\rm med}(t_1,t_1) = {\mathfrak S}_{t_1}^{+1/2} [\psi_{R}(t_1,t_1)] = {\mathfrak S}_{t_1}^{-1/2} [\psi_{L}(t_1,t_1)], \label{eq:med_wab_t1}
\end{align}
at $t=t_1$.
By repeating similar computations as in the previous subsection, one can show
\begin{align}
& {\cal S}_{t_1 + 0_+} [\psi_{R}(t_1,t_1)] = {\cal S}_{t_1 + 0_-}[\psi_{R}(t_1,t_1) R_1^{-1}] \nonumber\\
& {\cal S}_{t_1 + 0_+} [\psi_{L}(t_1,t_1) R_1] = {\cal S}_{t_1 + 0_-}[\psi_{L}(t_1,t_1)] \nonumber\\
& {\mathfrak S}^{\nu}_{t_1}[\psi_{L,R}(t_1,t_1)] = \psi_{L,R}(t_1,t_1) R_1^{-\nu}, \qquad (\nu \in {\mathbb R})
\end{align}
where $R_1$ is the matrix given in Eq.~(\ref{eq:RinB}).
Here, we used the fact that $R_1$ consists only of $B_1$ and hence ${\mathfrak S}^{\nu}_{t_1}[R_1]=R_1$.
From Eq.~(\ref{eq:med_wab_t1}), one finds 
\begin{align}
& \psi_{\rm med}(t_1,t_1) = \psi_{R}(t_1,t_1) R_1^{-1/2} = \psi_{L}(t_1,t_1) R_1^{+1/2}. \label{eq:psi_med_mov}
\end{align}
The matrix $R_0^{-1/2}$ in $G(0)$ in Eq.~(\ref{eq:G0_R0}) corresponds to $t=t_0$ in Fig.~\ref{fig:cycles1}.

The form in Eq.~(\ref{eq:psi_med_mov}) remains valid for $t$ away from $t_1$ until crossing the next Stokes curve, i.e.,
\begin{align}
& \psi_{\rm med}(t,t_1) = \psi_{R}(t,t_1) R_1^{-1/2} = \psi_{L}(t,t_1) R_1^{+1/2}, \qquad t_0 < t < t_2,
\label{eq:psi_med_t}
\end{align}
where $t_{0}, t_2 \in {\mathbb R}$ are points on the real axis intersecting adjacent Stokes curves.
We note that, in the main text, instead of $t_1$, we choose $t_0 = 0$ for the normalization point. Thus, the median-resummed versions of the monodromy matrix and the wave function have slightly different expressions by the Voros multiplier (the normalization matrix).

\section{Symmetry properties of the effective Hamiltonian}\label{appendix:symH}

The system discussed in Sec.~\ref{sec:4} possesses time-reversal symmetry, under which the Hamiltonian is invariant as
\begin{align}
\sigma_3 \, H(-t)^\ast \sigma_3 = H(t).
\label{eq:TS_inv}
\end{align}
In this appendix, we show that the effective Hamiltonian with the time-reversal symmetry~\eqref{eq:TS_inv} has an invariant property: 
\begin{align}
\sigma_3 \, H_{\rm eff}^\ast \, \sigma_3 = H_{\rm eff}.
\end{align}
To show this, we first prove the invariance of the time-evolution unitary matrix under time reversal: 
\begin{align}
\sigma_3 \, U(-t,0)^\ast \sigma_3 = U(t,0),
\label{eq:TR_U}
\end{align}
where $U(t,0)$ is the time-evolution unitary matrix $U(t,t_0)$ from $t_0=0$ to $t$, which satisfies 
\begin{align}
i \hbar \partial_t U(t,0) = H(t) U(t,0) ~~~ \mbox{with} ~~~ U(0,0) = \mathbf 1.
\label{eq:eq_U}
\end{align}
We can prove Eq.~\eqref{eq:TR_U} 
by showing that $\sigma_3 \, U(-t,0)^\ast \sigma_3$ satisfies Eq.~\eqref{eq:eq_U}: 
\begin{align}
i \hbar \partial_t (\sigma_3 \, U(-t,0)^\ast \sigma_3) = \sigma_3 (- i \hbar \partial_t U(-t,0) )^\ast \sigma_3 &= \sigma_3 ( H(-t) U(-t,0) )^\ast \sigma_3 \notag \\
&= (\sigma_3 \, H(-t)^\ast \sigma_3) (\sigma_3 \, U(-t,0)^\ast \sigma_3 ) \notag \\
&= H(t) (\sigma_3 \, U(-t,0)^\ast \sigma_3 ),
\end{align}
with the initial condition $\sigma_3 \, U(0,0)^\ast \sigma_3 = \mathbf 1$. Since the solution to the first-order linear differential equation with a given initial condition is unique, we conclude that the time-evolution unitary matrix is invariant under the time-reversal: $\sigma_3 \, U(-t,0)^\ast \sigma_3 = U(t,0)$. 

Next, we rewrite the time reversal invariance of the time-evolution unitary matrix~\eqref{eq:eq_U} as
\begin{align}
\sigma_3 \, U(T,T-t)^{\rm T} \sigma_3 = U(t,0), 
\end{align}
where we have used the periodicity $U(t+T,t_0+T)=U(t,t_0)$ and the inversion $U(t_0,t)^\dagger = U(t,t_0)$. 
By setting $t=T$ in this relation, we find that $U(T,0)$ satisfies
\begin{align}
\sigma_3 \, U(T,0)^{\rm T} \sigma_3 = U(T,0).
\end{align}
This implies that the effective Hamiltonian, defined as $H_{\rm eff} = i\hbar \log U(T,0)$, satisfies 
\begin{align}
\sigma_3 \, H_{\rm eff}^{\rm T} \, \sigma_3 = H_{\rm eff}. 
\end{align}
Given that the effective Hamiltonian is a hermitian matrix $H_{\rm eff}^{\rm T}=H_{\rm eff}^\ast$, it follows that the effective Hamiltonian satisfies the condition: $\sigma_3 \, H_{\rm eff}^\ast \, \sigma_3 = H_{\rm eff}$. 

\section{Quasi-energy in systems with elliptic oscillation}
\label{appendix:elliptic}
In this appendix, we extend the exact-WKB analysis to a one-parameter generalization of the system discussed in Sec.~\ref{sec:4}, where the time-dependent functions are given by
\begin{align}
f_1 =  v \cos \chi \sin \omega t , \hspace{5mm}
f_2 = -v \sin \chi \cos \omega t , \hspace{5mm}
f_3 = \Delta.
\end{align}
When $\chi=0$, this reduces to the system given in Eq.~\eqref{eq:example1}, where the oscillation in $\mathbb R \ni (f_1,f_2,f_3)$ follows a linear trajectory along the $f_1$-axis. For a generic value of $\chi$, the oscillation traces an elliptical path on the plane $f_3=\Delta$. In particular, at $\chi = \pi/4$, the oscillation becomes circular, and the system is exactly solvable. 
The case $\chi = \pi/4$ corresponds to the Hamiltonian obtained by applying the rotating-wave approximation to the original system in Eq.~\eqref{eq:example1}. 
This highlights the connection between the exact-WKB analysis and conventional approximations used in driven two-level systems, such as the Rabi problem.

Since $Q_0$ takes the same functional form as Eq.~\eqref{eq:Q_0etc},   
\begin{align}
Q_0(t) = - \tilde \Delta^2 - \tilde v^2 \sin^2 (\omega t) ~~~ \mbox{with~~~ $\tilde \Delta = \Delta^2 + v^2 \sin^2 \chi$ ~~ and ~~ $\tilde v = v \cos (2 \chi)$}, 
\end{align}
the Stokes graph remains identical to  Fig.~\ref{fig:stokes_undeformed}, 
allowing us to apply the monodromy-matrix formula~\eqref{eq:Mex} directly. 
The only differences arise in the expressions of $P_n$ and $B_n$, which are modified as follows: 
\begin{align}
P_n^{\pm \frac{1}{2}} &= \exp \left[ \pm \frac{2i \tilde \Delta}{\hbar \omega} E\left(- \frac{\tilde v^2}{\tilde \Delta^2} \right) \pm i \gamma\left(- \frac{\pi}{2} \right) \right] \left[ 1 \mp \frac{i\hbar \omega}{12\tilde \Delta} \tilde{\mathcal F} \left( \frac{\pi}{2} \right)  + \mathcal O(\hbar^2) \right], \\
B_n^{\pm \frac{1}{2}} &= \exp \left[ \pm \frac{2i \tilde \Delta}{\hbar \omega} E\left( {\rm arcsin} \frac{i\tilde \Delta}{\tilde v}, - \frac{\tilde v^2}{\tilde \Delta^2} \right) \pm i \gamma\left({\rm arcsin} \frac{i\tilde \Delta}{\tilde v} \right) \mp (-1)^n \frac{\pi i}{2} \right] \notag \\
& \hspace{48mm} \times \left[ 1 \mp \frac{i\hbar \omega}{12\tilde \Delta} \tilde{\mathcal F} \left( {\rm arcsin} \frac{i\tilde \Delta}{\tilde v} \right) + \mathcal O(\hbar^2) \right], \notag
\end{align}
where
\begin{align}
\tilde{\mathcal F}(x) &= F\left(x, - \frac{\tilde v^2}{\tilde \Delta^2} \right) - \left( 1 +\frac{v^2}{\tilde \Delta^2 + \tilde v^2} \sin^2 \chi \right) E\left(x, - \frac{\tilde v^2}{\tilde \Delta^2} \right), \\
\gamma (x) &= \frac{\Delta \cot \chi}{\tilde \Delta}  \,  \Pi\left(1-\cot^2\chi,x,-\frac{\tilde v^2}{\tilde \Delta^2}\right),
\end{align}
and $\Pi(n,x,m)$ is the elliptic integral of the third kind:  
\begin{align}
\Pi(n,x,m) = \int_0^x \frac{1}{1-n \sin^2 z}\frac{1}{\sqrt{1-m \sin^2 z}} dz.
\end{align}
The quasi-energy can be determined from Eq.~\eqref{eq:D_theta}. 
Fig.~\ref{fig:theta_npcorrected_chi}-(a) shows
the quasi-energy at $\chi = -\pi/8$, incorporating the first perturbative and non-perturbative corrections obtained from the exact-WKB analysis. As in Sec.~\ref{sec:4}, the results exhibit a good agreement with the numerical results.  
Fig.~\ref{fig:theta_npcorrected_chi}-(b) shows the quasi-energy at $\chi=-\pi/4$. In this case, non-perturbative gaps around $T=T_\ast^{(i)}~(i=1,2,\cdots)$ are absent in both numerical and exact-WKB results. 

At $\chi=-\pi/4$, the system becomes exactly solvable, allowing us to determine the time-evolution unitary matrix $U(t)$ explicitly: 
\begin{align}
    U(t) = 
    e^{-\frac{i}{2} \sigma_3 \omega t} \,
    \Xi \,
    e^{-\frac{i}{2} \sigma_3 \Omega t} \,
    \Xi^{-1},
\end{align}
where  
\begin{align}
    \Xi =
    \begin{pmatrix}
    i \sqrt{\frac{\hbar(\Omega-\omega)-2\Delta}{2\hbar \Omega}} & \sqrt{\frac{\hbar(\Omega+\omega)+2\Delta}{2\hbar \Omega}} \\
    \sqrt{\frac{\hbar(\Omega+\omega)+2\Delta}{2\hbar \Omega}} & 
    i \sqrt{\frac{\hbar(\Omega-\omega)-2\Delta}{2\hbar \Omega}}
    \end{pmatrix}, \hspace{10mm}
    \Omega = \frac{\sqrt{2v^2+(2\Delta+\hbar \omega)^2}}{\hbar}.
\end{align}
\begin{figure}[t]
\begin{minipage}{0.45\linewidth}
\centering
\includegraphics[width=70mm, page=22, bb=0 100 780 750]{floquet_fig4.pdf}
\caption*{(a) $\chi=-\pi/8$}
\end{minipage}
~
\begin{minipage}{0.45\linewidth}
\centering
\includegraphics[width=70mm, page=23, bb=0 100 780 750]{floquet_fig4.pdf}
\caption*{(b) $\chi=-\pi/4$}
\end{minipage}
\caption{Comparison of numerical results (dots) and the exact-WKB results (solid line) for the quasi-energy $\theta = \epsilon T/\hbar$, including next-to-leading-order perturbative and non-perturbative corrections. Parameters are set to $\Delta=1, v = 1.8, \hbar=1$ and (a) $\chi=-\pi/8$, (b) $\chi=-\pi/4$.}
\label{fig:theta_npcorrected_chi}
\end{figure}
The quasi-energy can be exactly determined from the condition $D(\theta)=\det (U(T)-e^{i\theta} \mathbf 1)=0$, where $D(\theta)$ is given by
\begin{align}
D(\theta) = 2 e^{i \theta} \left[ \cos \theta + \cos \left( \frac{\Omega T}{2} \right)\right].
\end{align}
A key observation is that these exact results do not contain any non-perturbative terms such as $e^{-c/(\hbar \omega)}$, indicating the absence of non-perturbative corrections at $\chi = -\pi/4$. This can be understood from the fact that, at $\chi=-\pi/4$, the parameter $\tilde v$ vanishes, leading to the disappearance of the time dependence in $Q_0(t)$. The absence of non-perturbative terms further supports the robustness of the exact-WKB formulation.

At $T=T_\ast$, the exact-WKB result truncated at the next-leading $\hbar$ in the perturbative sector does not give an accurate approximation at $\chi = -\pi/4$, since higher-order $\hbar$ corrections become more important in this case.
In this regime, the oscillation of the state vector exactly follows a well-known Rabi oscillation pattern, with the frequency $\omega_\ast = 2\pi/T_\ast$ that coincides with the resonance frequency observed in the standard Rabi oscillation.
This provides a natural connection between the exact-WKB analysis and the conventional Rabi oscillation framework, where the transition dynamics is fully described by a simple analytical solution.

\bibliographystyle{utphys}
\bibliography{EWKB.bib,Floquet.bib}

\end{document}